\documentclass[onecolumn]{article}

\usepackage[utf8]{inputenc}
\usepackage[T1]{fontenc}

\usepackage{amsmath}
\usepackage{amssymb}

\usepackage{graphicx} 
\usepackage{url}

\usepackage[normalem]{ulem}
\usepackage{comment}

\usepackage{xcolor}

\usepackage{upgreek}

\usepackage[a4paper,margin=2.7cm]{geometry}

\usepackage{booktabs}   
\usepackage{makecell}   
\usepackage{array}      

\usepackage[hidelinks]{hyperref}
\IfFileExists{orcidlink.sty}{\usepackage{orcidlink}}{\newcommand{\orcidlink}[1]{}}
\newcommand{\orcidauthorA}{0000-0001-9325-4672}
\newcommand{\orcidauthorB}{0000-0003-4536-4644}
\newcommand{\orcidauthorC}{0000-0001-8261-6236}
\newcommand{\orcidauthorD}{0009-0007-2835-2963}
\newcommand{\orcidauthorE}{0000-0002-6955-0321}
\newcommand{\orcidauthorF}{0009-0001-4240-6362}
\newcommand{\orcidauthorG}{0000-0002-2549-4401}
\newcommand{\orcidauthorH}{0000-0002-7839-2951}
\newcommand{\orcidauthorI}{0000-0002-2586-1021}
\newcommand{\orcidauthorL}{0000-0003-3808-963X}
\newcommand{\orcidauthorM}{0000-0003-2080-9010}
\newcommand{\orcidauthorN}{0000-0002-2921-2475}
\newcommand{\orcidauthorO}{0000-0003-0348-092X}

\newcommand{\orcidA}{\orcidlink{\orcidauthorA}}
\newcommand{\orcidB}{\orcidlink{\orcidauthorB}}
\newcommand{\orcidC}{\orcidlink{\orcidauthorC}}
\newcommand{\orcidD}{\orcidlink{\orcidauthorD}}
\newcommand{\orcidE}{\orcidlink{\orcidauthorE}}
\newcommand{\orcidF}{\orcidlink{\orcidauthorF}}
\newcommand{\orcidG}{\orcidlink{\orcidauthorG}}
\newcommand{\orcidH}{\orcidlink{\orcidauthorH}}
\newcommand{\orcidI}{\orcidlink{\orcidauthorI}}
\newcommand{\orcidL}{\orcidlink{\orcidauthorL}}
\newcommand{\orcidM}{\orcidlink{\orcidauthorM}}
\newcommand{\orcidN}{\orcidlink{\orcidauthorN}}
\newcommand{\orcidO}{\orcidlink{\orcidauthorO}}

\title{Development of a Cherenkov-Based Time-of-Flight\\ Detector Using Silicon Photomultipliers}

\author{%
Liliana~Congedo$^{1}$\orcidB{}, Giuseppe~De~Robertis$^{1}$\orcidC{}, Antonio~Di~Mauro$^{2}$\orcidO{}, Mario~Giliberti$^{1,3}$\orcidD{},\\
Francesco~Licciulli$^{1}$\orcidE{}, Antonio~Liguori$^{1,3}$\orcidF{}, Rocco~Liotino$^{1,3}$, Leonarda~Lorusso$^{1}$\orcidG{},\\
Mario~Nicola~Mazziotta$^{1,*}$\orcidA{}, Eugenio~Nappi$^{1}$\orcidM{}, Nicola~Nicassio$^{1,2,3,*}$\orcidH{},\\
Giuliana~Panzarini$^{1}$\orcidI{}, Roberta~Pillera$^{1}$\orcidL{} and Giacomo~Volpe$^{1,3}$\orcidN{}%
}
\date{} 

\begin{document}
\maketitle

\begin{center}
\small
$^{1}$\ Istituto Nazionale di Fisica Nucleare (INFN), Sezione di Bari, via Orabona 4, I-70126 Bari, Italy\\
$^{2}$\ CERN, European Organization for Nuclear Research, Esplanade des Particules 1, 1211 Geneva, Switzerland\\
$^{3}$\ Dipartimento di Fisica dell'Universit\`a e del Politecnico di Bari, via Amendola 173, I-70126 Bari, Italy
\end{center}

\noindent\textbf{Correspondence:}
\href{mailto:mazziotta@ba.infn.it}{mazziotta@ba.infn.it};
\href{mailto:Nicola.Nicassio@ba.infn.it}{Nicola.Nicassio@ba.infn.it}

\begin{abstract}
The aim of this work is to develop high precision Time-of-Flight (TOF) devices based on high refractive index solid Cherenkov radiators read out by silicon photomultipliers (SiPMs).
Cherenkov light is prompt and therefore ideal for reaching the intrinsic timing limits of TOF systems. By utilizing a thin, high-refractive-index radiator
a nearly instantaneous signal is generated by particles exceeding the Cherenkov threshold.  In order to achieve the ultimate time resolution, we carried out a rigorous optimization of the radiator material and geometry, alongside the efficiency of the optical coupling to the SiPM sensors.  The key factors limiting the time resolution were characterized by comprehensive Monte Carlo simulations, subsequently validated against experimental beam test data.
We assembled small-scale prototypes instrumented with various Hamamatsu SiPM arrays sensors with pitches ranging from 1.3 to 3 mm coupled with various window materials, such as fused silica and MgF$_2$, featuring various thickness values. The prototypes were successfully tested in beam test campaigns at the CERN-PS T10 beam line. The data were collected with a complete chain of front-end and readout electronics based on either the Petiroc 2A or the Radioroc 2 interfaced to a picoTDC to measure charges and times.  By comparing the time measurements with two SiPM arrays we were able to measure a time resolution better than 33.2 ps at the full system level with a charged particle detection efficiency of~100\%.
Our results demonstrate the expected performance benchmarks for the charged particle detection efficiency and time resolution and highlight the potential of the developed Cherenkov-based TOF detectors for next-generation particle identification systems.
\end{abstract}





\section{Introduction}
\label{sec:intro}

The detection of Cherenkov radiation is a well-established technique for charged particle identification (PID) across broad momentum ranges, most notably in Ring-Imaging Cherenkov (RICH) detectors~\cite{Nappi:2005mz}. 
The generation of Cherenkov photons occurs when a charged particle traverses a medium with a refractive index $n$ at a speed exceeding the phase velocity of light in the medium, $\beta > 1/n$. 
Photons are promptly emitted ($\approx$~ps) along a cone with an opening angle $\theta_C$ defined by $\cos \theta_C = (n(\lambda)\beta)^{-1}$.

The intrinsic promptness of the Cherenkov effect is also advantageous for applications in TOF devices.
It has already been shown~\cite{Credo:2004qgy,Inami:2006cp,Krizan:2008zz,Vavra:2009qpk,Albrow:2012ha} that for charged particles traversing a radiator slab made of a high refractive index material, such as fused silica (SiO$_2$), at a speed exceeding the threshold for Cherenkov emission, the resulting photons generate a fast, well-timed signal in photon sensors like Microchannel Plate Photomultipliers (MCP-PMTs).

A very promising photon sensor option to MCP-PMTs, in the Cherenkov-based TOF devices, are Silicon Photomultipliers (SiPMs) for  their excellent key metrics as the high intrinsic single photon time resolution (SPTR) and photon detection efficiency (PDE) in the visible and near-ultraviolet (NUV) region of the spectrum, the insensitivity to magnetic fields, and the low material budget~\cite{Gundacker:2020cnv}.

Previous studies have already investigated the use of SiPMs for the direct detection and timing of incident charged particles~\cite{Paper_bolognesi_Carnesecchi}.   
In such cases, the signals originate from Cherenkov photons emitted in the SiPM protective resin layer.
However, these signals are typically confined to the single SiPM where the interaction occurs. This confinement, combined with the limited packaging factor of SiPMs in arrays, significantly reduces their overall efficiency for the direct detection of the impinging charged particles. 

To overcome this limitation, we explored the approach of coupling directly to a SiPM array a thin slab of transparent material (window), acting as a Cherenkov radiator for the incident charged particles, and resulting in clusters of contiguous fired SiPMs in the array. 
With a proper optimization of the radiator material and thickness, the geometry of the SiPM array, the optical couplings between the window and the SiPM array, and the front-end electronics, the proposed approach enables the achievement of time resolutions down to tens of ps with a 100\% charged-particle detection efficiency. 

In this paper, we discuss the key factors affecting the time resolution and detection efficiency of the proposed approach by comparing the results obtained for various window materials and thicknesses, SiPM layouts, and off-the-shelf read-out electronic components.
First, a comprehensive Monte Carlo simulation framework is presented, with emphasis on the interplay between radiator geometry, photon transport, and the intrinsic timing characteristics of the SiPMs.
Subsequently, a characterization and optimization study of the SiPM anti-reflective coating is reported, aimed at quantifying and mitigating signal-photon losses induced by reflections within the relevant range of photon polarization, wavelength, and incidence angle of the impinging charged particles.
Finally, the experimental setup of  dedicated beam-test campaigns is described, outlining the prototype architecture and benchmarking the measured performance against simulation results.
This discussion highlights the impact of key detector parameters on the achieved overall time resolution, thereby establishing a reference for the development of next-generation Cherenkov-based timing systems for charged-particle identification.

\section{Materials and Methods}
\label{sec:meth}

\begin{figure}[!t]
    \centering
    \includegraphics[clip, scale=0.8]{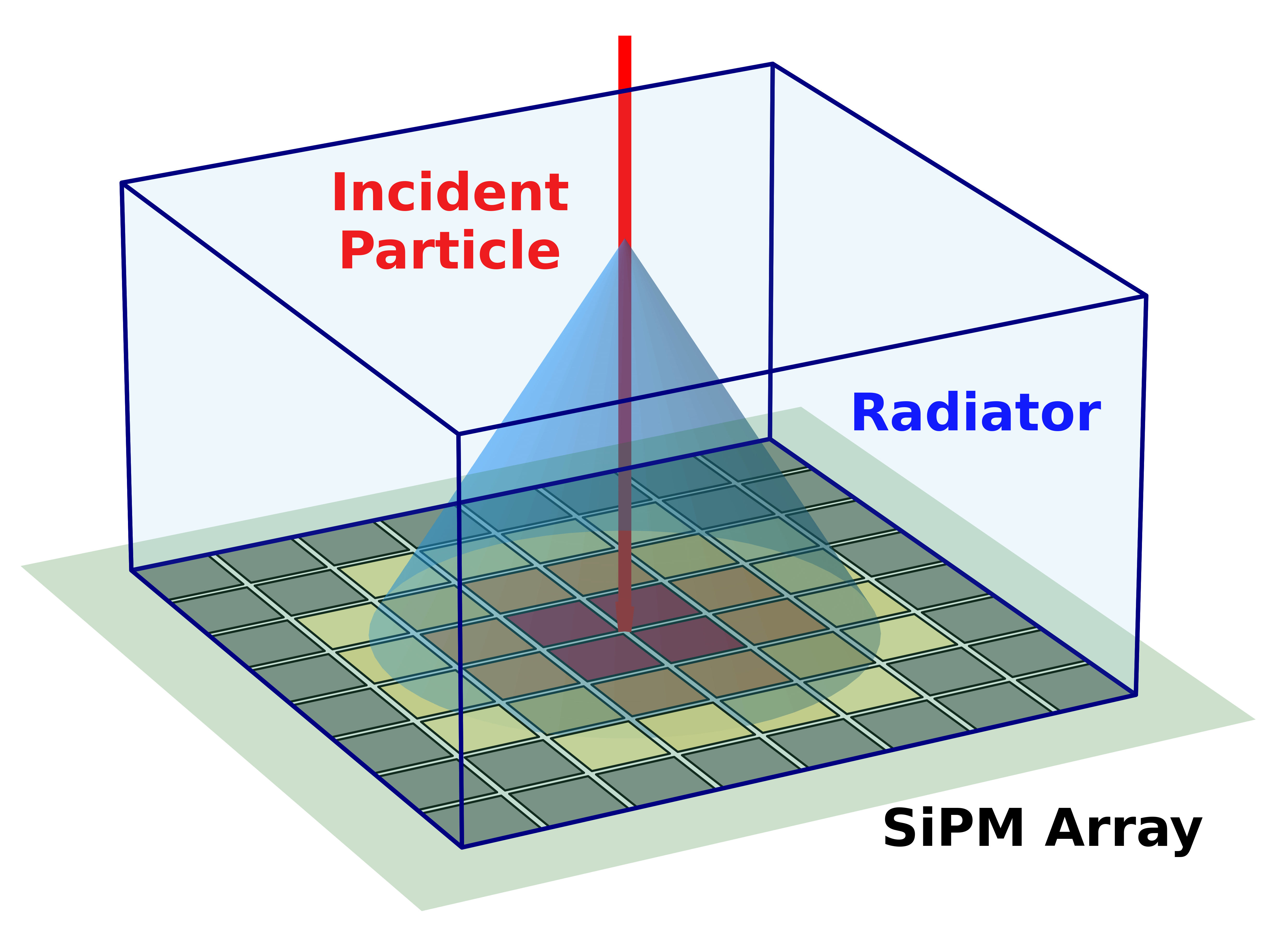}
    \caption{Three-dimensional model of the radiator-sensor coupling. The incident particle track is shown passing through the radiator. The Cherenkov radiation is projected onto the SiPM array layer, highlighting the spatial correlation between the particle trajectory and the fired pixels. The red-orange area corresponds to the core of the emission, assuming negligible reflections at the radiator boundaries.}
    \label{fig:ChRadSiPM}
\end{figure}

The principle of operation of the proposed Cherenkov-based timing approach is illustrated in Figure~\ref{fig:ChRadSiPM}, showing the radiator coupled to the SiPM array and the Cherenkov cone from a charged particle traversing the radiator at a speed exceeding the threshold for Cherenkov emission.
The emitted photons result in a
localized cluster of hits in the array of SiPMs. 
The topology of these clusters in terms of radius, average number of fired SiPMs, and distribution of photoelectrons is determined by the slab refractive index, transmittance and thickness, as well as by the SiPM granularity and PDE.

With a proper tuning of the radiator refractive index and thickness, the spatial extension of the Cherenkov cone effectively mitigates the impact of dead areas between adjacent SiPMs in the array and between the Single-Photon Avalanche Diodes (SPADs) in the same SiPM, ensuring a 100\% detection efficiency across the entire sensitive surface.

The track timing is then obtained by combining the timestamps of the SiPMs in the cluster. The precision of each individual measurement is determined by the intrinsic resolution of the SiPMs and the contributions of the electronics jitter, the  time-to-digital converter (TDC) resolution and the time reference system.
This method offers two advantages.
\begin{itemize}
    \item In the cluster core, where many photoelectrons are expected, the pixel with maximum charge yields the best intrinsic time resolution and the lowest time jitter.
    \item By combining the timestamps of those SiPMs in the cluster core with a sufficiently high number of photoelectrons, the overall time precision could be further improved.
\end{itemize}
As a result, track time resolutions at the level of a few tens of ps can be achieved.  
 
\subsection{Candidate radiator materials}

The choice of the radiator material is primarily driven by the particle species and momentum range of interest for TOF measurements, typically requiring high detection efficiency down to the lowest possible momenta. In the proposed Cherenkov-based timing approach, this requirement is primarily addressed by minimizing the Cherenkov emission threshold for the different particle species, ensuring that particles reach Cherenkov angle saturation at the lowest possible momenta, while maximizing the emission angle.

For a particle with mass $m$, the momentum threshold in a material with refractive index $n$ is $p_{\text{th}}=m/\sqrt{n^2-1}$.
Materials with high refractive indices are characterized by low momentum thresholds $p_{\text{th}}$, large Cherenkov emission angles $\theta_\text{c}$, and high photon yields $N_\text{ph}$, which scale as $N_\text{ph} \propto \sin^2 \theta_ \text{c}$, thus leading to large clusters. 

Possible materials for the sensor entrance window are NaF ($n \approx 1.33$), MgF$_2$ ($n \approx 1.38$), and SiO$_2$ ($n \approx 1.47$). These materials provide excellent NUV transparency, good radiation hardness, and favorable optical coupling to common SiPM protective coatings, typically made of either silicone ($n \approx 1.41$) or epoxy ($n \approx 1.55$) resins, using a thin layer (<100 $\upmu$m) of optical grease or adhesive for gluing the window to the SiPM array.
The quoted refractive index values are reported at a wavelength of $400$ nm. For radiators based on these materials, the Cherenkov‑emission threshold is below $1~\text{GeV}/c$ for protons, and substantially lower for lighter~species.

Our choice for SiO$_2$ as radiator material was driven by several key features. First, because it allows for a precise control of the thickness over the radiator surface, which is a critical parameter in tuning the Cherenkov cone geometry and cluster size on the SiPM array.
Moreover, its excellent radiation hardness ensures stable performance even in the high-luminosity environments typical of modern particle accelerators, preventing the degradation of transparency.

Although materials with $n>1.6$ enable Cherenkov emission by particles with a lower momenta than those traversing fused silica radiators, their use for this application has several drawbacks. The optical coupling to the SiPM is more challenging due to the large mismatch of the refractive indices of the various interfaces. The resulting effect is a severe loss of photons because of the multiple reflections and a larger background represented by photons experiencing total-internal reflections.

\subsection{Contributions to the time resolution}

\label{sec:contributions_time_resolution}

The total time resolution at the single-SiPM level results from the convolution of several physical and instrumental contributions, including the Cherenkov emission process, signal formation in the SiPMs, and the response of the electronics~\cite{Mazziotta:2026vor,VINKE2009188_contributi_sigma}:
\begin{equation}
    \sigma_\text{t}^2 = \sigma_{\text{geom}}^2 + \sigma_{\text{SiPM}}^2 + \sigma_{\text{FE}}^2 + \sigma_{\text{TDC}}^2
    \label{eq:risoluzione_generica}
\end{equation}

\begin{itemize}
\item \textbf{Geometric Spread}, $\sigma_{\text{geom}}$: Photons travel different paths from the emission point to the sensor. For a radiator of thickness $d$ and normally incident charged particles, the maximum time spread is $\Delta t_{\text{max}} = d\,(\beta^2 n^2 - 1)/\beta c$, leading to a contribution to the resolution $\sigma_{\text{geom}} \propto d/\sqrt{12}$. This term also accounts for the chromatic dispersion of the radiator in the spectral region to which the SiPMs are sensitive, with the wavelength dependence of $n(\lambda)$ broadening the arrival time distribution.
\item \textbf{Intrinsic SiPM Jitter}, $\sigma_{\text{SiPM}}$: 
This contribution originates from the stochastic nature of photoelectron creation, charge multiplication~\cite{Riegler:2021xtq}, and the spread in the charge transit time across multiple SPADs.
For SiPMs with a given SPTR, the intrinsic  time resolution depends on the number of photoelectrons $N_{\text{PE}}$ as $\sigma_{\text{SiPM}} \approx \text{SPTR} / \sqrt{N_{\text{PE}}} \propto 1/\sqrt{d}$. Both the SPTR and $N_{\text{PE}}$ improve with the increasing operation overvoltage. 
Moreover, SPTR depends on the total capacitance of the SiPM, which is proportional to the area of the photo-sensitive part of the SiPM and
the number of SPADs. 
\item \textbf{Front-End Jitter}, $\sigma_{\text{FE}}$:
This term accounts for the time uncertainty introduced by the front-end electronics in the timestamp reconstruction. It is primarily driven by the electronic noise and by the finite signal rise time: noise causes fluctuations in the time pick-off (e.g.\ threshold crossing), and the effect is larger when the signal slope at the pick-off point is smaller. Since the signal amplitude and slope scale with the collected charge, this contribution scales approximately as $\sigma_{\text{FE}} \propto 1/N_{\text{PE}} \propto 1/d$.
\item \textbf{TDC Quantization Uncertainty}, $\sigma_{\text{TDC}}$: This term accounts for the finite time binning of the TDC. Assuming a uniform quantization error within one bin, the corresponding contribution is $\sigma_{\text{TDC}}=\mathrm{LSB}/\sqrt{12}$, where LSB is the least significant bit.
\end{itemize}
While a thinner radiator minimizes the geometric spread ($\sigma_\text{geom} \propto d$), it reduces the $N_{\text{PE}}$, resulting in a degradation of the $\sigma_{\text{SiPM}}$ and $\sigma_{\text{FE}}$. Additional instrumental contributions, such as the finite precision of the reference time ($t_0$) and the clock stability, can also broaden the measured time distributions and therefore affect the overall resolution.

When the Cherekov photons emitted in the radiator fire more SiPMs in the array, $N_{\text{SiPM}}$, the independent measurement of the time for each SiPM will enable to improve the time resolution ideally by a factor $1/\sqrt{N_{\text{SiPM}}}$. However, when combining the timestamps from several SiPMs with different $N_{\text{PE}}$, time walk effects introduced by leading-edge discrimination must be corrected, since signals with smaller amplitude exhibit a reduced slew rate and therefore cross the discriminator threshold at a later time, biasing the measured timestamp.
After applying the time-walk correction, the particle time of arrival can be estimated as the average of the corrected timestamps measured by the SiPMs in the cluster.

\subsection{Monte Carlo simulation optimization studies}

\label{sec:Monte_Carlo_simulations_contributions}

We implemented a comprehensive Monte Carlo simulation for design optimization studies. This section illustrates the results obtained assuming SiO$_2$~\cite{1965JOSA...55.1205M} as radiator material for ultra-relativistic ($\beta\approx1$) particles at normal incidence. 
We generated Cherenkov photons in the wavelength interval from 260 to 900~nm using Frank-Tamm formula (Eq.~34.44 of Ref.~\cite{ParticleDataGroup:2024cfk}).
The PDE of Hamamatsu S13360 SiPMs~\cite{s13660-3075CS} and a SPTR of 100 ps was assumed for the sensors in the array.
Pixel sizes of $1.3\times1.3$~mm$^2$, $2\times2$~mm$^2$ and $3\times3$~mm$^2$ with a 200~$\upmu$m gap between each pixel were considered.
SPAD pitches of 50~$\upmu$m and  75~$\upmu$m were assumed, corresponding to a PDE of 40\% and 50\%, respectively, at a wavelength of 450~nm for operation at an overvoltage of $3$~V.
The contribution from the readout chain, $\sigma_{\text{FE}}$, $\sigma_{\text{TDC}}$, and other instrumental effects, was modeled as $\sigma_{\text{ele}} = 50\;\text{ps}/N_{\text{PE}} \oplus 20\;\text{ps}$.

\begin{figure}[!t]
 \centering
 \includegraphics[width=0.49\columnwidth]{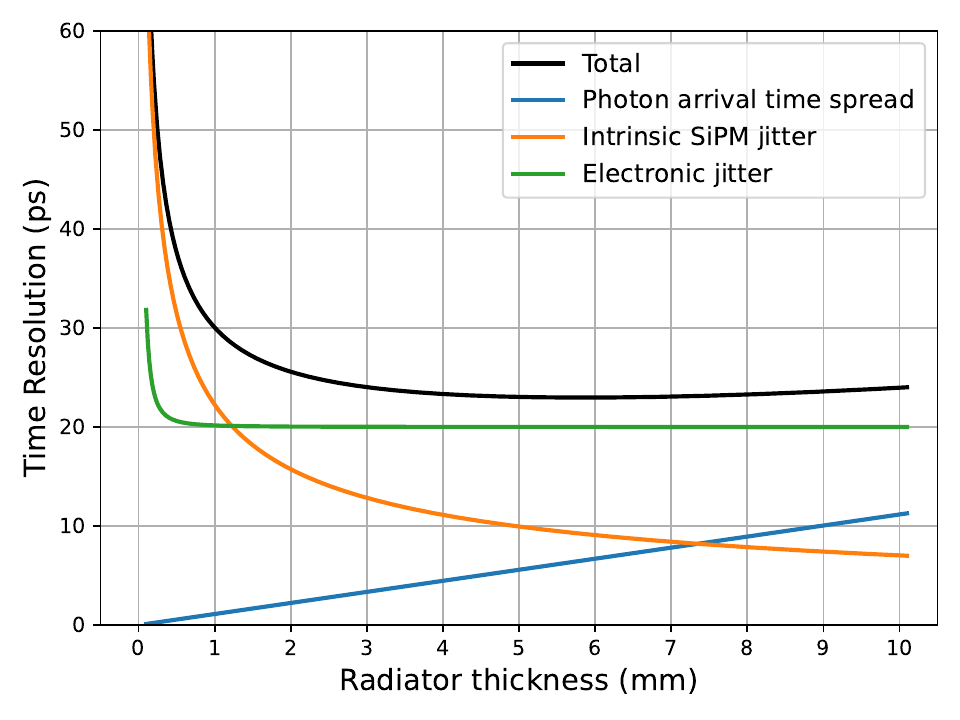}
 \includegraphics[width=0.49\columnwidth]{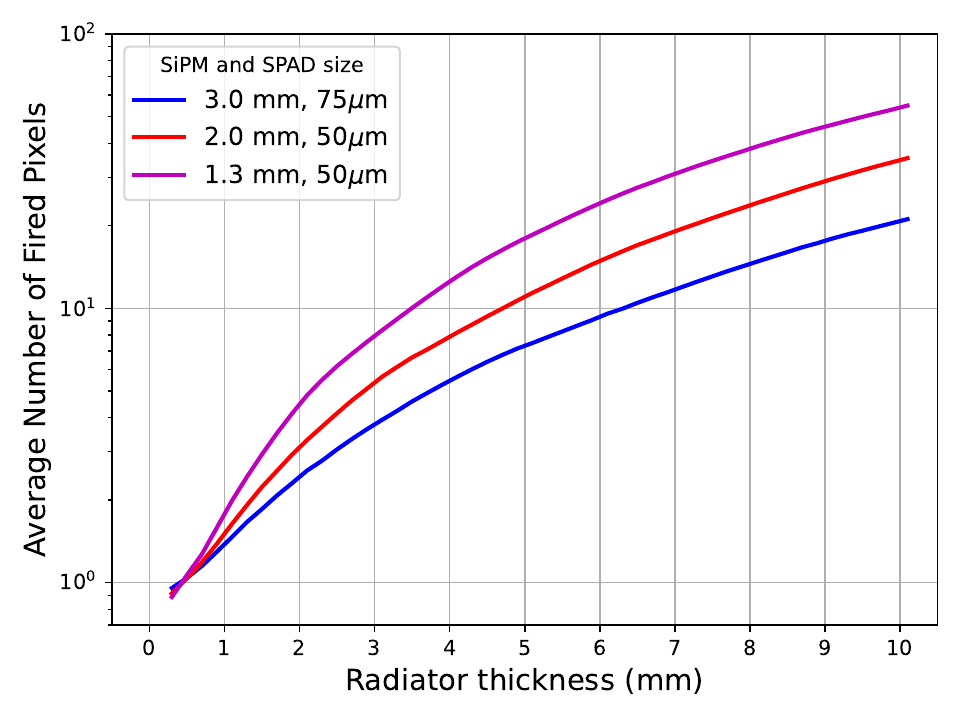}
 \includegraphics[width=0.49\columnwidth]{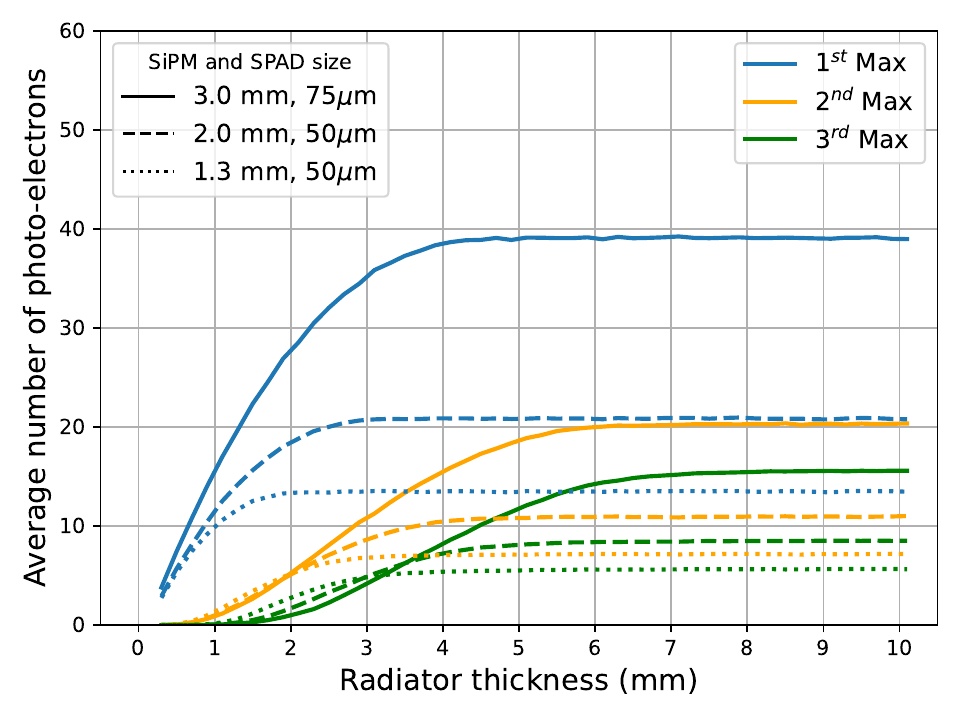}
 \includegraphics[width=0.49\columnwidth]{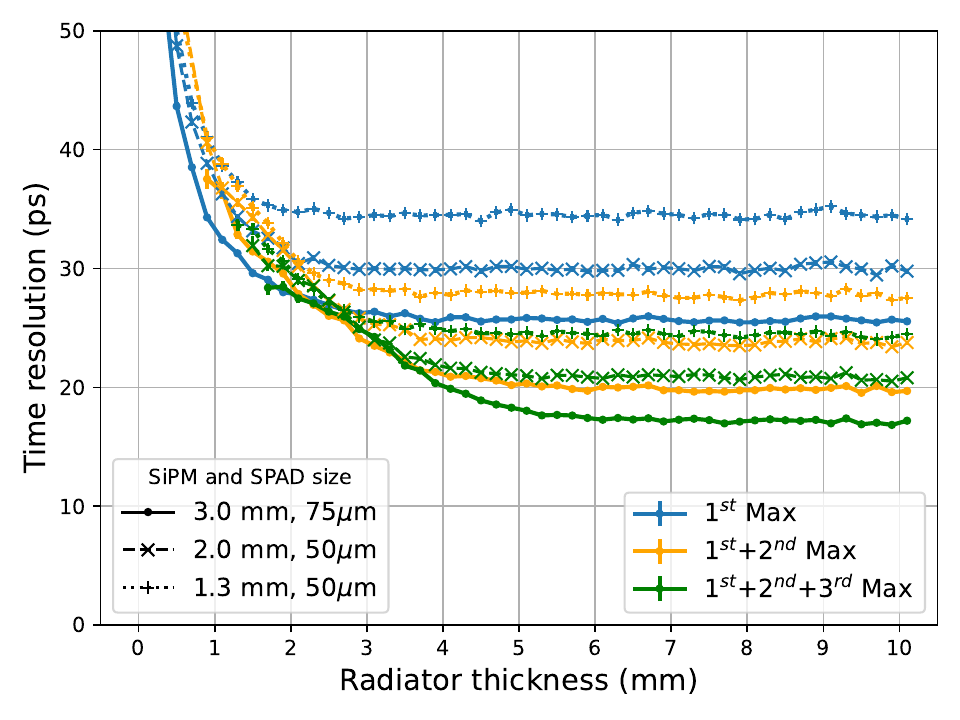}
 \caption{Top left: Expected time resolution as a function of the SiO$_2$ radiator thickness. The colored curves represent the individual contributions to the total time resolution. 
 Top right: Average number of fired pixels as a function of the radiator thickness. Bottom left: Average number of photoelectrons collected in the three pixels with the highest hit probability, shown as a function of the radiator thickness. Bottom right: Total time resolution as a function of the radiator thickness, computed using the timestamps from the three highest‑charge channels. Three SiPM configurations are investigated: 3 mm active area with a $75\;\upmu\text{m}$ SPAD size, 2 mm active area with a $50\;\upmu\text{m}$ SPAD size, and 1.3 mm active area with a $50\;\upmu\text{m}$ SPAD size.}
 \label{fig:sigt_d}
\end{figure}

The top-left panel of Figure~\ref{fig:sigt_d} presents the total time resolution and the individual contributions as a function of radiator thickness, including the geometric spread, intrinsic SiPM jitter, the contribution of the electronics. The calculation assumes an infinitely extended SiPM active area for the collection of photons and a SPAD pitch of $75\;\upmu\text{m}$. This dependence highlights the intrinsic design trade-off: reducing the radiator thickness decreases the geometric time spread ($\sigma_{\text{geom}} \propto d$), yet simultaneously lowers the number of detected photons, thereby worsening the photoelectron-statistics term ($\sigma_{\text{SiPM}} \propto 1/\sqrt{d}$) and increasing the $N_{\text{PE}}$-dependent electronic jitter. 

The top-right panel of Figure~\ref{fig:sigt_d} reports the average number of fired pixels as a function of radiator thickness for the SiPM technologies under consideration accounting for the actual SiPM size for photon collection. The bottom-left panel of Figure~\ref{fig:sigt_d} shows the corresponding average number of photoelectrons collected in the three pixels with the highest signal amplitude. The bottom-right panel of Figure~\ref{fig:sigt_d} presents the resulting total time resolution, obtained by averaging the photon-arrival times from those three pixels.
The results indicate that a total time resolution better than $40\;\text{ps}$ can be achieved for radiator thicknesses of $d \ge 1\;\text{mm}$ with the SiPM devices and electronic parameters considered in this study. It is worth noting that the ultimate performance is limited by the timing capabilities of the readout electronics implemented in the simulation.

\subsection{Photon reflections at the SiPM interface}

\label{sec:reflection_background}

Photon reflections at the interface between the SiPM and the radiator have an impact on the overall PDE, which directly influences the number of produced photoelectrons, generating fluctuations in the photo-statistics and therefore in the timing resolution. 
At the same time, photons reflected by a SiPM might reach and be detected by farther SiPMs potentially resulting in delayed and displaced background hits.

\begin{figure}[!t] 
\centering
{\includegraphics[width=0.99\columnwidth]{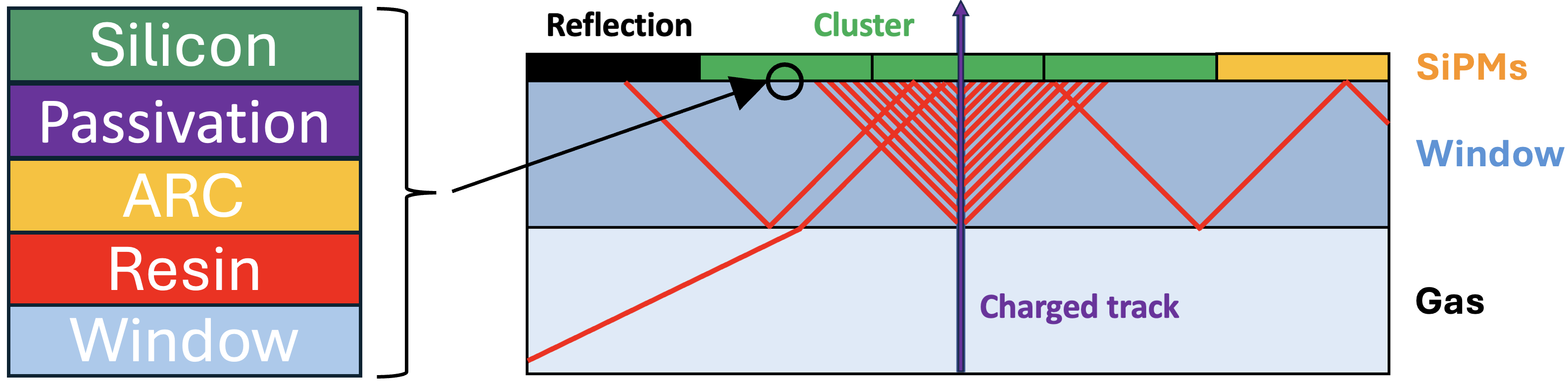}}
\caption{Schematic view of the detector stack, including the environmental gas, the window, and the SiPMs. The SiPM structure includes the protective resin, the ARC, the passivation layer, and silicon.}
\label{plot_ARC_schematic}
\end{figure}

Fresnel reflections arise at the interfaces between the environmental gas in the expansion gap, the sensor window, and the successive layers of the SiPM structure, including the protective resin, the anti-reflective coating (ARC), the passivation layer, and silicon, as illustrated in Figure~\ref{plot_ARC_schematic}. Diffuse reflections are also expected because of the microstructures on the SiPM surface, such as quenching resistors, trenches and traces used to connect the SPADs, as well as from regions between adjacent SiPMs in the array~\cite{WANG2020164171}. 
Although the background from reflected photons can be suppressed by operating at a sufficiently high threshold, optimizing the optical coupling between the radiator window and the SiPM array could enhance the detection of the Cherenkov photons.

Since little information regarding the layout and material composition of ARCs is available from the SiPM vendors, to characterize the SiPM reflectance, we performed dedicated measurements with commercial arrays featuring various pixel pitches and active areas, as well as various protective resins and radiator materials.
We measured the corresponding total, diffuse and specular reflectance.
In addition, we developed a simple model (Appendix~\ref{appending_model_reflections}) to explain and compare the various contributions to the measurement results. 
Using this model, we also extrapolated the wavelength dependence of the reflectance resulting from the built-in ARC of the SiPMs in the tested arrays.

The specifications of the tested arrays for reflectance measurements are summarized in Table~\ref{table_SiPM_reflectance_samples}. 
\begin{table}[!t]
\centering
\small            
\setlength{\tabcolsep}{4pt}   
\renewcommand{\arraystretch}{1.05} 
\begin{tabular}{l c c c c c c c c} 
\toprule
\makecell{\makecell{Array\\model}} & \makecell{Number\\of SiPMs} & 
\makecell{Array\\size\\(mm$^2$)} &
\makecell{SiPM\\size\\(mm$^2$)} &
\makecell{SPAD\\pitch\\(\textmu m)} &
\makecell{Resin\\layer} &
\makecell{Window\\material}  & \makecell{FF\\(\%)} & \makecell{IF\\(\%)} \\
\midrule
\makecell{S13361\\2050} & $8\times 8$ & 17.8$\times$17.8 & $2\times2$ & 50 & \makecell{Epoxy}    & —  & 74 & 83 \\
\makecell{S13361\\3050} & $8\times 8$ & 25.8$\times$25.8 & $3\times3$ & 50  & \makecell{Epoxy}    & —  & 74 & 88 \\
\makecell{S13361\\3075} & $8\times 8$ & 25.8$\times$25.8 & $3\times3$ & 75 & \makecell{Epoxy}    & — & 82 &  88 \\
\makecell{S13361\\3075} & $8\times 8$ & 25.8$\times$25.8 & $3\times3$ & 75 & \makecell{Silicone} & \makecell{SiO$_2$} & 82 & 88 \\
\makecell{S13361\\3075} & $8\times 8$ & 25.8$\times$25.8 & $3\times3$ & 75 & \makecell{Silicone} & \makecell{High-n} & 82 & 88 \\
\bottomrule
\end{tabular}
\caption{Nominal specifications of the tested SiPM arrays for reflectance measurements.}
\label{table_SiPM_reflectance_samples}
\end{table}
The tested arrays include four $8\times8$ Hamamatsu S13361 Series arrays with SiPMs having an  area of $3\times3$~mm$^2$ featuring a pixel pitch of 50~\textmu m or 75~\textmu m, and a protective layer made of epoxy resin or silicone resin. 
On top of the arrays with silicone resin, a 1~mm thick SiO$_2$ window or a glass with a refractive index of 1.84 at 400~nm, hereafter referred to as high-n glass, was glued.
An additional $2\times 2$~mm$^2$ array with epoxy resin and a 50~\textmu m SPAD pitch was also tested. 
The corresponding fill factor (FF), defined as the fraction of active area in a SiPM and the integration factor (IF), corresponding to the fraction of the array area covered by SiPMs, are also reported in Table~\ref{table_SiPM_reflectance_samples}.

Reflectance measurements were performed using an Agilent 900 External Diffuse Reflectance Accessory (DRA) integrating sphere coupled to an Agilent Cary 4000 UV-Vis spectrophotometer.
For each array we performed direct measurements of the total reflectance $R_{\text{tot}}$ and diffuse reflectance $R_{\text{diff}}$ in the wavelength range from 200 to 800~nm in steps of 1~nm at 8$^\circ$ incidence angle.
We then extrapolated the corresponding specular reflectance $R_{\text{spec}}$ as:
\begin{equation}
    R_{\text{spec}}= R_{\text{tot}} - R_{\text{diff}}\;. 
\end{equation}
We assigned an uncertainty of 0.05\% to the measured $R_{\text{tot}}$ and $R_{\text{diff}}$, and an additional factor~of~\(\sqrt{2}\), resulting from  error propagation, to $R_{\text{spec}}$. Results are reported in Section~\ref{sec:sipm_reflectance_results}.

\subsection{Beam test measurement campaigns}
\label{sec:res_bt}

We carried out a series of dedicated beam test campaigns at the CERN-PS T10 beamline~\cite{vanDijk:2025ggb} to validate the proposed charged-particle timing approach.
The primary goal was the characterization of the topology of the Cherenkov clusters by charged particles, and the evaluation of the efficiency and time resolution of a complete SiPM–electronics chain. 

\subsubsection{Tested SiPM arrays}
Following the simulation studies discussed in Section~\ref{sec:Monte_Carlo_simulations_contributions}, we tested \(8\times8\) Hamamatsu S13361 SiPM arrays featuring various pixel sizes (1.3~mm, 2.0~mm, and 3.0~mm) and pixel pitches (1.5~mm, 2.2~mm, and 3.2~mm), as well as two different protective resins (epoxy and silicone) with a thickness of about 100~$\upmu$m.

The SiPMs were optically coupled to windows made of either SiO$_2$ or MgF$_2$, with nominal thicknesses of 1 mm or 2 mm. The S13361-3075 arrays equipped with 1‑mm‑thick SiO$_2$ or MgF$_2$ windows were purchased directly from Hamamatsu with the pre-bonded optical window.
Conversely, the 2‑mm‑thick SiO$_2$ windows were bonded in‑house using an optical cement.
Measurements with bare arrays were also performed for comparison.
In this work, we focus on the results obtained with the arrays reported in~Table~\ref{table_SiPM_testbeam}.

All the measurements reported in the following were performed with the T10 negatively charged beam at 10~GeV/$c$ momentum.
The beam composition consisted of approximately 95\% pions, 3\% electrons, and 2\% kaons~\cite{vanDijk:2025ggb}.
For the radiator materials considered in this work, all the beam particle species were in the Cherenkov angle saturation regime.

\begin{table}[!t]
\centering
\small
\setlength{\tabcolsep}{4pt}
\renewcommand{\arraystretch}{1.05}
\begin{tabular}{l c c c c c c c c} 
\toprule
\makecell{Array\\model} &
\makecell{Number\\of SiPMs} &
\makecell{Array\\size\\(mm$^2$)} &
\makecell{SiPM\\pitch\\(mm$^2$)} &
\makecell{SiPM\\size\\(mm$^2$)} &
\makecell{SPAD\\pitch\\(\textmu m)} &
\makecell{Resin\\layer} &
\makecell{Window\\material} &
\makecell{Window\\thickness\\(mm)}\\
\midrule
\makecell{S13361\\1350} & $8\times 8$ & 12.2$\times$12.2 & $1.5\times1.5$ & $1.3\times1.3$ & 50 & \makecell{Epoxy}    & SiO$_{2}$  &  2 \\
\makecell{S13361\\2050} & $8\times 8$ & 17.8$\times$17.8 & $2.2\times2.2$ & $2\times2$ & 50 & \makecell{Epoxy}    & SiO$_{2}$  & 1\\
\makecell{S13361\\3075} & $8\times 8$ & 25.8$\times$25.8 & $3.2\times3.2$ & $3\times3$ & 75 & \makecell{Silicone} & SiO$_{2}$  & 1 \\
\makecell{S13361\\3075} & $8\times 8$ & 25.8$\times$25.8 & $3.2\times3.2$ & $3\times3$ & 75 & \makecell{Silicone} & MgF$_{2}$  & 1 \\
\makecell{S13361\\3075} & $8\times 8$ & 25.8$\times$25.8 & $3.2\times3.2$ & $3\times3$ & 75 & \makecell{Epoxy} & —  & — \\
\bottomrule
\end{tabular}
\caption{Nominal specifications of the SiPM arrays tested in the beam campaigns.}
\label{table_SiPM_testbeam}
\end{table}

\subsubsection{The SiPM array telescope}

\begin{figure}[!t]
    \centering
    \includegraphics[width=1.0\linewidth]{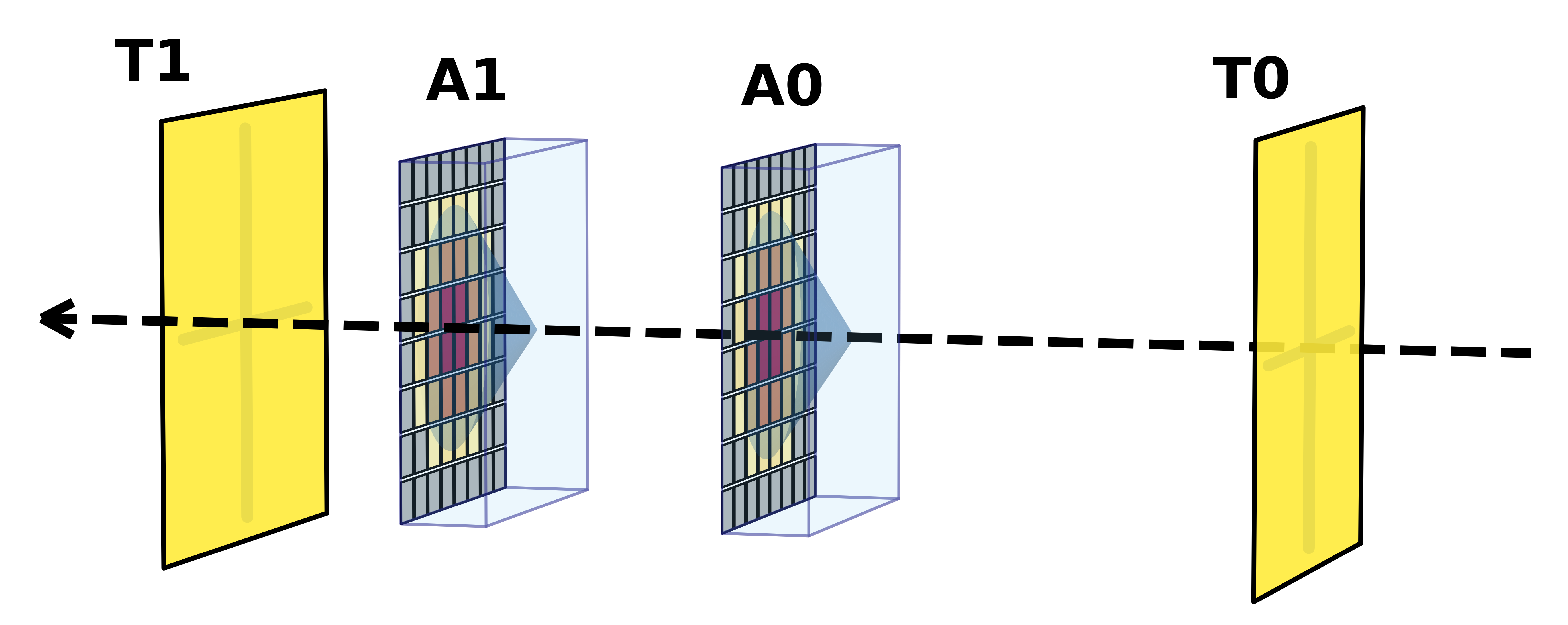}
    \includegraphics[width=1.0\linewidth]{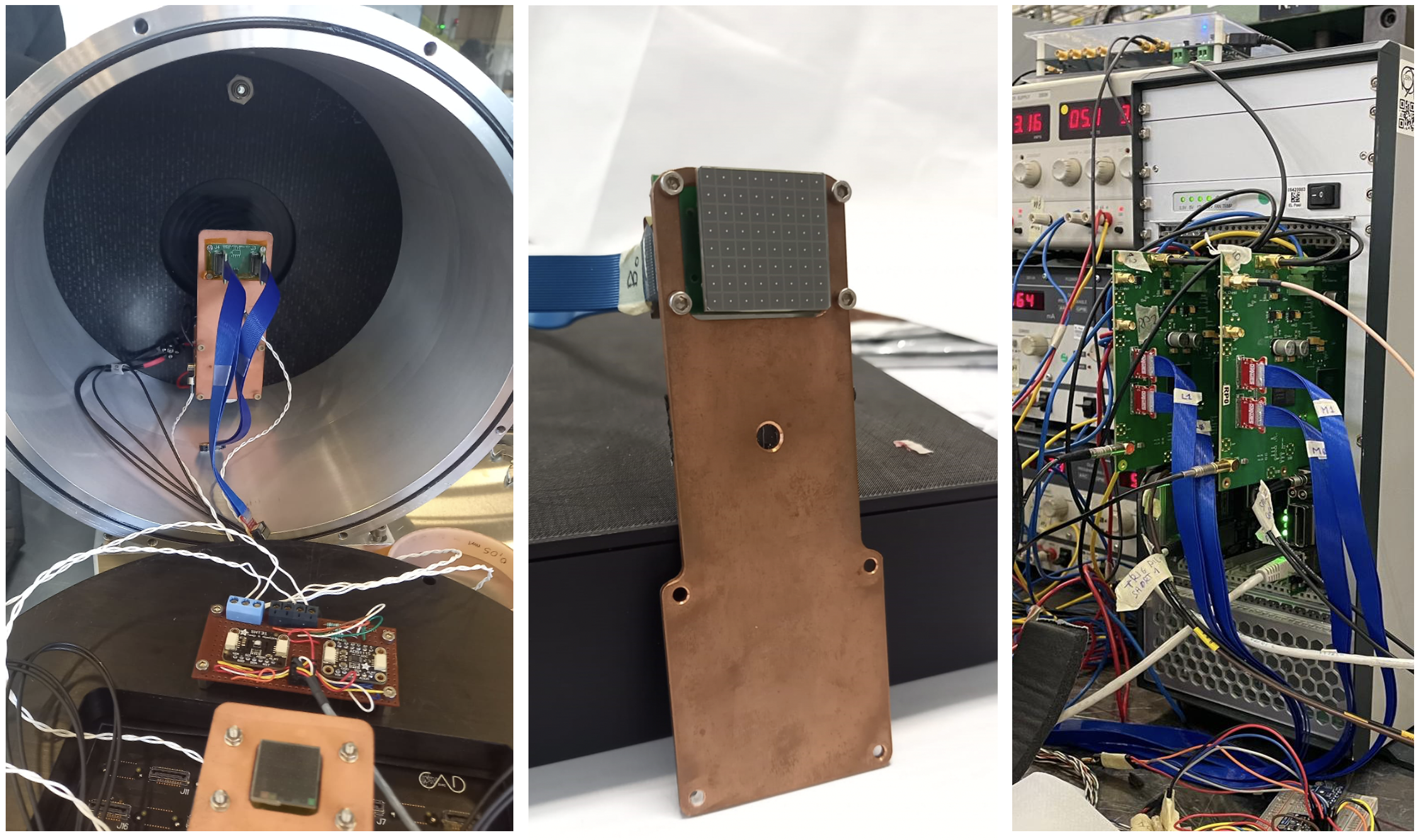}
    \caption{Top: Schematic view of the beam test setup. Bottom: Photographs of the cylindrical vessel housing the arrays A0 and A1, and connection to the Front-End boards.}
    \label{fig:setup_testbeam}
\end{figure}

The setup is illustrated in Figure~\ref{fig:setup_testbeam}. 
A telescope of four detectors arranged along the beamline was used.
Two X-Y tracker modules, hereafter denoted as T0 and T1, based on staggered round plastic scintillating fibers
coupled with Hamamatsu S13552 128-channel SiPM arrays~\cite{S13552} were mounted upstream and downstream the SiPM arrays at a distance of about 125 cm for triggering and tracking purposes~\cite{Mazziotta:2022vow,Pillera:2023hzp,Cerasole:2025tiz}. 
Two SiPM arrays, hereafter referred as A0 and A1, both facing the beam, were mounted inside a cylindrical vessel placed between the tracker modules, at a separation of about 28~cm, with A0 located upstream of A1.
We tested various combinations of the arrays  listed in Table~\ref{table_SiPM_testbeam}.
This enabled the characterization of the timing performance by comparing the times of arrival measured by the A0 and A1 arrays, using one of the two as a reference time for the other.

\subsubsection{SiPM cooling and temperature monitoring}
To ensure mechanical stability and efficient thermal management, the SiPM arrays were mounted on custom-designed printed circuit boards (carrier boards), which were fixed to copper baseplates.
The SiPMs were cooled down to  0~$^\circ$C  through a dedicated cooling system incorporating water chillers and Peltier cells coupled to the baseplates to reduce the SiPM dark count rate (DCR). 
The temperatures of the baseplates and the carrier boards were monitored using analog temperature sensors TT4-10KC3-T125-M5-500~\cite{ntc10k} assembled on the edges of the baseplates and read out using a Raspberry Pi 3~\cite{raspberrypi3modelb} with ADS1115 16-bit ADCs~\cite{ads1115}, and 1-wire digital temperature sensors DS18B20~\cite{ds18b20} mounted on each carrier board, respectively.
The temperature remained stable with maximum observed fluctuations of about  $1~^\circ$C throughout all runs and for all tested configurations.
The vessel was flushed with argon gas to reduce the humidity and maintain dew point values well below the minimum operating temperature inside the vessels, thereby preventing condensation.
Humidity sensors SHT31-D~\cite{sht31} were used to monitor both the ambient and vessel humidity.

\subsubsection{Cabling of the SiPMs to the front-end}
All SiPM arrays were bonded on one side of the carrier board, routing the channels to Samtec LSHM-120 multi-channel connectors~\cite{lshm} bonded on the opposite side of the carrier board. 
The 128 channels of each S13552 array of T0 and T1 were arranged into 32 groups (OR4), each consisting of four adjacent channels, resulting in a readout pitch of $1\times1.625$ mm$^2$. Each of those groups of channels was routed to a single LSHM-120 multi-channel connector.
The S13361 arrays at the inside the vessel were instead plugged with their Samtec ST4-40~\cite{st4} connectors to different carrier boards equipped with Samtec SS4-40~\cite{ss4} connectors and all channels were routed to two LSHM-120 multi-channel connectors, each handling 32 channels.
For the arrays reading the views of T0 and T1, the Samtec LSHM-120 multi-channel connectors were connected directly to the Front-End Boards (FEBs) by means of high-speed 50~$\Omega$  multi-channel Samtec HLCD-40 cables~\cite{hlcd}. For  A0 and A1, the Samtec LSHM-120 multi-channel connectors were first connected to a feed-through PCB connector using Samtec HLCD-40 cables, and then an additional cable was plugged for each array to route the analogue signals from the feed-through PCB connector to the FEBs placed outside.
The SiPM bias voltage was provided through four of the HLCD channels by CAEN A7585D bias modules~\cite{a7585}.
A HLCD channel was dedicated to the readout of the 1-wire digital temperature sensor.

\subsubsection{SiPM readout electronics}
We tested two different front-end architectures for the A0 and A1 arrays, one based on the Petiroc~2A ASIC~\cite{petiroc2a} and one based on the Radioroc~2 ASIC~\cite{Saleem:2023pwt} coupled to the picoTDC~\cite{Altruda:2023qoh}, in order to assess the contribution of the electronics to the overall time resolution. 
The SiPMs were interfaced to the front-end electronics through custom FEBs, hereafter referred to as the Petiroc board and the RadioPico board.
The arrays reading the X and Y views of T0 and T1 were connected to two different Petiroc FEBs for all the measurements reported in this work.
In contrast, the A0 and A1 arrays  were read out by the same Petiroc FEB for measurements performed with the Petiroc~2A, and by two RadioPico FEBs for measurements performed with the Radioroc~2 and picoTDC.

\subsubsection*{The Petiroc board}

The Petiroc board, developed by the Electronics Workshop of the INFN unit of Bari, is equipped with four Petiroc 2A ASICs developed by Omega-Weeroc~\cite{petiroc2a}. 
The Petiroc 2A,  is a 32-channel ASIC providing per-channel arrival time and charge readout. Each channel integrates a 10-bit analog-to-digital converter (ADC) for charge measurement and a 10-bit time-to-digital converter (TDC) with 40 ps LSB for timing. A nominal time jitter smaller than 150 ps FWHM is expected for larger than two PE-signals for operation at 0.5 PE threshold (ASIC only). 
Four LSHM-120 Samtec connectors are used to route the analogue SiPM signals to each ASIC. Besides the front-end ASICs, the Petiroc board hosts a CAEN A7585D SiPM voltage supply module~\cite{a7585} to bias the SiPMs and a Kintex-7 FPGA mounted on a Mercury+ KX2 module~\cite{b_Xilinx_FPGA}. The board and DAQ infrastructure is based on the MOSAIC system~\cite{DeRobertis:2018vls}. There are six NIM input and output lines dedicated to external trigger operation and synchronous multi-board operation.
Further~details~can~be~found~in~\cite{Mazziotta:2022vow}.
   
\subsubsection*{The RadioPico board}

The RadioPico board is equipped with one Radioroc 2 ASIC developed by Weeroc \cite{weeroc_radioroc2}, coupled with one picoTDC ASIC~\cite{cern_picoTDC_site}, developed by CERN, connected to a MOSAIC readout board~\cite{DeRobertis:2018vls}.
The ASICs have 64 channels, allowing the readout of a $8\times8$ SiPM array with each board.
Two LSHM-120 Samtec multi-channel connectors are used to route 32 SiPMs each to the Radioroc 2. The signals entering Radioroc 2 are fed to the time-preamplifier and discriminated by setting a threshold. The ASIC is configured to provide a differential output digital signal of each input channel, with an expected jitter as low as 35 ps FWHM on a single PE (ASIC only). This signal is fed as input to the picoTDC, working with a 40 MHz differential clock provided through dedicated SMA connectors by an external Skyworks SI5341-D-EVB differential clock board~\cite{si5341_d_evb}, enabling a LSB as low as 3.05 ps. The picoTDC is operated in ToT mode, allowing the measurement of both the time of arrival (ToA) and the time over threshold (ToT), serving as a proxy for the charge.
For each external trigger, hits detected within a programmable time window and matched to the trigger with a programmable latency were acquired by the MOSAIC board.
External bias to the SiPMs is provided by a CAEN A7585D module through an additional LEMO connector on the mezzanine.

\subsubsection{Overvoltage and threshold}

\label{sec:vov_and_th}

In our measurements, the S13361 SiPMs exhibit an average breakdown voltage of approximately 52 V at 25~$^\circ$C, with a maximum channel‑to‑channel variation of $\pm$ 0.1 V and a temperature coefficient of about 54 mV/$^\circ$C. 
Since the Petiroc 2A and Radioroc 2 ASICs provide only per‑channel fine trimming of the bias, the temperature‑dependent variation of the overvoltage is compensated by adjusting the global HV offset through the external supply module (CAEN A7585D). This configuration ensures stable overvoltage conditions and uniform gain across all channels over the full operating temperature range.

Data were collected at various overvoltages and discriminator thresholds.
The results reported in the following refer to configurations in which the arrays were operated at an overvoltage of about 6~V for all tested A0 and A1 combinations, with the thresholds set as described below.
For acquisitions with the Petiroc board, the threshold was set to approximately three PEs for all S13361 arrays equipped with a window, and between one and two PEs for arrays without a window. For acquisitions with the RadioPico board, the threshold was set to one PE for all tested configurations.
At the considered overvoltage of 6~V and a temperature of $0~^\circ$C, the measured DCR was between 20 and 30~kHz/mm$^2$ for arrays with a 50~$\upmu$m SPAD pitch and between 30 and 45~kHz/mm$^2$ for arrays with a 75~$\upmu$m SPAD pitch at the single-PE level.
When the threshold was set to three~PEs, the DCR was reduced to less than 0.1~kHz/mm$^2$ for all arrays.

\subsubsection{Event selection}

The signals from the fiber tracker were used for external triggering. The trigger was then distributed to the other Petiroc and RadioPico boards. 
We selected events requiring that the SiPM channels with the
maximum charge observed in both the X and Y views of T0 and T1 had at least six photoelectrons.
We also required times of those channels to be within a 8~ns window to suppress the background due to wrong or ghost hits in multi-particle event.
In addition, after a standard alignment procedure based on the minimization of the residuals, only events with tracker hits within fiducial regions defined according to the SiPM sizes of the A0 and A1 arrays were considered.
These regions were chosen to ensure that the particle traversed both arrays.

\subsubsection{Charge and time calibration}

\label{sec:calibration}

For the arrays read out with the Petiroc board, a 550~ns data-acquisition time window was opened for each trigger event.
Within this window, the system recorded the fired channels together with their corresponding ADC and TDC measurements.
Charge calibration was performed starting from the measured ADC counts. For each event we evaluated the actual ADC counts of individual channels by subtracting the corresponding pedestal values. We identified the peaks in the ADC charge distribution corresponding to each number of photoelectrons. Then we calculated the ADC-to-PEs calibration constants~\cite{Cerasole:2025tiz}.
Time calibration was subsequently performed. The hit time-of-arrival in the Petiroc 2A ASIC is determined in two-step. A coarse-time (CT) counter with a 40 MHz clock is used as reference and digitized with a 9-bits counter. A ramp-based Time-to-Amplitude Converter (TAC) is used to interpolate the fine-time (FT) between two coarse time edges and digitized with a 10-bits counter. For each Petiroc 2A channel the discriminator output is used as start for the TAC ramp that is stopped by the following coarse counter signal clock running at 40~MHz. Based on the shape of the measured CT and FT distributions, h$_{\text{CT}}$ and h$_{\text{FT}}$, we reconstructed the time information t(ns) as follows:
\begin{equation}
    \text{t(ns)} = \text{P}_{40} \times \left(  \text{CT} +  1 - \sum_{i=0}^{\text{FT}} \text{h}_{\text{FT}}(i) \bigg{/} \sum_{i=0}^{\text{1023}} \text{h}_{\text{FT}}(i)\right) 
\end{equation}
where $\text{P}_{40} = 25$~ns is the period of the reference 40 MHz clock.

\begin{figure}[!t]
    \centering
    \includegraphics[width=0.49\linewidth, trim=0mm 0mm 15mm 0mm,clip]{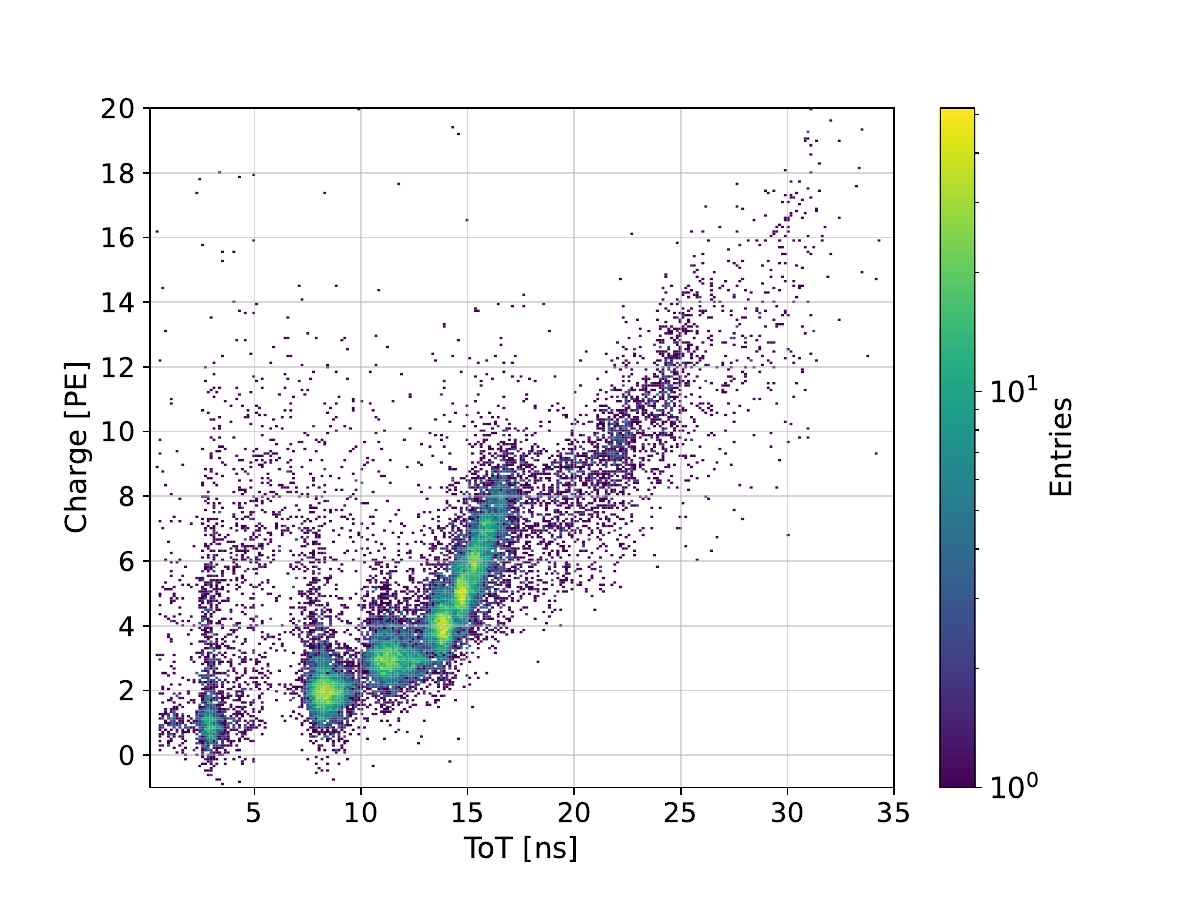}
    \includegraphics[width=0.49\linewidth, trim=5mm 0mm 15mm 0mm,clip]{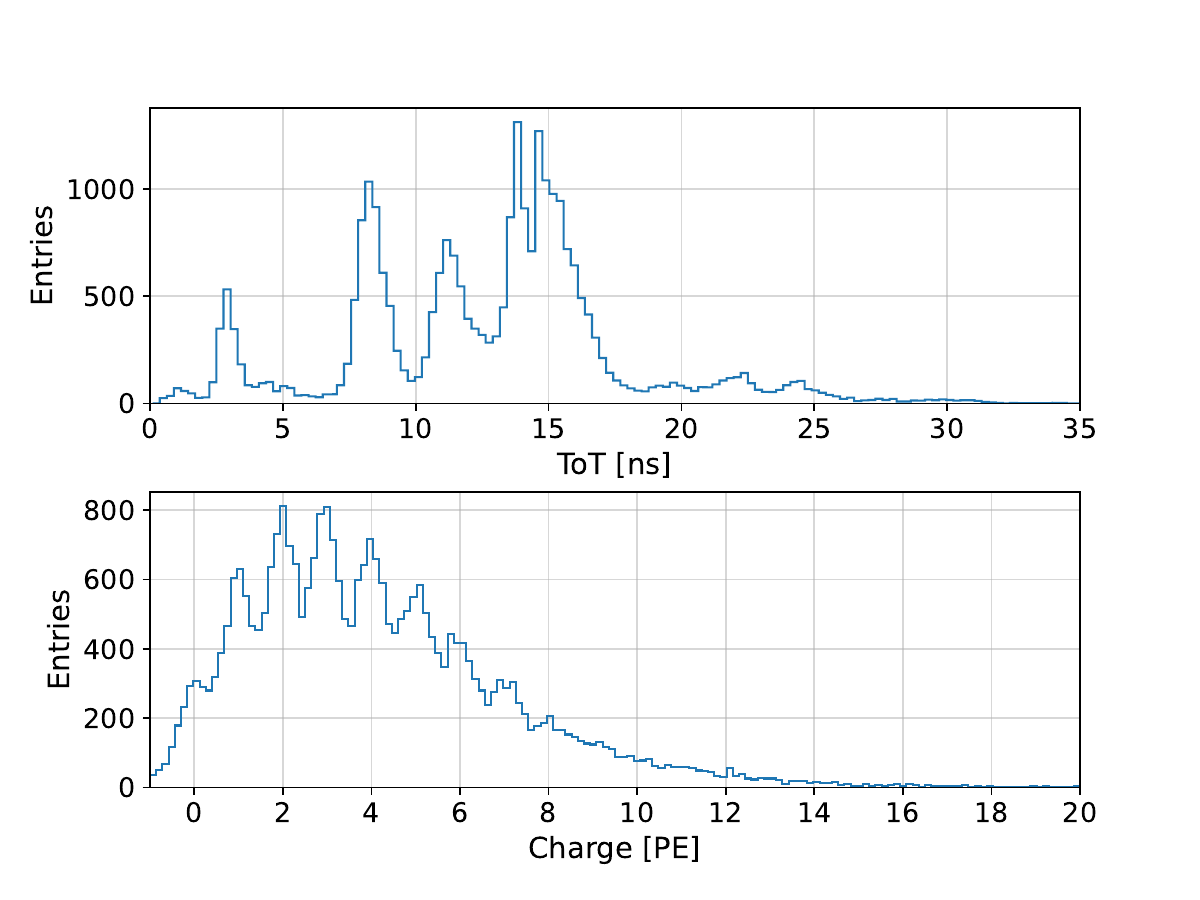}
    \caption{ToT measured from the digital output signal of the Radioroc 2, which is given as input to the picoTDC, versus shaper charge measured from the analog probe output of the shaper (left panel). The measurements were performed with an oscilloscope while illuminating a single SiPM with a laser. The one-dimensional projections of the ToT and charge are displayed.}
    \label{fig:tot_vs_charge}
\end{figure}

For the arrays read out with the RadioPico board, the Radioroc 2 time-preamplifier was configured to operate at its maximum gain, while picoTDC was configured with a ToA LSB of 3.05 ps and a ToT LSB of 195 ps to achieve best time resolution.
Data acquisition was performed in triggered mode with a 200 ns acquisition window, and the ToA was measured relative to an external trigger. Upon each trigger, with a programmable latency, all digital hits recorded by the picoTDC in the acquisition window are read out by the FEB–MOSAIC system, allowing multiple hits per channel within a single event.
The ToA and ToT values were computed from the corresponding ToA and ToT counters, implemented with 16-bit and 11-bit resolution, respectively, assuming the nominal ToA and ToT LSB values.
Although the ToT provides a proxy for the deposited charge, the relationship between ToT and number of PEs is not linear and does not exhibit a strictly monotonic one-to-one correspondence.
Figure~\ref{fig:tot_vs_charge} shows the charge distribution measured with an oscilloscope and the corresponding ToT distributions for the same set of events acquired under laser illumination.
This is particularly important for defining accurate time-walk corrections, especially when pixels with small charge are used to calculate some cluster parameters, such as the average timing.
For this reason, the cluster analyses reported in this work were mainly performed using Petiroc data, while RadioPico data were limited to SiPMs with the maximum ToT.

\section{Results}
\label{sec:res}

\subsection{SiPM reflectance measurements and ARC optimization studies}

\label{sec:sipm_reflectance_results}

\begin{figure}[!t]
\centering
\includegraphics[width=0.49\columnwidth]{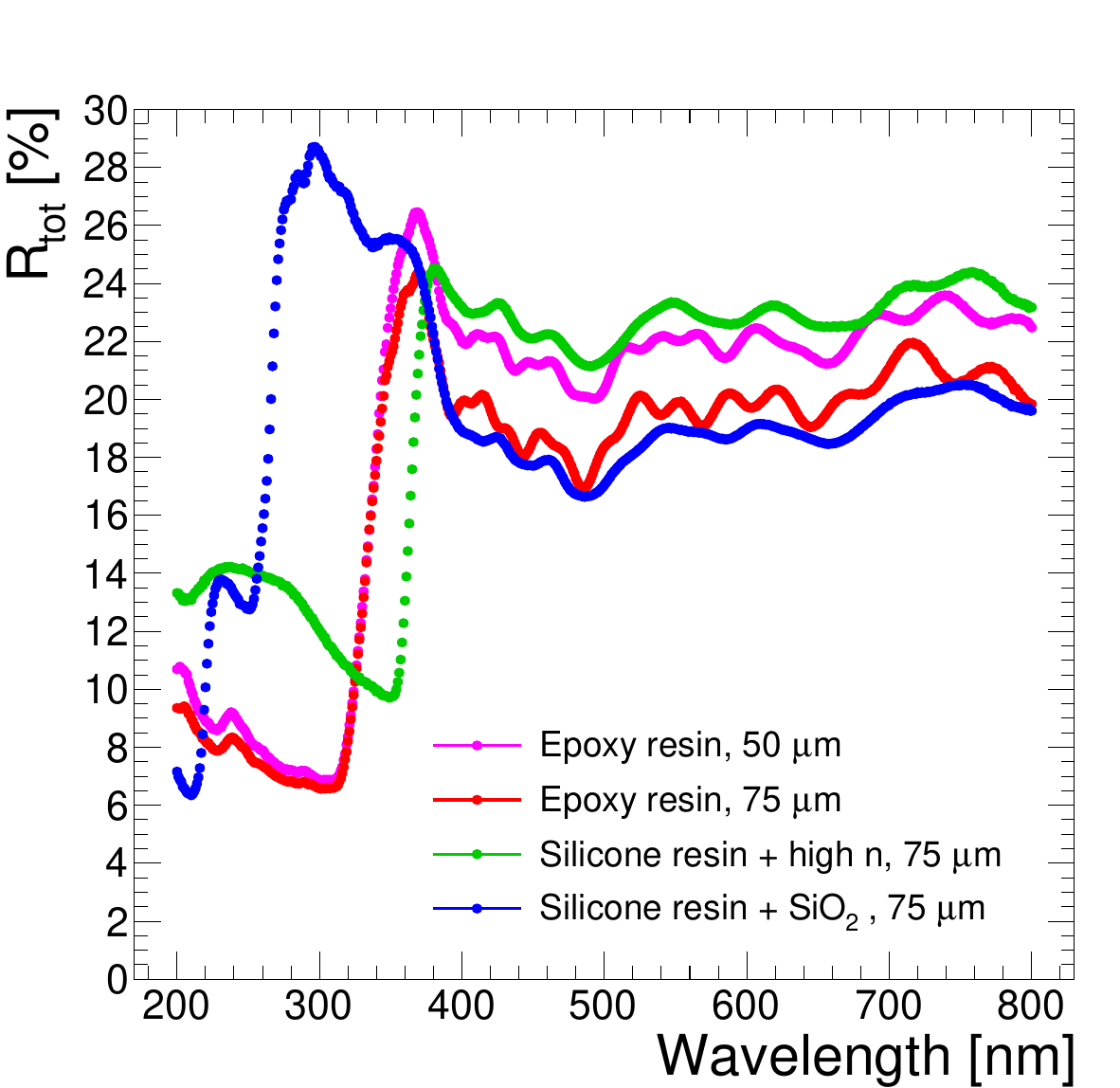}
\includegraphics[width=0.49\columnwidth]{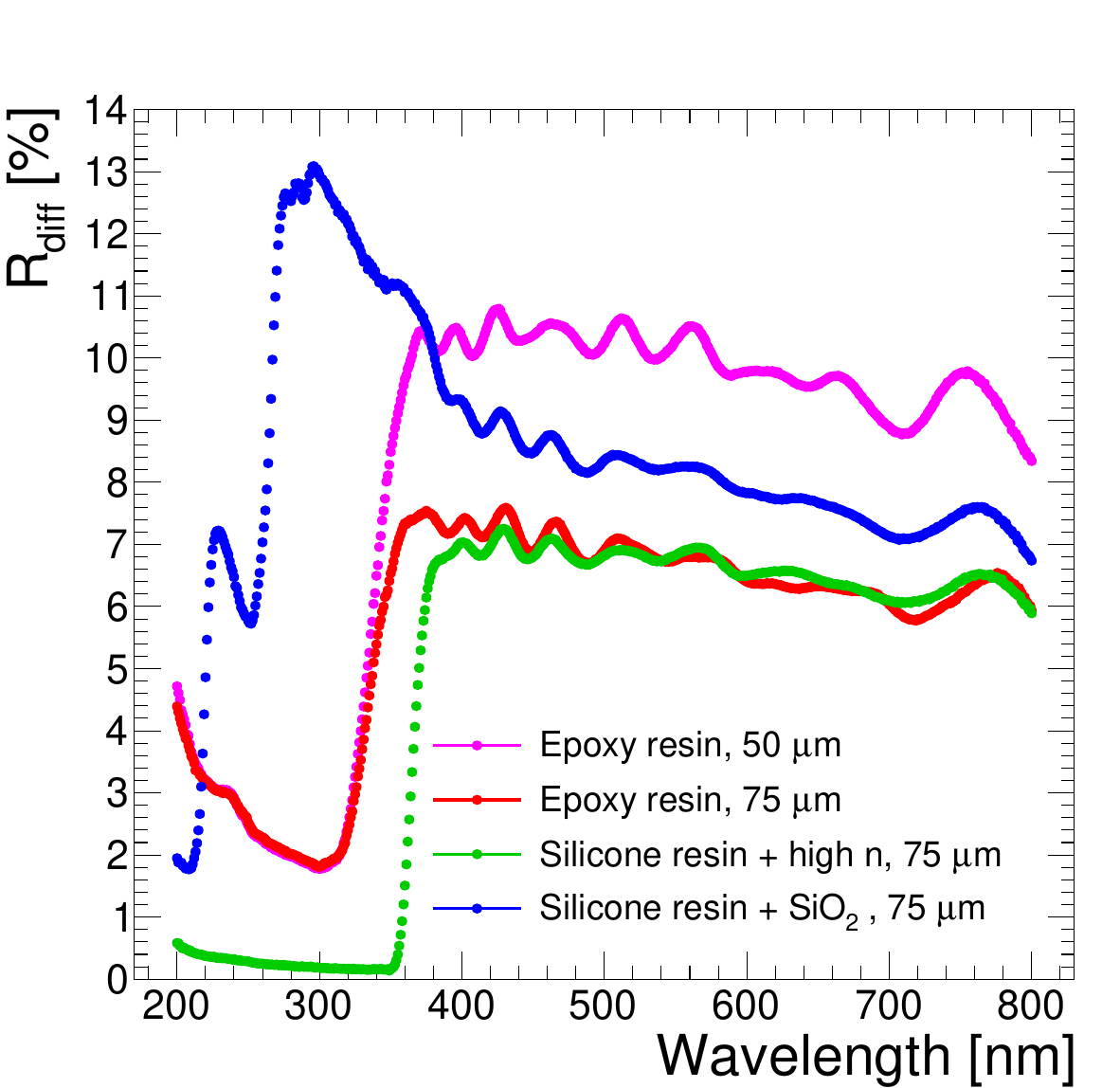}
\includegraphics[width=0.49\columnwidth]{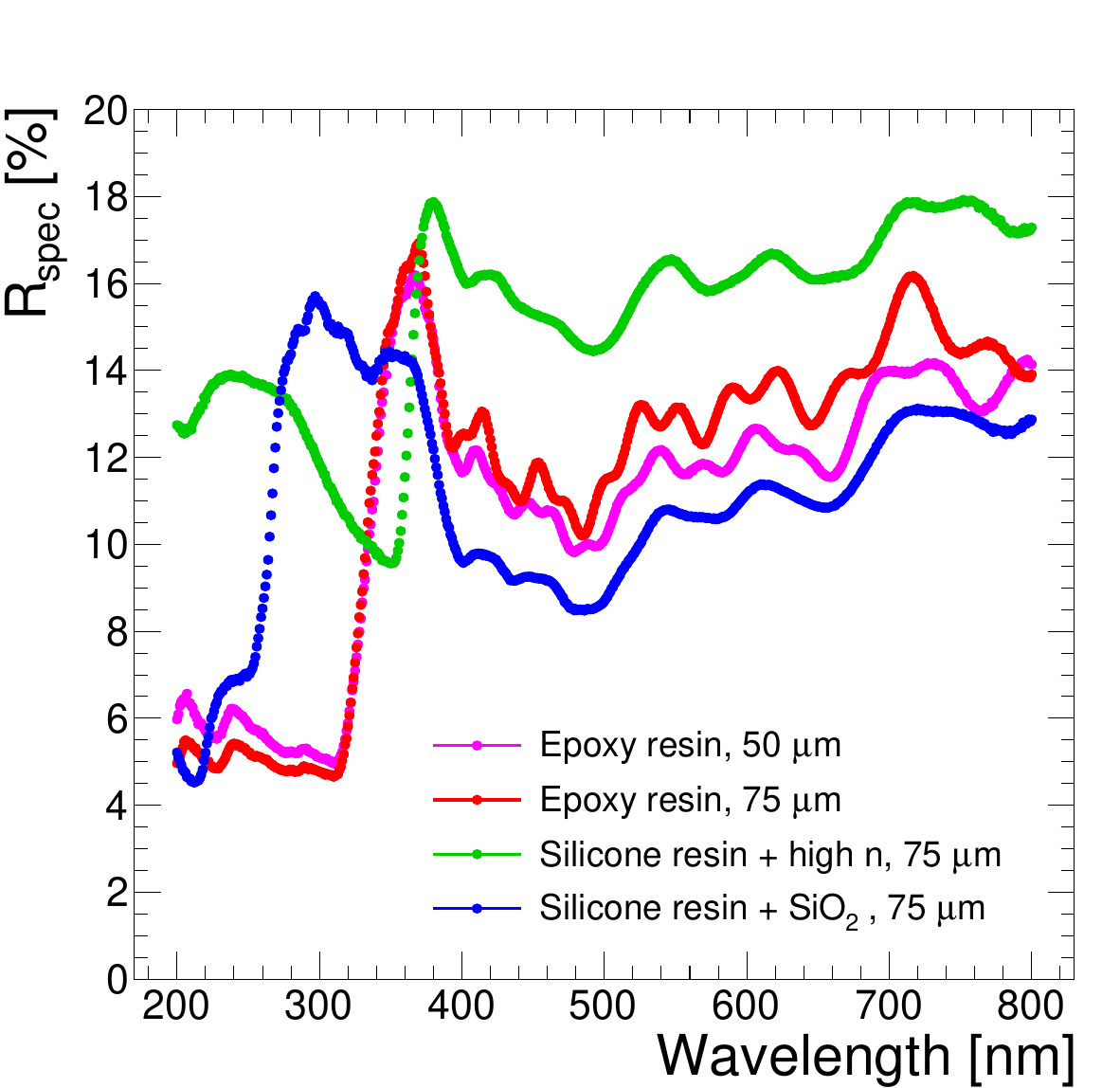}
\includegraphics[width=0.49\columnwidth]{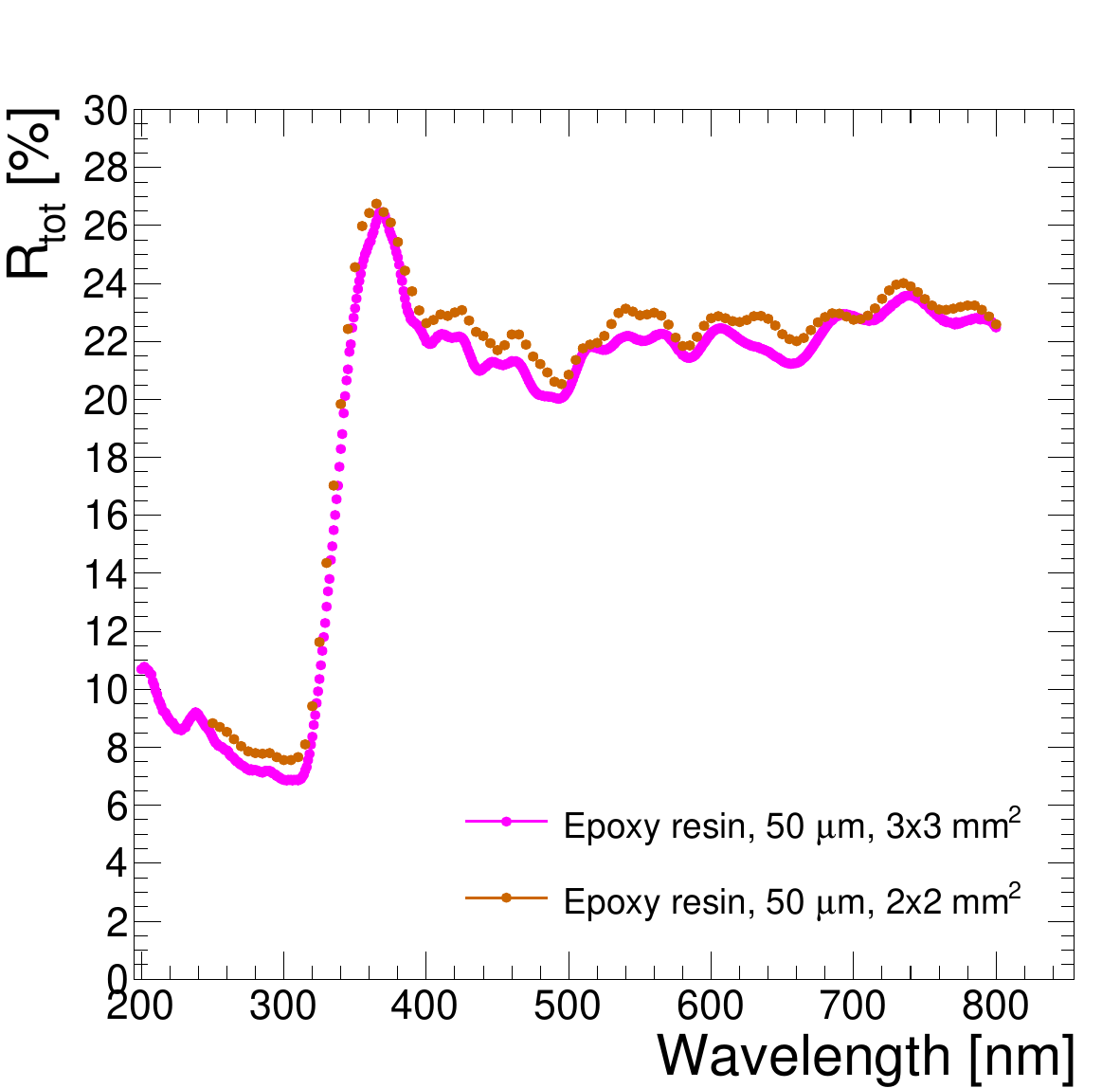}
\caption{
Measured total reflectance (top left), diffuse reflectance (top right), and specular reflectance (bottom left) for Hamamatsu S13360 series $8\times8$ arrays of $3\times3$~mm$^2$ SiPMs, spanning different SPAD pitches, protective resins, and window materials (see Table~\ref{table_SiPM_reflectance_samples}). The bottom right panel compares total reflectance for arrays with $3\times3$~mm$^2$ and $2\times2$~mm$^2$ active area that share the same 50~\textmu m SPAD pitch and the same protective layer made of epoxy resin.
} 
\label{figure_SiPM_reflectance_basics}
\end{figure}

The measured total, diffuse and specular reflectance values as a function of wavelength for the arrays in Table~\ref{table_SiPM_reflectance_samples} are shown in Figure~\ref{figure_SiPM_reflectance_basics}. The following features can be highlighted.
Oscillation structures are observed for all tested SiPMs. They are mainly generated by thin-film interference effects at the upper and lower boundaries of the ARC stacks, as thin as about tens of~nm, on the surfaces of the SiPMs~\cite{WANG2020164171}. Since the protective layers and the additional entrance windows are much larger than the wavelengths of the incident light, about 100~$\upmu$m and 1~mm respectively, their contributions to the oscillations are negligible. 
Total reflectance values are between 18\% and 24\% for wavelengths larger than 380~nm for all tested arrays. They include the contributions of a diffuse reflectance between 6\% and 10\% and a specular reflectance between 8\% and 18\%.
For wavelengths below 380~nm, the reflectance curves of the tested arrays drop sharply due to the absorption in the protective layer or, when present, in the SiPM entrance window. The pronounced cutoffs at 380~nm, 360~nm, and 280~nm correspond to regions where absorption becomes dominant in high-n glass, epoxy resin, and silicone resin, respectively.
By contrast, the SiO$_2$ window is expected to cut off at shorter wavelengths, outside the range of interest.

The two arrays with epoxy resin and pitches of 50~\textmu m exhibit similar specular reflectance curves. However, the total reflectance is higher for the 50~~\textmu m SPAD arrays because the lower FF increases the diffuse contribution to about 10\% compared with about 7\% for the 75~\textmu m arrays.
The slight difference in reflectance between the two arrays with $2\times2~\mathrm{mm}^2$ and $3\times3~\mathrm{mm}^2$ SiPMs, both mounting epoxy resin and using $50~\upmu\mathrm{m}$ SPADs, is consistent with their different IF of the SiPMs in the arrays, as reported in Table~\ref{table_SiPM_reflectance_samples}. 

The array coupled to the SiO$_2$ window by using silicone resin shows the lowest cut off,  at about 280~nm,
which is driven by absorption in silicone resin. This array also shows the lowest total reflection values at wavelengths larger than 380~nm. In particular, it exhibits the lowest specular reflectance and a diffuse reflectance not larger than 1\% with respect to the other tested arrays.
This array thus presents optimal features for TOF applications.

\begin{figure}[!t]
\centering
\includegraphics[width=0.69\columnwidth]{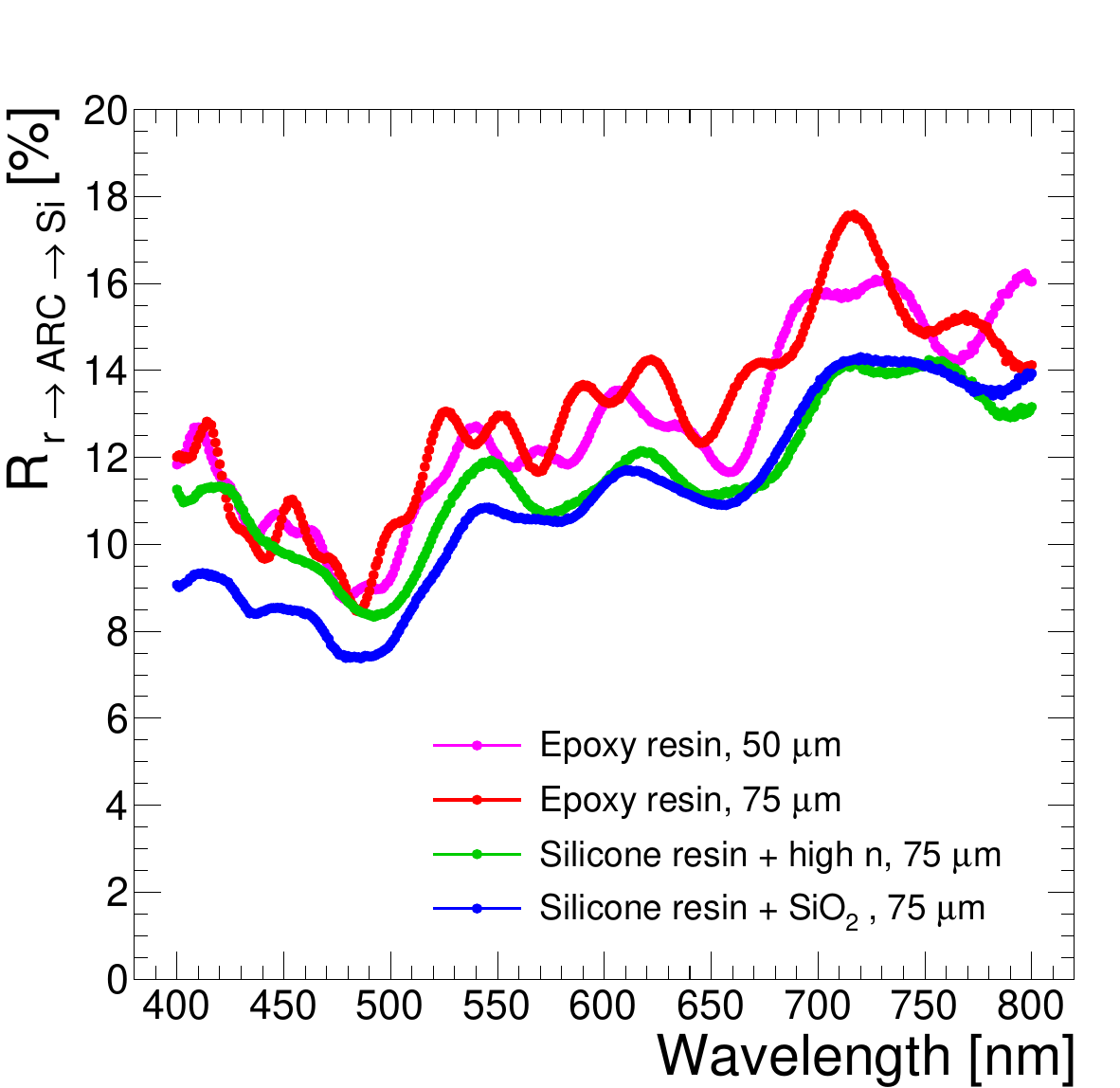}
\caption{Reflectance at the SiPM resin-ARC-passivation-silicon interfaces $R_{\text{r}\rightarrow\text{ARC}\rightarrow\text{Si}}$ extrapolated from the measured specular reflectance using Eq.~\ref{equation_ARC_extrapolation} (Appendix A) for the tested arrays as a function of wavelength.
} 
\label{figure_SiPM_reflectance_ARC}
\end{figure} 

The array coupled to the high-n glass shows the highest total and specular reflectance due to stronger Fresnel reflections upon entrance, with about 9\% probability, driven by the larger refractive index difference between air and the high-n material. Conversely, the coupling of SiO$_2$ windows to either silicone or epoxy resins presents a smaller refractive index difference, which leads to lower reflectance, about 4\%, at the air interface.
For the same reason, the observed diffuse reflectance for the high-n glass array is smaller than for the other arrays since a significantly larger fraction is already reflected specularly upon entrance and therefore does not even reach the surface microstructures that generate diffuse contributions.
Given these characteristics, this array is not optimal for the proposed TOF application.

Using the SiPM reflection model reported in Appendix~\ref{appending_model_reflections}, from the measured specular reflectance of the tested arrays, we extrapolated the overall reflectance resulting from the interfaces between the SiPM protective resin and ARC, between the ARC and passivation layer, and between the passivation layer and silicon $R_{\text{r}\rightarrow\text{ARC}\rightarrow\text{Si}}$. In particular, using the $\mathrm{IF}$ and $\mathrm{FF}$ values in Table~\ref{table_SiPM_reflectance_samples}, together with the nominal refractive indices of SiO$_2$, high-n glass, silicone resin, and epoxy resin, we extracted the $R_{\text{r}\rightarrow\text{ARC}\rightarrow\text{Si}}$ curves shown in Figure~\ref{figure_SiPM_reflectance_ARC} for wavelengths larger than 400~nm at 8$^\circ$ incidence angle.

Across all tested SiPM arrays, the extrapolated specular reflectance at the ARC interface falls in the range 8–14\%. 
The measured $R_{\text{r}\rightarrow\text{ARC}\rightarrow\text{Si}}$ includes both the intrinsic properties of the ARC and contribution of the resin, which cannot be separated on the basis of the available data.
In particular, as expected, the extrapolated curves in Figure~\ref{figure_SiPM_reflectance_ARC} group into two families, blue and green for silicone resin and magenta and red for epoxy resin. Coupling with silicone resin yields a systematically lower reflectance by about 2\% relative to epoxy resin. This reduces interface losses at the considered 8$^\circ$ incidence angle. 
However, it is not known whether the silicone and epoxy variants employ the same ARC, and differences in the ARC design cannot be excluded.
Nevertheless, the agreement of the curves within each resin family, independent of the window material or its absence, together with the differing $\mathrm{FF}$ values across arrays, provides additional validation of the proposed reflection model.

Although the tested commercial SiPMs demonstrated optimal reflectance at the interfaces with the built-in ARC, even lower reflectance values could be achieved by optimizing the ARC  accounting for the for the wavelength-dependent Cherenkov spectrum, scaling as $\text{d}N/\text{d}\lambda\propto\lambda^{-2}$, the polarization of Cherenkov photons, featuring inherent polarization perpendicular to the surface of the cone, and the SiPM spectral response. In such a way it is possible to maximize photon collection over the relevant photon incidence-angle distributions by reducing reflections and increasing the effective SiPM PDE. 

\begin{figure}[!t]
\centering
\includegraphics[width=0.495\linewidth,trim=0mm 0mm 0mm 8mm,clip]{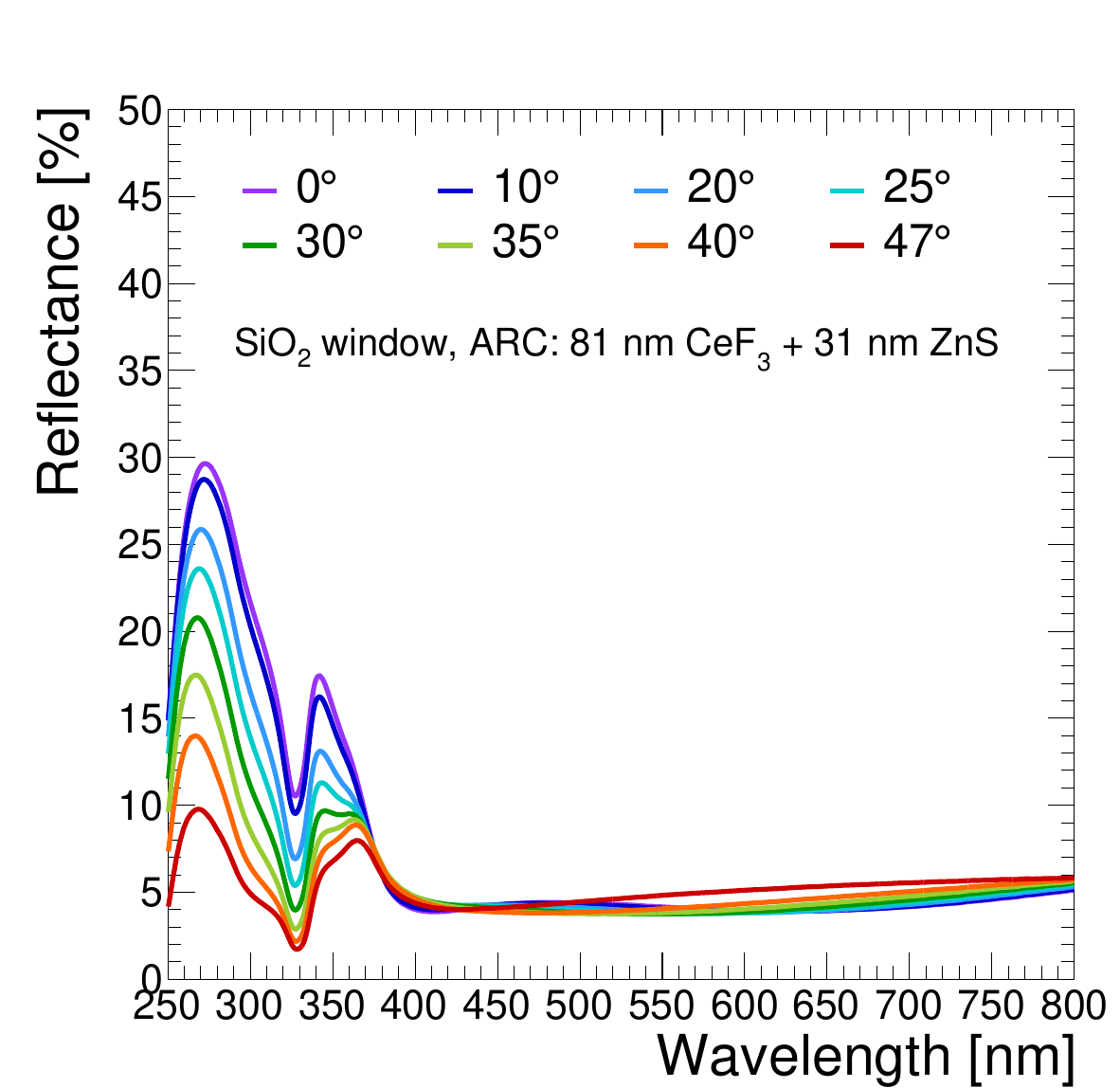}
\includegraphics[width=0.495\linewidth,trim=0mm 0mm 0mm 8mm,clip]{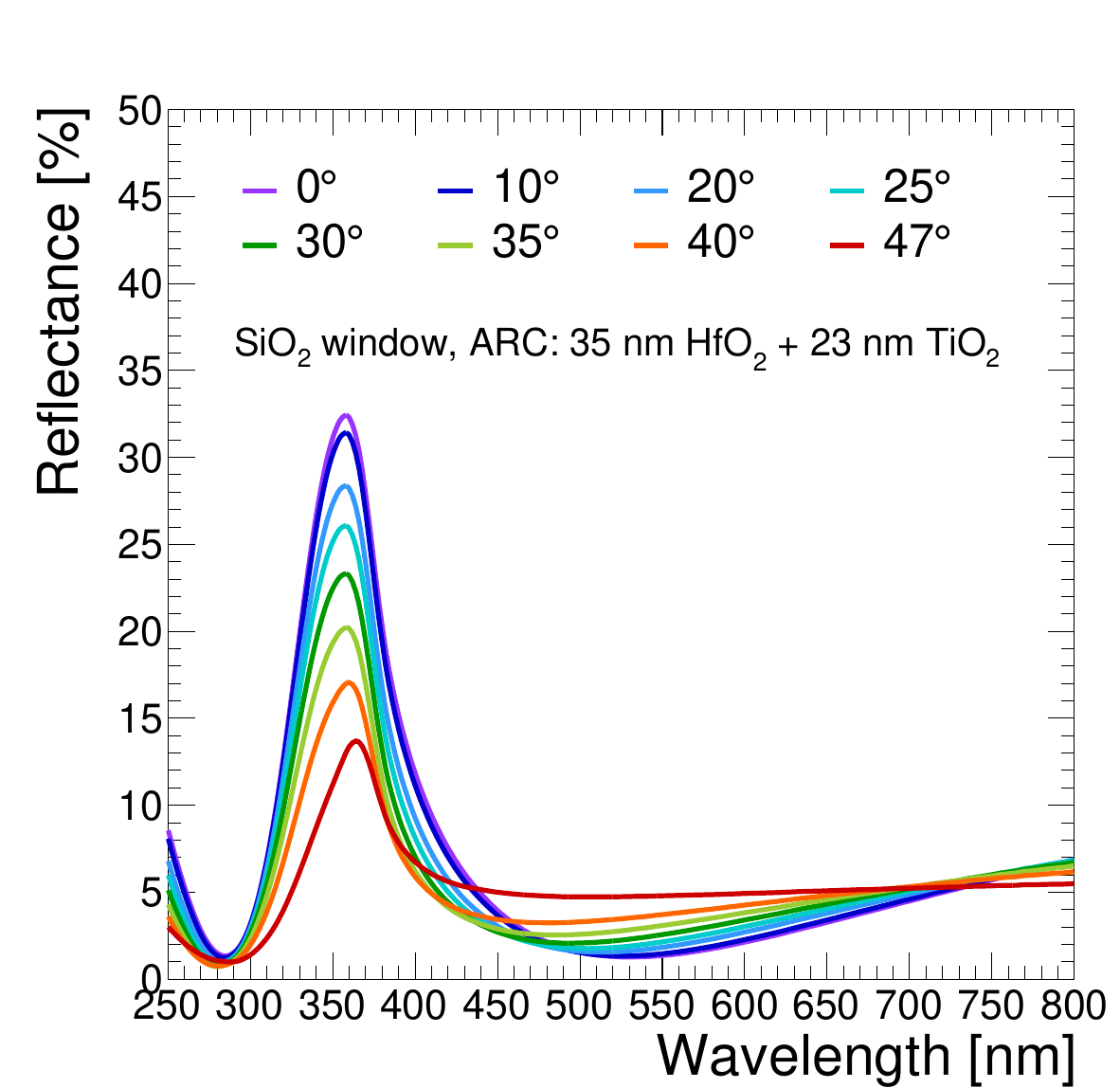}
\caption{Expected reflectance as a function of wavelength for a custom ARCs made of 81~nm CeF$_3$ + 31~nm ZnS (left) and 35~nm HfO$_2$ + 23~nm TiO$_2$ (right) optimized for Cherenkov photons emitted by normally incident charged particles at the Cherenkov angle saturation in a SiO$_2$ window ($\theta_C\approx47^\circ$). A 100~$\upmu$m thick silicone resin protective layer and a 10~nm thick SiO$_2$ passivation layer on top of silicon are assumed.
The different curves correspond to various photon emission angles
to illustrate the reflectance for photons emitted in the regime between the Cherenkov threshold and saturation.
} 
\label{figure_SiPM_reflectance_ARC_optimization_with_sim}
\end{figure} 

We developed an advanced simulation tool to optimize an arbitrary $N$-layer SiPM ARC, based on the polarization-dependent transfer-matrix formalism for thin films (see Chapter~10 of~\cite{steck2006classical}). The tool accounted for the upstream window material and SiPM protective resin, as well as for the downstream passivation layer and silicon. To include absorption in silicon and other materials, we used the general formalism with a complex refractive index, $\tilde{n}=n+i\,\kappa$, where $n$ is the usual refractive index and $\kappa$ is the extinction coefficient.
The ARC affects reflection through thin film interference. For layers with thickness larger than the photon wavelength, such as the $\approx100~\upmu$m  thick typical resin layers or the mm thick windows, coherence is lost and the corresponding contribution to the reflectance is averaged over the photon propagation phase. 

Figure~\ref{figure_SiPM_reflectance_ARC_optimization_with_sim} shows the overall reflectance for two examples of a 2-layer ARC aimed at maximizing the overall number of detected photons for normal incident tracks at saturation
assuming a 1~mm thick SiO$_2$ window, a  100~$\upmu$m thick silicone resin protective layer and a 10~nm thick SiO$_2$ passivation layer. We optimized the thicknesses of the ARC layers assuming the spectral response of Hamamatsu S13360-3075CS SiPMs~\cite{s13660-3075CS}.
In particular, the left panel shows the results  for an ARC made of a 81 nm thick layer of CeF$_3$~\cite{ARC_CeF3_ref} and a 31~nm thick layer of ZnS~\cite{ARC_ZnS_ref}, while the  right panel shows the results  for a 2-layer ARC made of a 35~nm thick layer of HfO$_2$~\cite{ARC_HfO2_ref} and a 23~nm layer of TiO$_2$~\cite{ARC_TiO2_ref}. 
Together with the reflectance curve at the saturated Cherenkov emission angle in SiO$_2$ ($\theta_C\approx47^\circ$), the curves corresponding to lower photon emission angles with the expected polarization (P-polarization) are also shown to illustrate the effect expected for photons emitted by charged particles in the regime between the Cherenkov threshold and saturation.

Reflectance values of about 5\% are obtained with the considered ARCs throughout the wavelength region of interest.
Even lower reflectances are expected with more complex ARCs. 
Comparing with the values in Figure~\ref{figure_SiPM_reflectance_ARC}, these simulations indicate that custom ARCs tailored to the specific application discussed in this work could, in principle, reduce the overall reflectance by about a factor of two with respect to commercial ARCs.

\subsection{Window cluster topology and charge}

\label{sec:cluster_size}

\begin{figure}[!t]
    \centering
    \includegraphics[width=\linewidth,trim=0mm 0mm 0mm 0mm,clip]{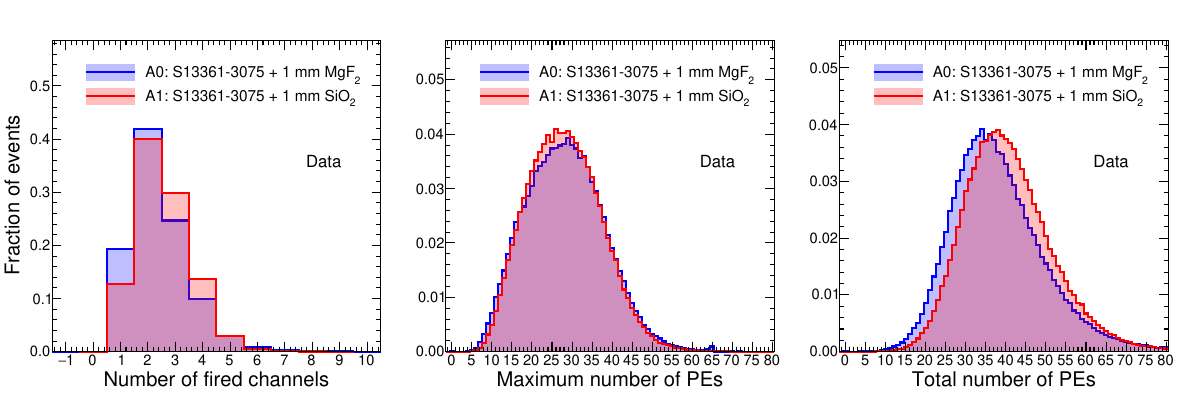}
    \includegraphics[width=\linewidth,trim=0mm 0mm 0mm 0mm,clip]{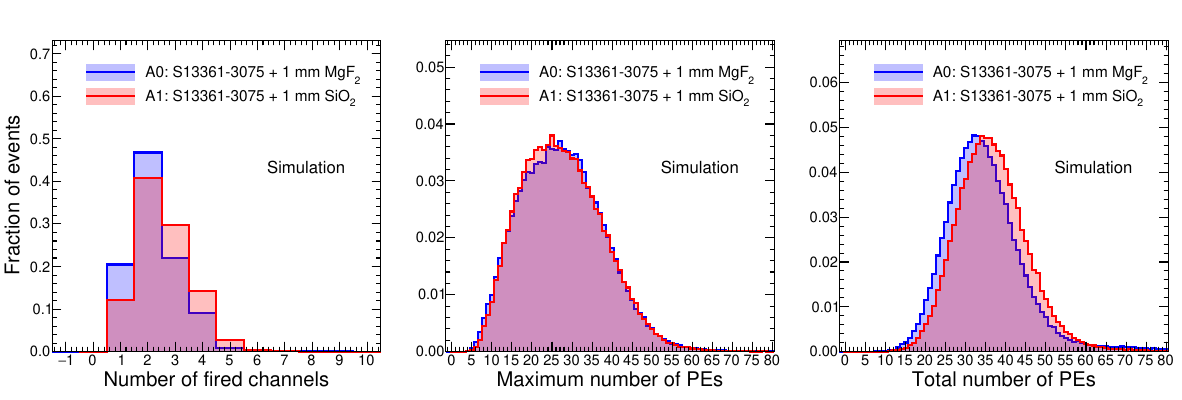}
    \caption{
    Measured (top panels) and simulated (bottom panels) distributions of the number of SiPMs (left), the maximum charge in a single SiPM (middle) and the total number of photoelectrons (right) in the clusters for runs taken with the negatively charged beam at 10 GeV/$c$ using Petiroc boards. The plots refer to the configuration with two S13361-3075 arrays coupled to a 1 mm thick SiO$_2$ window (A1) and a 1 mm thick MgF$_2$ window  (A0). 
    }
    \label{fig:cell_and_charge_Petiroc_3x3vs3x3}
\end{figure}

For a given combination of A0 and A1, we studied the topology of the clusters resulting from the impinging charged particles. 

\begin{figure}[!hbt]
    \centering
    \includegraphics[width=1.0\linewidth,trim=0mm 0mm 0mm 0mm,clip]{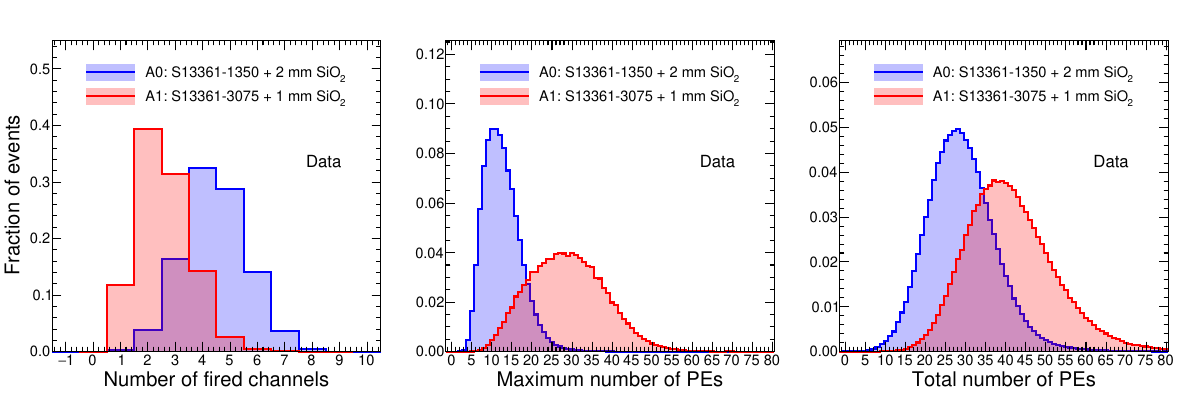}
    \includegraphics[width=1.0\linewidth,trim=0mm 0mm 0mm 0mm,clip]{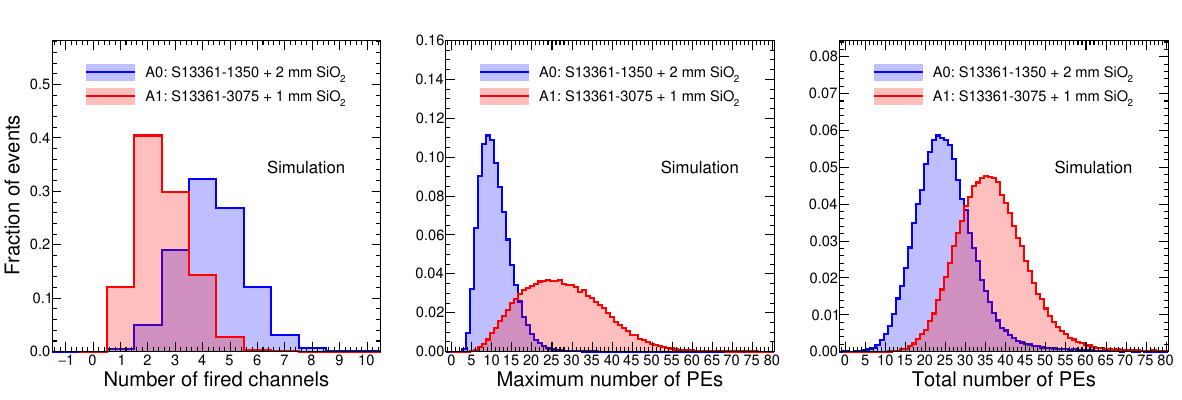}
    \caption{
    Measured (top) and simulated (bottom) distributions of the number of fired channels (left), the maximum charge in a single SiPM (middle) and the total number of photoelectrons (right) in the clusters for runs taken with the negatively charged beam at 10 GeV/$c$ using Petiroc boards. The plots refer to the configuration with the S13361-3075 array coupled to a 1 mm thick SiO$_2$ window (A1) and the S13361-1350 array coupled to a 2 mm thick SiO$_2$ window (A0).}
    \label{fig:cell_and_charge_Petiroc_3x3vs1.3x1.3}
\end{figure}

The top panels of Figures~\ref{fig:cell_and_charge_Petiroc_3x3vs3x3} and \ref{fig:cell_and_charge_Petiroc_3x3vs1.3x1.3} show the measured number of fired channels, the maximum number of
photoelectrons in a single SiPM in the clusters, and the total number of photoelectrons in the clusters for
two different configurations tested  with the Petiroc boards.
In both cases, the same S13361-3075 array coupled to a 1 mm thick SiO$_2$ window  was used as A1. The S13361-3075 array coupled to a 1~mm MgF$_2$ thick window  and the S13361-1350 array coupled to a 2\,mm thick SiO$_2$ window, respectively, were used as A0.

The bottom panels of the same figures show the corresponding expectations from the full simulation of the setup in the Geant4 framework~\cite{GEANT4:2002zbu}. The simulation includes a detailed description of the relevant processes, such as optical transport with reflections, the production and propagation of secondary particles, and the response of the SiPMs in terms of DCR, PDE and cross-talk. For the dark-count rate, we use the measured values. For the PDE and the cross-talk probability, we assume the nominal PDE as a function of photon wavelength and the nominal cross-talk probability at the operating overvoltage~\cite{s13660-3075CS}.
In addition, the simulation also accounts for the emission of Cherenkov photons by electrons and pions at 10~GeV/$c$ momentum in the argon circulated through the vessel.
Overall, the data are reasonably well reproduced by the simulation.

Under the operating conditions used in terms of overvoltage and threshold, the arrays equipped with $3\times3$~mm$^{2}$ SiPMs with 75~$\upmu$m SPAD pitch exhibit on average 2-3 fired channels per cluster, with a mean maximum charge of about 30~PEs and a mean total charge of 35-40~PEs for 1~mm thick SiO$_2$ and MgF$_2$ windows. 
As expected, the arrays with SiO$_2$ and MgF$_2$ windows exhibit very similar behaviour, given their very close refractive indices and optical transparency. The slight excess observed with SiO$_2$ in terms of number of fired channels and total charge is mostly attributable to the detection of extra Cherenkov photons emitted by 10~GeV/$c$ pions and electrons in the Ar volume at angles smaller than $1.5^\circ$.
In particular, simulations show that about 4-5 Cherenkov photons from the 28~cm Ar region between the two arrays  are detected on average in A1, compared to about 1 detected photon on average in A0 from the 8~cm gap between the vessel entrance and A0.

Despite the $3$~PE threshold for the $3\times3$~mm$^{2}$ SiPM arrays, about 3\% of the events recorded with 1~mm thick windows exhibit clusters with five or six fired channels. If only direct signal photons emitted in the window contributed, no more than four channels would be above the threshold. This excess is well reproduced by the simulation once additional contributions are included, namely photons undergoing multiple reflections before detection, photons produced in the Ar volume (distributed over a cone with a radius of about 7~mm at A1 and 2~mm at A0), and correlated background from secondary particles. In addition, optical cross-talk generates correlated photoelectrons that can lift otherwise sub-threshold cells above threshold, enhancing the observed channel multiplicity to the observed values.

For the array with $1.3\times1.3$~mm$^{2}$ SiPMs with 50~$\upmu$m SPAD pitch and a 2~mm thick SiO$_2$ window, about 4-5 channels are fired, with a mean maximum charge of 10-11~PEs and a total charge of about 23-24~PEs, indicating significantly larger charge sharing as expected from the smaller pixel size.
Although these values are consistent with the expected larger charge sharing for a thicker window and a smaller SiPM active area, the total charge might appear low when compared to the 1~mm window case, since a 2~mm window is expected to produce approximately twice as many photons. This difference can be understood by accounting for the lower fill factor and PDE of the 50~$\upmu$m SPAD pitch case with respect to the 75~$\upmu$m case, the reduced geometric acceptance of the $1.3\times1.3$~mm$^{2}$ active area relative to the $1.5\times1.5$~mm$^{2}$ SiPM pitch compared to the $3\times3$~mm$^{2}$ $3.2\times3.2$~mm$^{2}$ SiPM pitch, the smaller cross-talk probability~\cite{s13660-3075CS}, and the reduced contribution from Ar photons in A0.
A significant contribution also arises from the different optical transmittance of the protective resins of the arrays, as highlighted in Section~\ref{sec:sipm_reflectance_results}. The epoxy resin of the S13361-1350 array absorbs photons with wavelengths shorter than about 360~nm, whereas the silicone resin of the S13361-3075 array exhibits high transmittance down to about 280~nm.
Finally, the finer segmentation distributes the photons over more channels with a smaller charge per channel, increasing the probability that some channels fall below the 3~PE threshold and do not contribute to the measured total charge. When all these effects are included, the measured values are consistent with expectations, as validated by simulation. 

\begin{figure}[!t]
\centering
\includegraphics[width=0.325\linewidth,trim=0mm 0mm 0mm 0mm,clip]{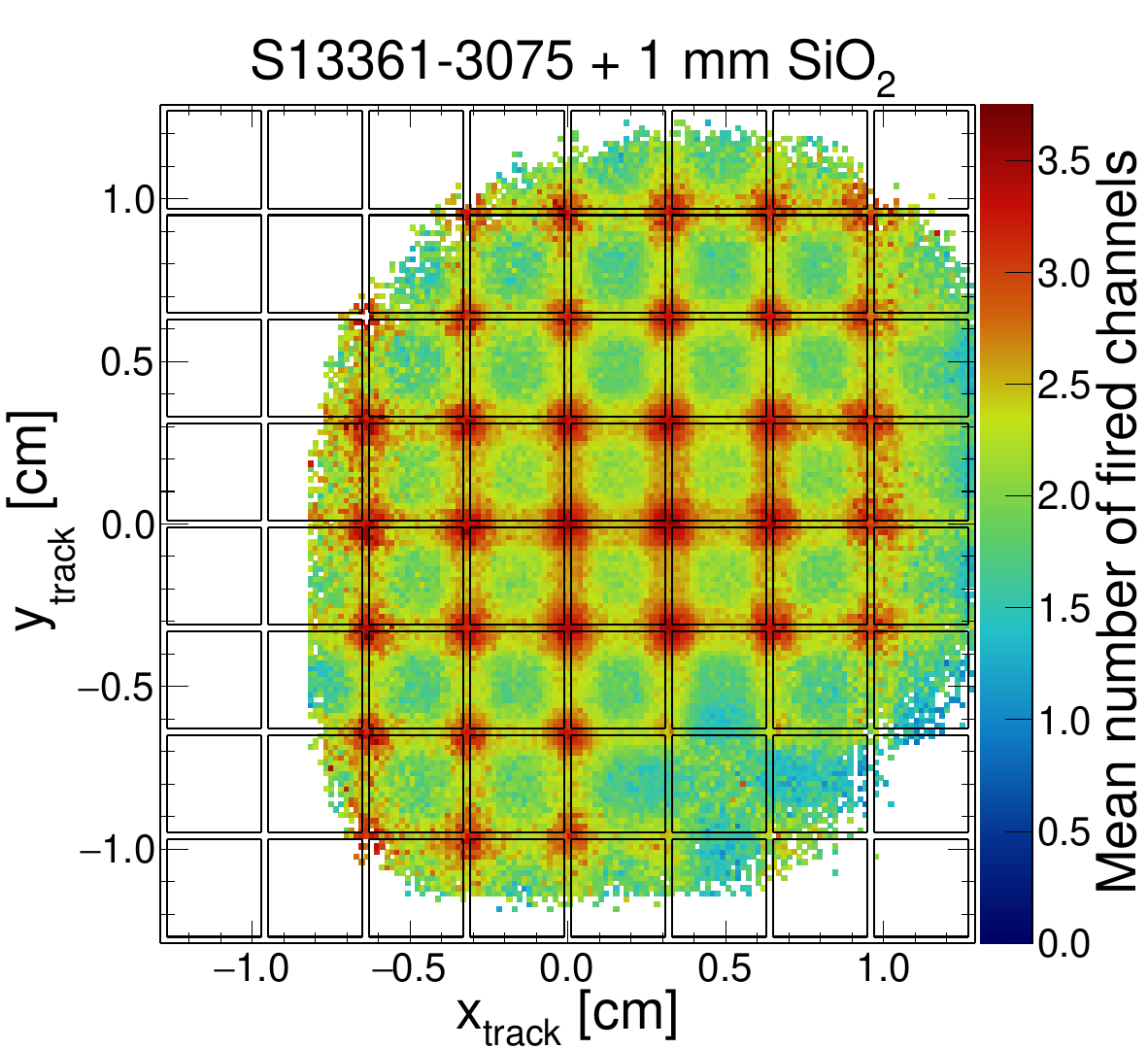}
\includegraphics[width=0.325\linewidth,trim=0mm 0mm 0mm 0mm,clip]{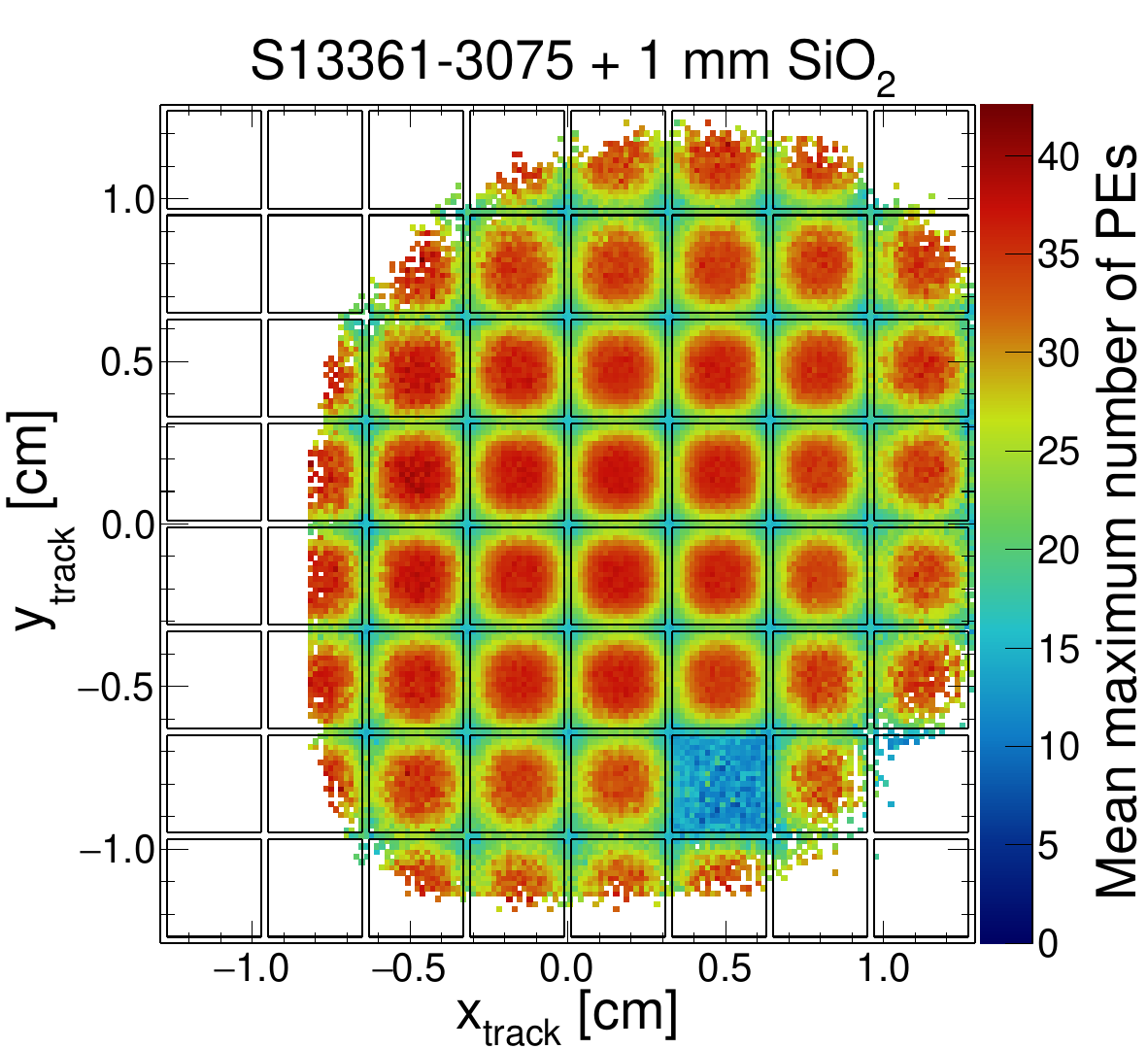}
\includegraphics[width=0.325\linewidth,trim=0mm 0mm 0mm 0mm,clip]{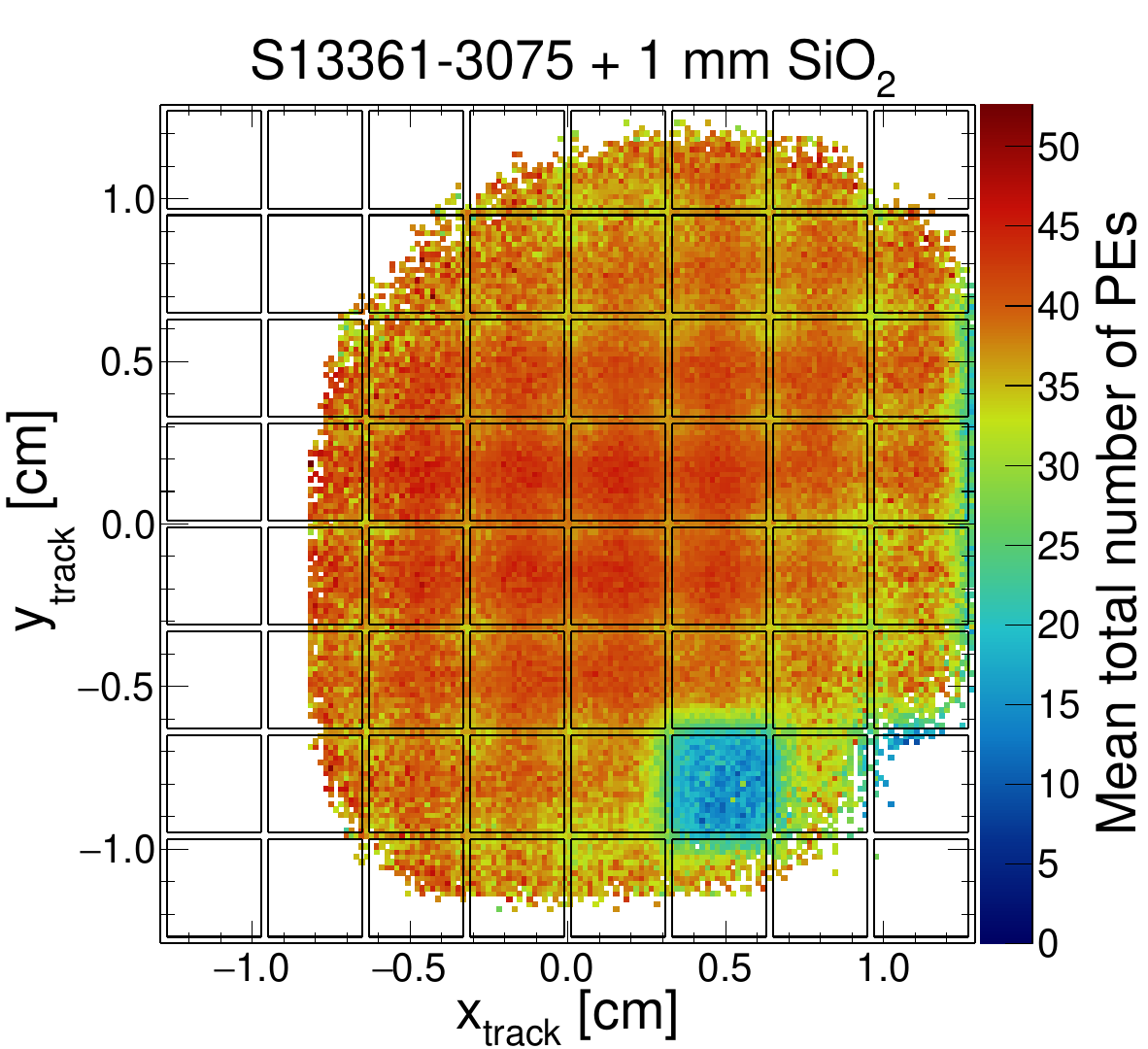}
\includegraphics[width=0.325\linewidth,trim=0mm 0mm 0mm 0mm,clip]{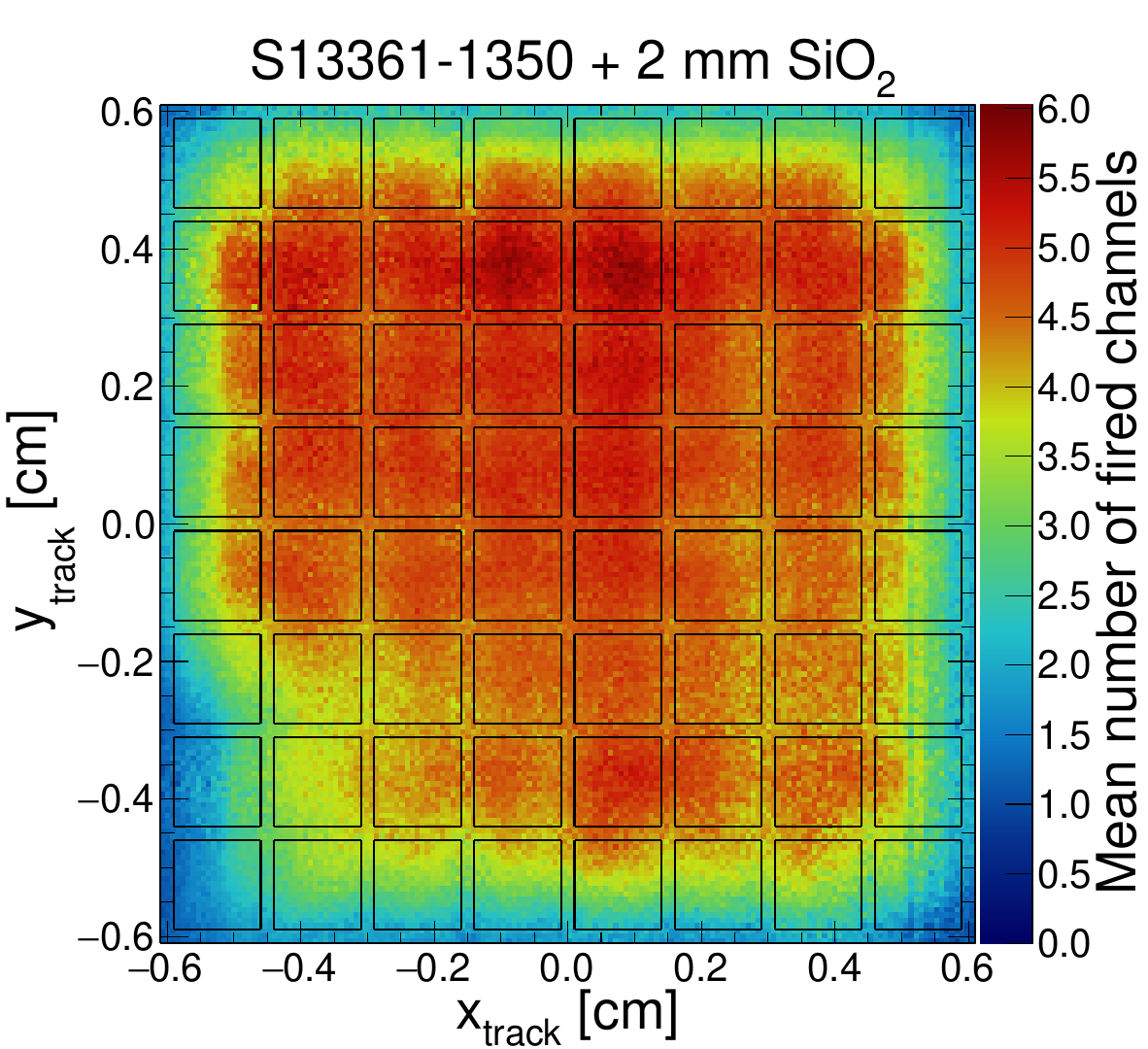}
\includegraphics[width=0.325\linewidth,trim=0mm 0mm 0mm 0mm,clip]{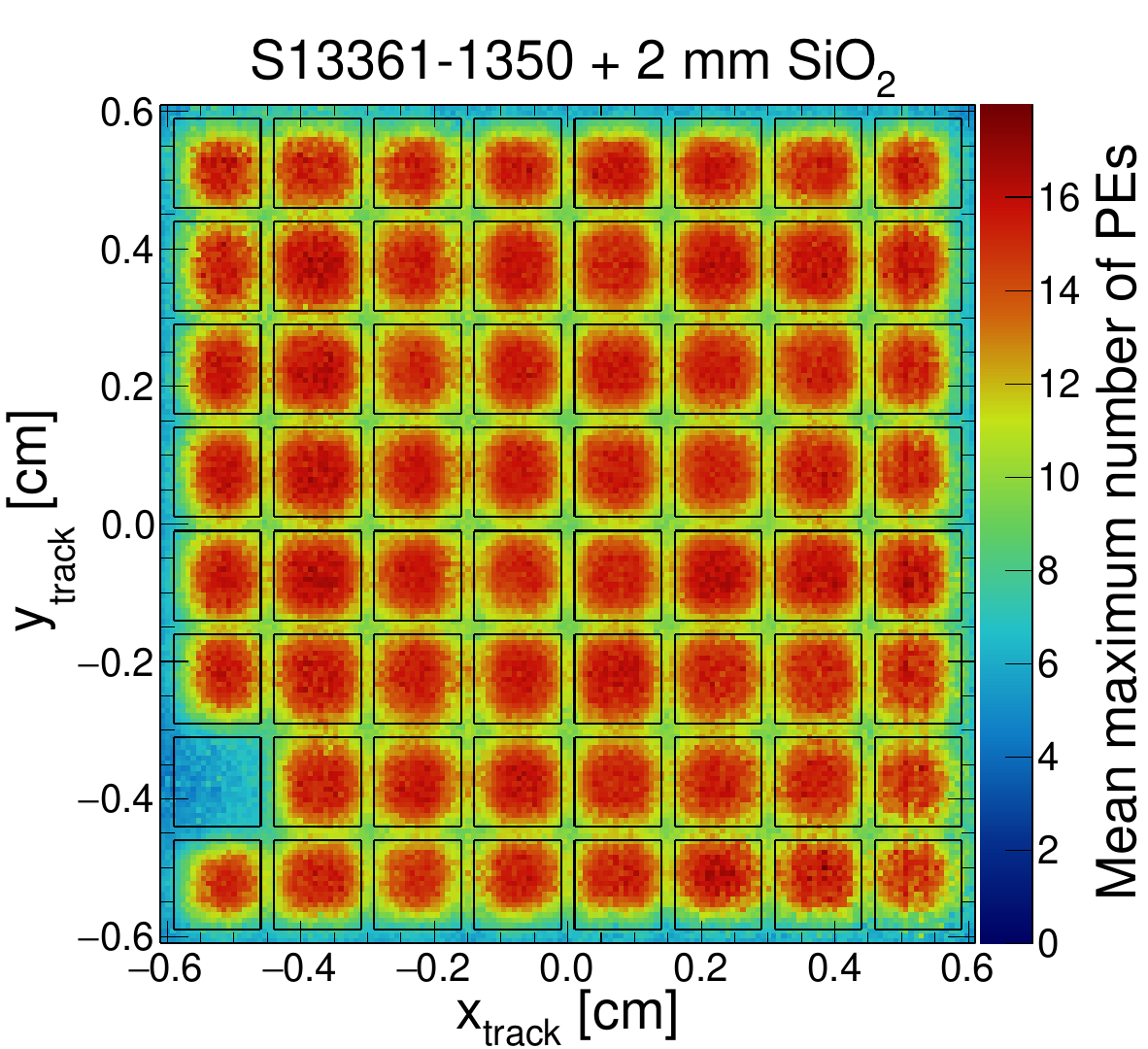}
\includegraphics[width=0.325\linewidth,trim=0mm 0mm 0mm 0mm,clip]{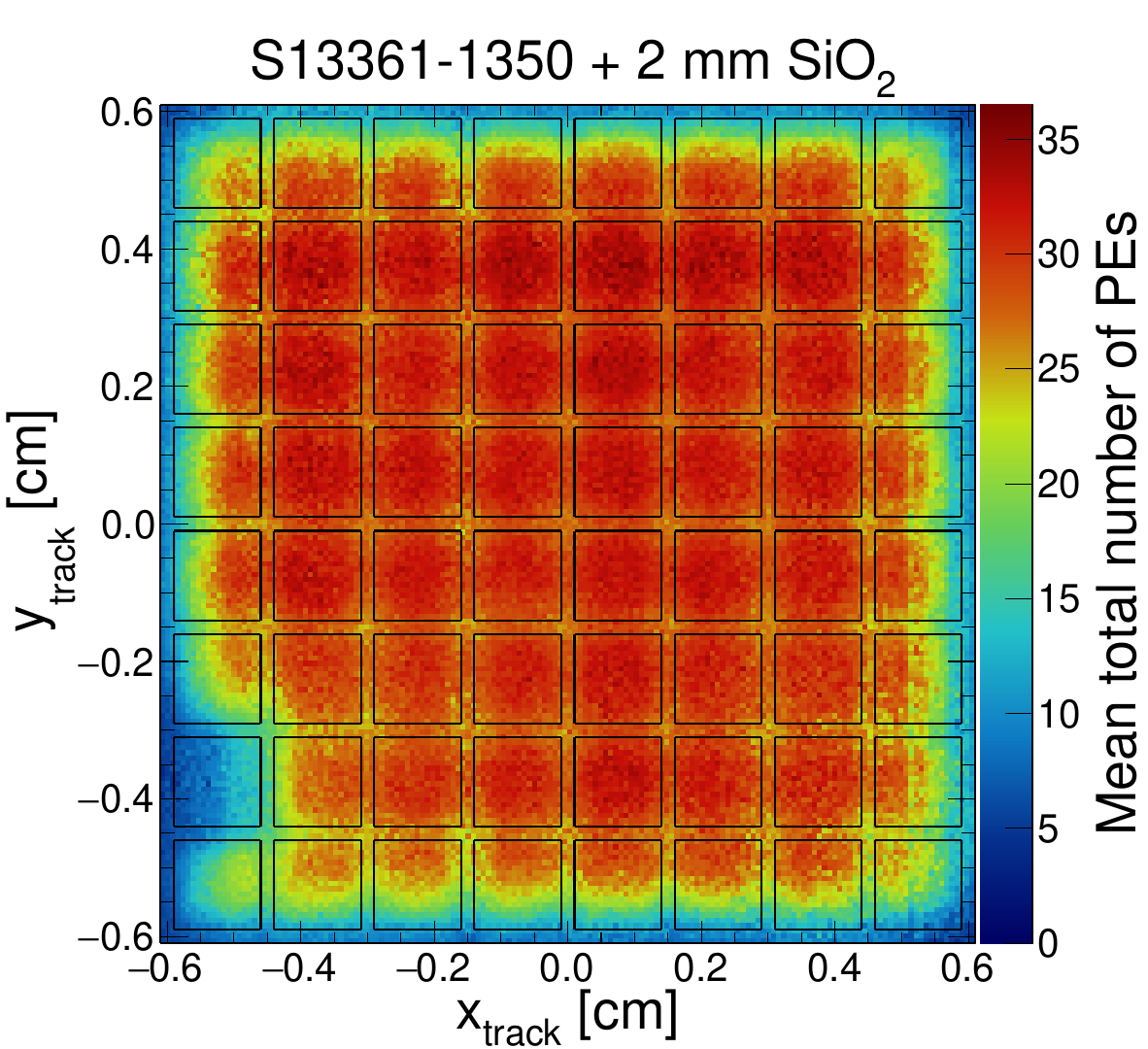}
\caption{Measured mean number of fired channels (left), mean maximum charge in a single SiPM (middle) and mean total number of photoelectrons (right) in the clusters as a function of the track impact point $(\text{x}_{\text{track}},\,\text{y}_{\text{track}})$ on the SiPM arrays for runs taken with the negatively charged beam at 10~GeV/$c$ using Petiroc boards. The plots refer to the configurations with the S13361-3075 array coupled to a 1~mm thick SiO$_2$ window (top) and the S13361-1350 array coupled to a 2~mm thick SiO$_2$ window  (bottom).
} 
\label{figure_Petiroc_scanXY_topology}
\end{figure} 

Figure~\ref{figure_Petiroc_scanXY_topology} shows the  mean number of fired channels in the clusters, the mean maximum charge in a single SiPM and the mean total number of photoelectrons in the charged track clusters measured with the Petiroc boards  as a function of the
track impact point ($\text{x}_\text{track},\,\text{y}_\text{track}$) for the S13361-3075 array coupled to a
1 mm thick SiO$_2$ window (top panels) and the S13361-1350 array coupled to a 2 mm thick SiO$_2$ window (bottom panels). Because of the presence of dead channels in the two arrays, we excluded the regions corresponding to those channels from the timing analysis.

The maps of the number of fired channels and of the cluster charge highlight how the available photoelectrons are shared among neighbouring SiPMs. For the S13361-3075 configuration, multi-channel clusters are predominantly observed when the track crosses boundaries between adjacent SiPMs. For the S13361-1350 configuration, instead, multiple channels are fired over the full impact region, reflecting the finer segmentation and the additional lateral spread induced by the thicker window. As expected, in both cases the mean total charge is approximately uniform across the array.

The maps of the mean maximum charge show that the largest values are reached for tracks incident near the centre of a SiPM, with a mean maximum up to about 40~PEs\ for the S13361-3075 array and about 17~PEs\ for the S13361-1350 array, as expected from considerations on the geometry.
Approaching the SiPM boundaries the maximum observed  charge decreases and the same charge is shared more equally among the neighbouring SiPMs. This is crucial for the timing performance achieved taking into account only the SiPM with maximum charge, or averaging the contributions of multiple SiPMs, as discussed in detail in Section~\ref{sec:timing_perforance}.

\begin{figure}[!t]
\centering
\includegraphics[width=0.49\linewidth,trim=0mm 0mm 0mm 0mm,clip]{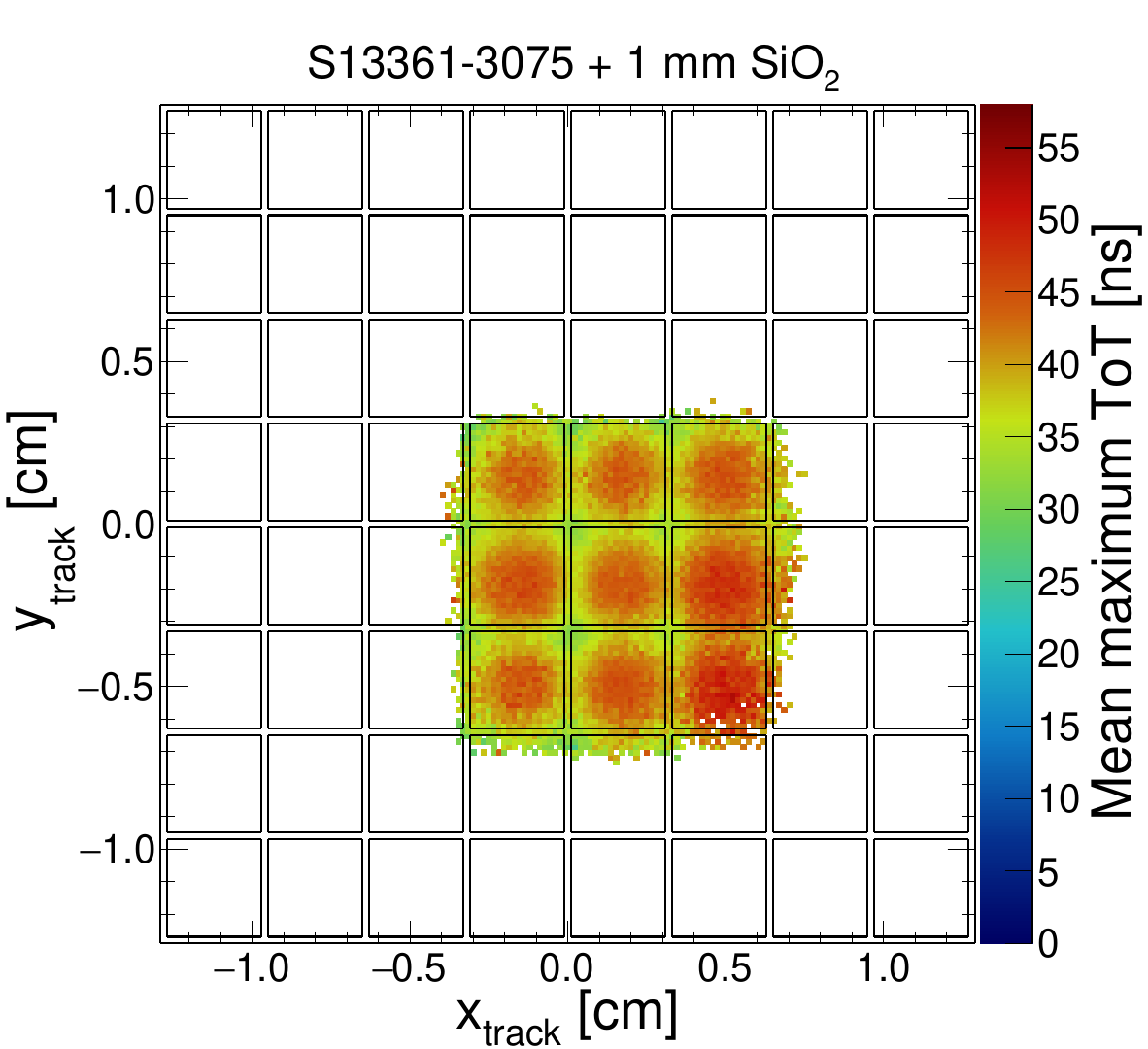}
\includegraphics[width=0.49\linewidth,trim=0mm 0mm 0mm 0mm,clip]{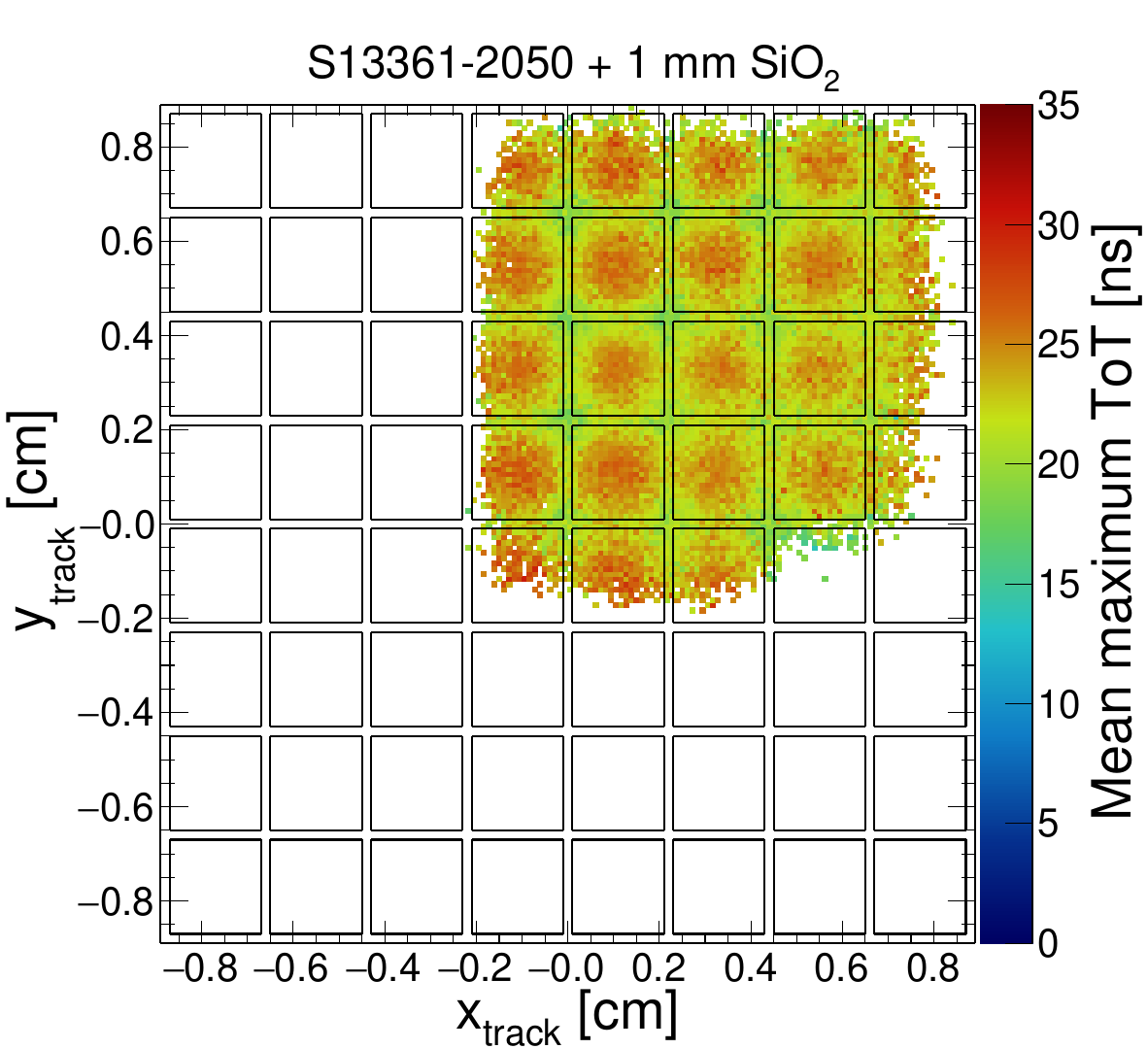}
\caption{Measured mean maximum ToT in a single SiPM in the clusters as a function of the track impact point $(\text{x}_{\text{track}},\,\text{y}_{\text{track}})$ on the SiPM arrays for runs taken with the negatively charged beam at 10~GeV/$c$ using RadioPico boards. The plots refer to the configurations with the S13361-3075 array coupled to a 1~mm thick SiO$_2$ window (left) and the S13361-2050 array coupled to a 1~mm thick SiO$_2$ window (right).}
\label{figure_Radioroc_scanXY_topology}
\end{figure}

To compare the response of the Petiroc board with the one of the RadioPico board, Figure~\ref{figure_Radioroc_scanXY_topology} shows the maps of the mean maximum ToT  in the charged track clusters measured with the RadioPico boards  as a function of the
track impact point ($\text{x}_\text{track},\,\text{y}_\text{track}$) for the S13361-3075 array coupled to a
1 mm thick SiO$_2$  window (left panel) and the S13361-2050 array coupled to a 1 mm thick SiO$_2$ window  (right panel).
The observed map follows the one of the maximum charge, with a mean maximum ToT of about 45-50~ns for the S13361-3075 array and 25-30~ns for the S13361-2050 array close to the center of the SiPMs.
Therefore, we considered the channel with the maximum ToT in the cluster as a reference for the characterization of the timing performance with the RadioPico boards.

\subsection{Charged particle detection efficiency}
\label{sec:efficiency}

\begin{figure}[!t]
    \centering
    \includegraphics[height=0.42\linewidth,width=0.495\linewidth,trim=0mm 0mm 14mm 5mm,clip]{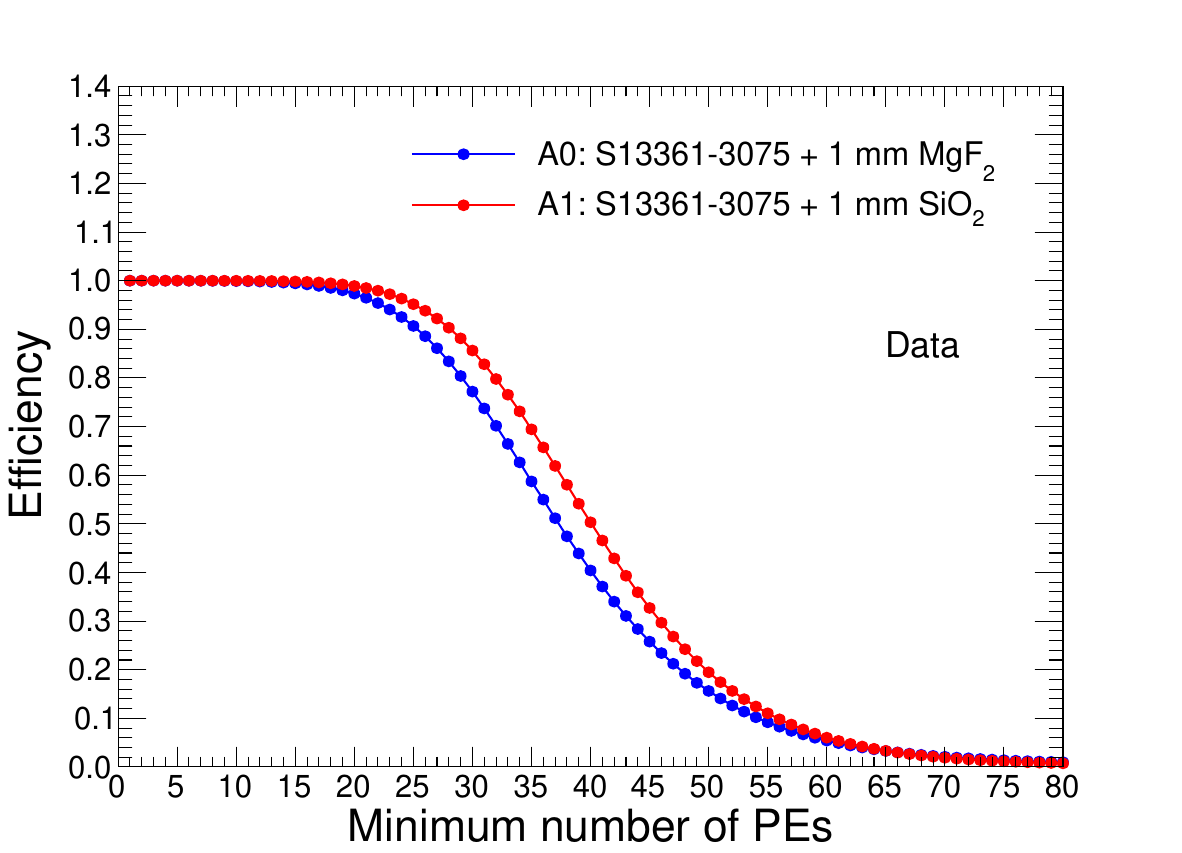}
    \includegraphics[height=0.42\linewidth,width=0.495\linewidth,trim=0mm 0mm 14mm 5mm,clip]{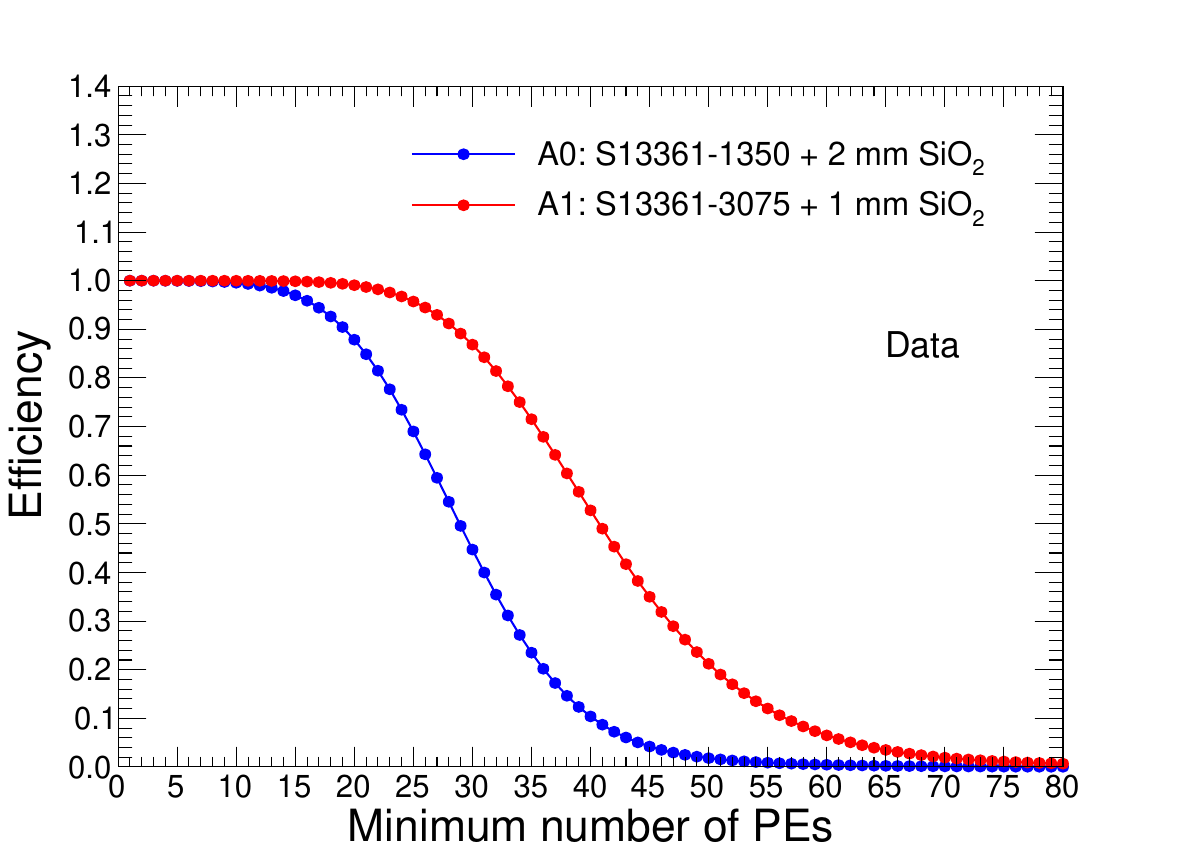}
    \includegraphics[height=0.42\linewidth,width=0.495\linewidth,trim=0mm 0mm 14mm 5mm,clip]{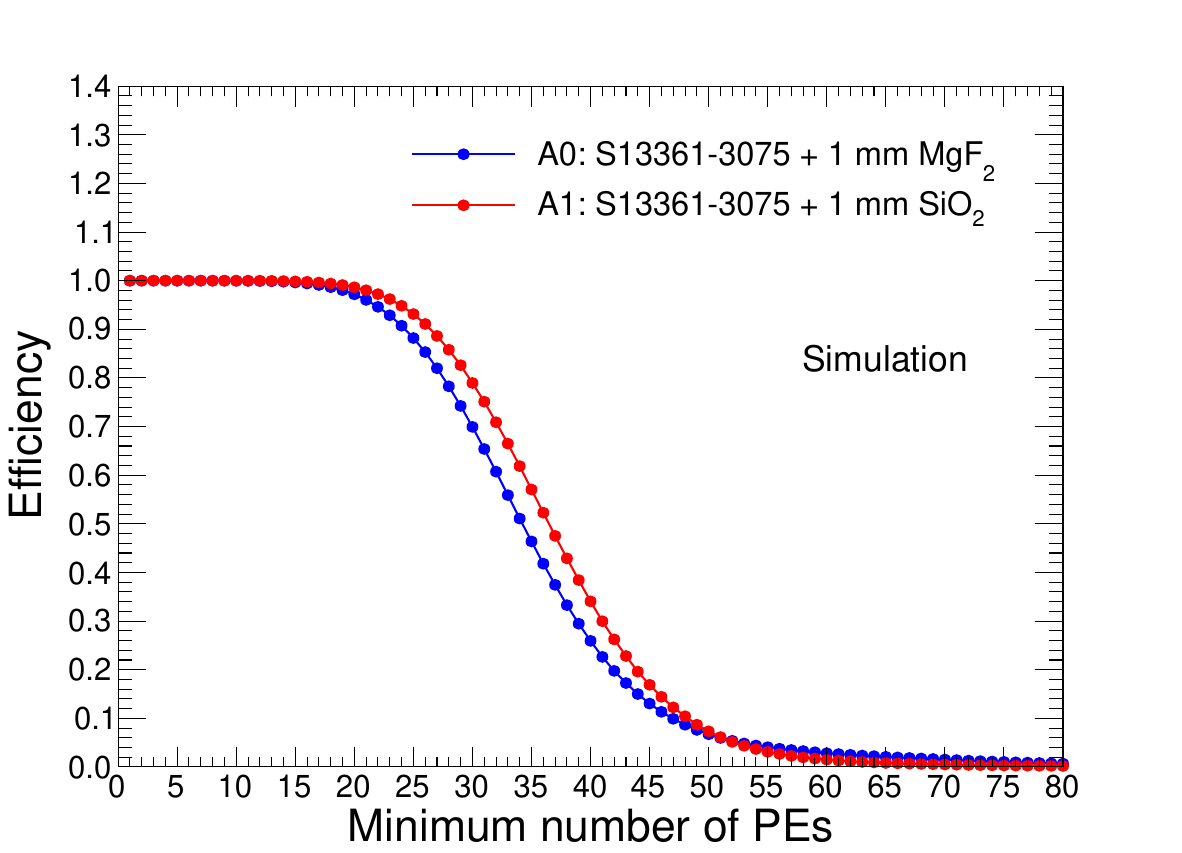}
    \includegraphics[height=0.42\linewidth,width=0.495\linewidth,trim=0mm 0mm 14mm 5mm,clip]{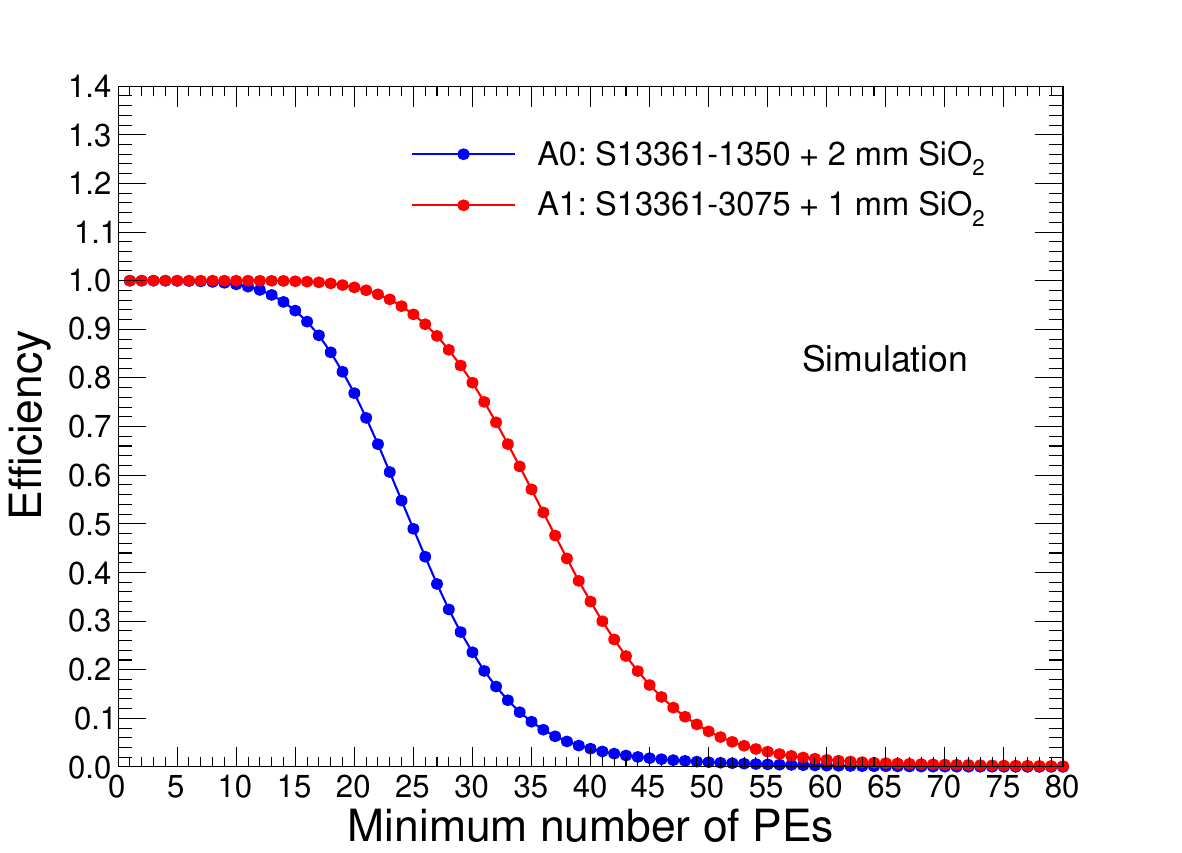}
    \caption{Measured (top) and simulated (bottom) charged-particle detection efficiency as a function of the minimum number of photoelectrons $(N_{\text{PE}} \geq1)$ required in the clusters for runs taken using Petiroc boards. The left and right panels refer to the configurations with the S13361-3075 array coupled to a 1 mm thick SiO$_2$ window (A1) and either the S13361-3075 coupled to a 1 mm thick MgF$_2$ window or the S13361-1350 coupled to a 2~mm thick SiO$_2$ window (A0).
    }
    \label{fig:all_efficiencies}
\end{figure}

For a given combination of A0 and A1, we evaluated the efficiency of A0 (A1) as the 4-fold/3-fold coincidence ratio, defined as the fraction of events with signals in both tracker planes and in A1 (A0) for which a cluster in the array under study met or exceeded the required minimum number of PEs.
The left panels of Figure~\ref{fig:all_efficiencies} show the measured and the simulated efficiency as a function of the minimum number of PEs required in the clusters for runs taken using Petiroc boards for the configuration with the S13361-3075 array coupled to a 1 mm thick SiO$_2$ window (A1) and the S13361-3075 array coupled to a 1 mm thick MgF$_2$ window (A0). The right panels of the same figure show the corresponding results with the S13361-1350 array coupled to a 2 mm thick SiO$_2$ window used as A0.

We found efficiency values above 99.9\% when requiring up to 14 and 7~PEs in the clusters of the S13361-3075 and S13361-1350 arrays, respectively, and above 99\% when requiring up to 12 and 20~PEs, respectively, and we found intermediate values for tests with the S13361-2050 arrays.
The measured values are consistent with the simulation predictions and directly follow from the considerations reported in Section~\ref{sec:cluster_size}.

These results demonstrate the high efficiency of the adopted timing strategy, while allowing operation at threshold values that can suppress the background from the SiPM DCR even in very high-radiation environments. This approach is therefore a solid option for timing applications in future high-energy physics experiments, where radiation tolerance is a primary requirement.

\subsection{Time resolution}

\label{sec:timing_perforance}

We performed timing studies by comparing the ToAs measured by the A0 and A1 arrays, using one of the two as a reference time for the other.
We corrected the raw time differences for channel- by-channel time offsets, due to the different routing of the channels, and the time-walk effect, accounting for signals with the same shape but increasing amplitudes systematically crossing the discriminator threshold at earlier times.
We subtracted the global time-of-flight offset as well.
We determined all the required corrections in a single step using dedicated calibration runs and a lookup-table procedure. In particular, for data taken with the Petiroc boards, we built, for each combination of A0 and A1, a profile of the time difference between signal hits for every A0-A1 SiPM pair $(i,\,j)$ and as a function of the number of photoelectrons in the two SiPMs $(q_i,\,q_j)$. 
We took the mean value of each time profile bin $\mu(i,j,q_i,q_j)$ as the corresponding lookup-table correction. 
Finally, in the timing analysis we corrected the raw time differences $\Delta t_{\text{raw}}$ by subtracting the lookup-table value associated with the considered SiPM pair and number of photoelectron pair as:
\begin{equation}
     \Delta t_{\text{corr}} (i,j, q_i, q_j) = \Delta t_{\text{raw}}(i,j) - \mu(i,j,q_i,q_j)\,.
\end{equation}
For the data taken with the RadioPico boards, we followed the same approach, based on pairs of ToT intervals $(\text{ToT}_i,\text{ToT}_j)$ rather than on photoelectron-count pairs. We used ToT bins corresponding to three ToT LSBs, i.e.\ 585~ps. We thus derived the corrections $\mu(i,j,\text{ToT}_i,\text{ToT}_j)$ and finally evaluated the corrected time differences as:

\begin{equation}
     \Delta t_{\text{corr}} (i,j, \text{ToT}_i, \text{ToT}_j) = \Delta t_{\text{raw}}(i,j) - \mu(i,j,\text{ToT}_i,\text{ToT}_j)\,.
\end{equation}

\subsubsection{Analysis using channels with maximum charge}

\begin{figure}[!t]
    \centering
    \includegraphics[height=0.42\linewidth,width=0.495\linewidth,trim=0mm 0mm 14mm 5mm,clip]{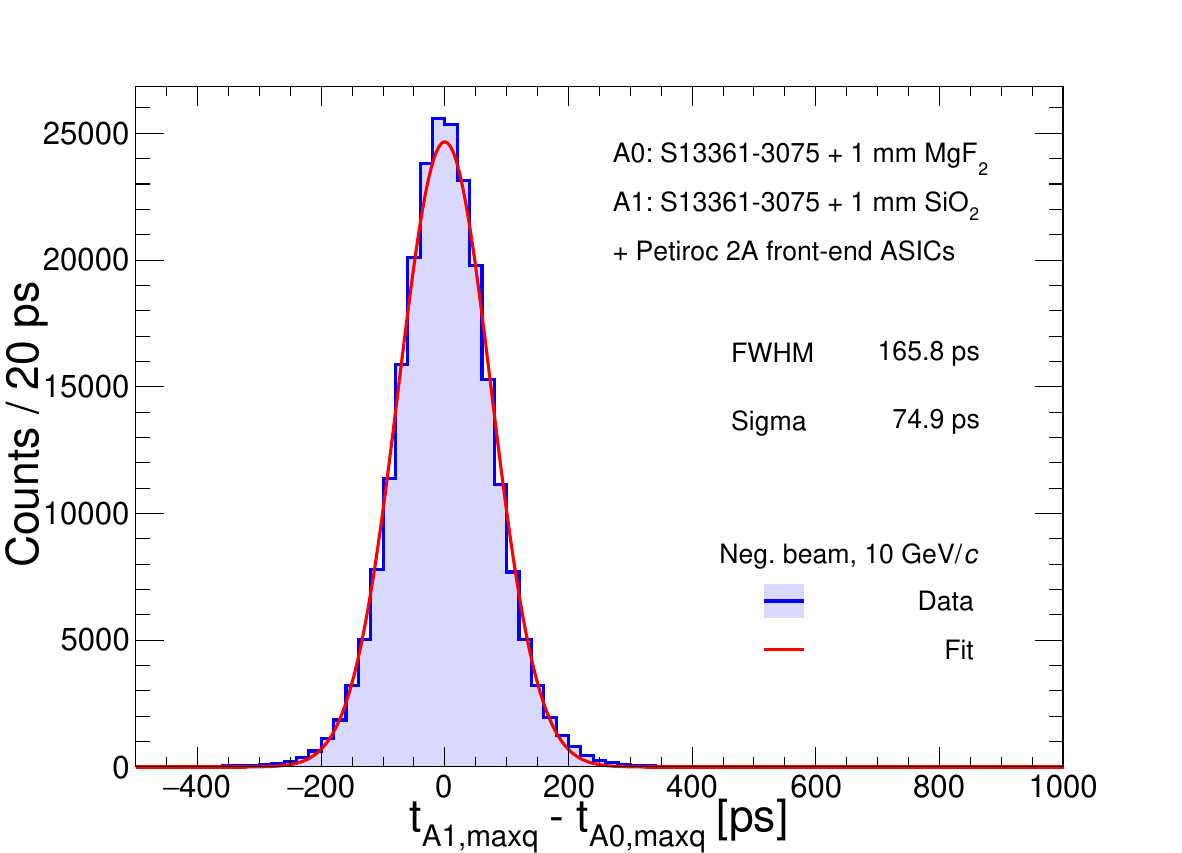}
    \includegraphics[height=0.42\linewidth,width=0.495\linewidth,trim=0mm 0mm 14mm 5mm,clip]{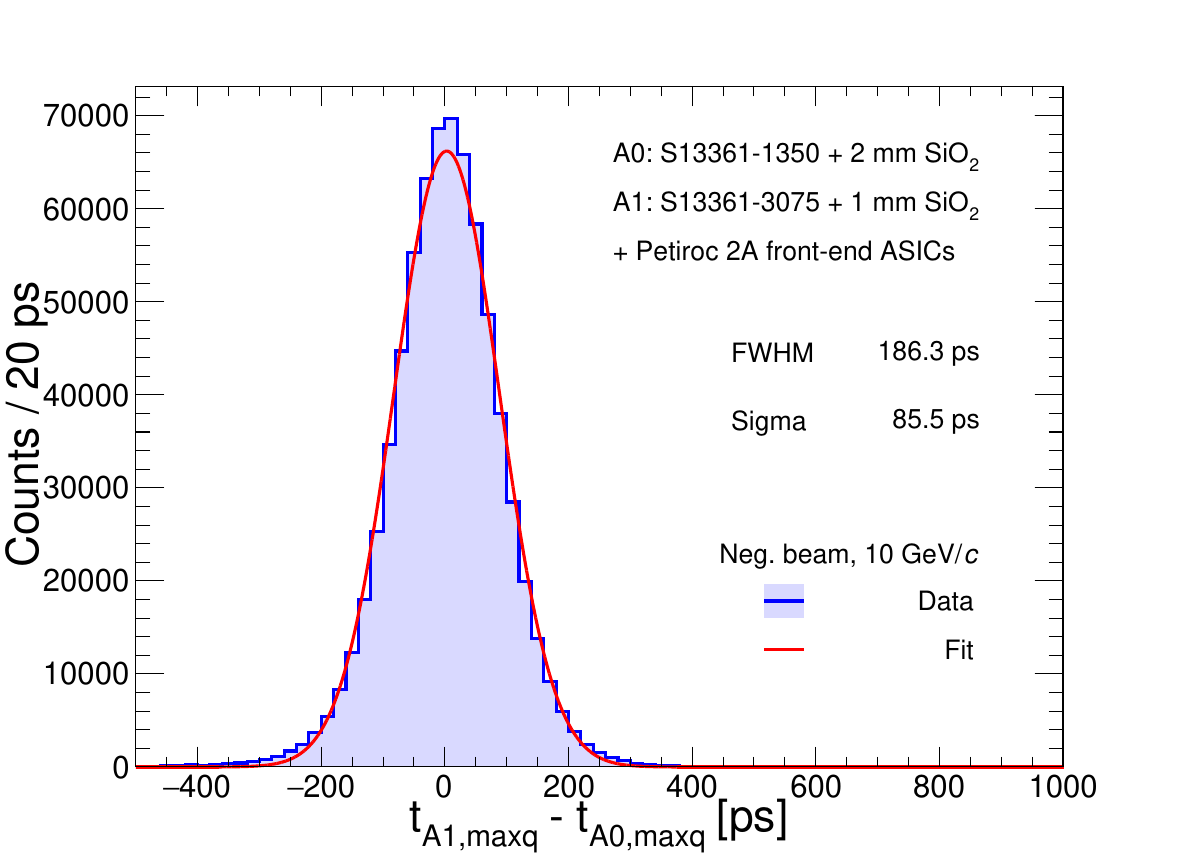}
    \caption{Measured distributions of the time difference between the SiPMs with maximum charge in the clusters for runs taken using Petiroc boards. The left and right panels refer to the configurations with the S13361-3075 array coupled to a 1 mm thick SiO$_2$ window (A1), and either the S13361-3075 array coupled to a 1 mm thick MgF$_2$ window or the S13361-1350 array coupled to a 2 mm thick SiO$_2$ window (A0). The Gaussian fit to the distributions is also shown.
    }
    \label{fig:timing_Petiroc_all}
\end{figure}

The left panel of Figure~\ref{fig:timing_Petiroc_all} shows the resulting distribution of the time difference between the SiPMs with maximum charge for the configuration with the S13361-3075 array coupled to a 1 mm thick SiO$_2$ window (A1) and the S13361-3075 array coupled to a 1 mm thick MgF$_2$ window (A0). The right panel of the same figure shows the corresponding distribution with the S13361-1350 array coupled to a 2 mm thick SiO$_2$ window used as A0.
The Gaussian fit to the distributions yields a resolution on the time difference $\sigma_{\Delta t,\text{maxq}}$ of about $74.9$~ps and $85.5$~ps, respectively. 

\begin{figure}[!t]
    \centering
    \includegraphics[height=0.425\linewidth,width=0.495\linewidth,trim=0mm 0mm 6mm 8mm,clip]{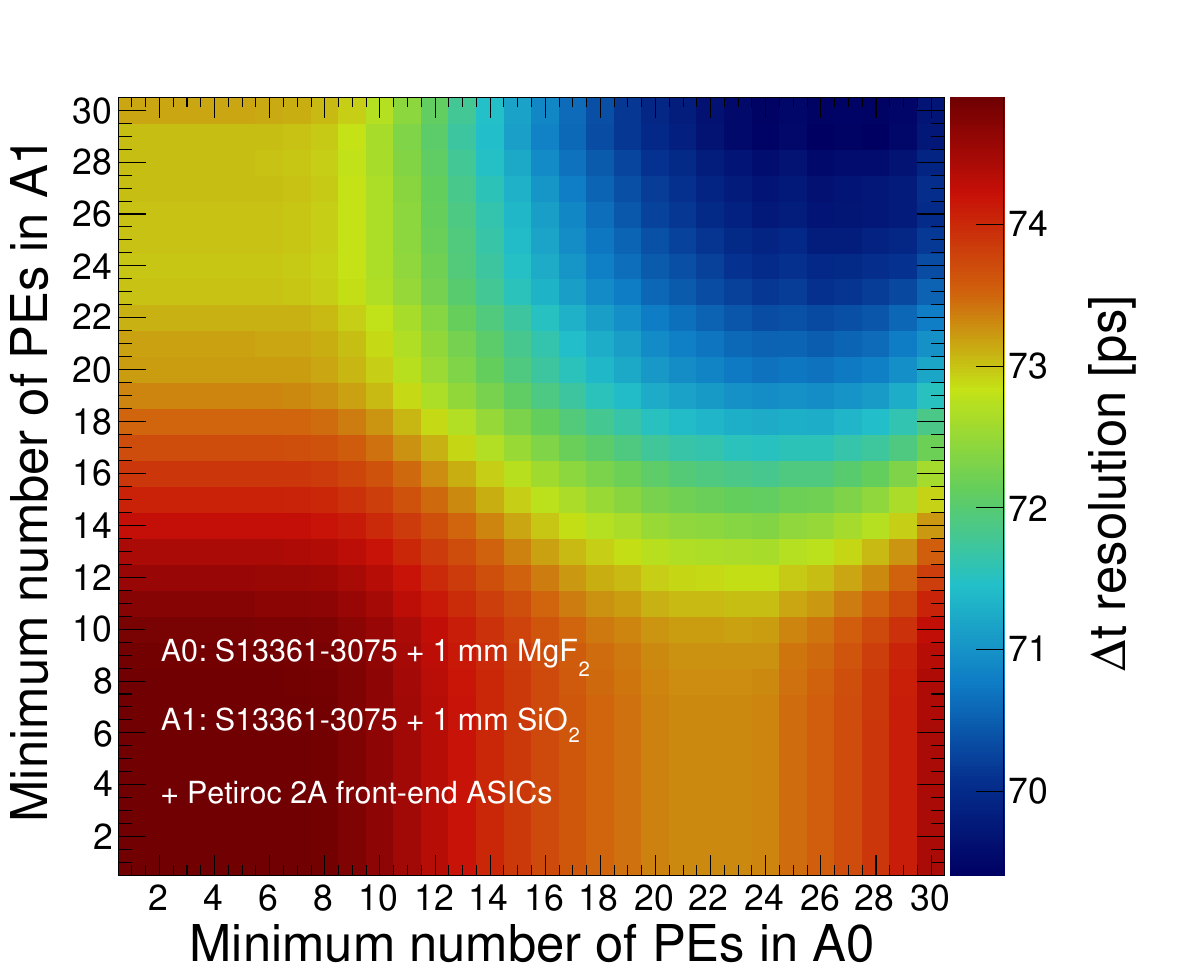}
    \includegraphics[height=0.42\linewidth,width=0.495\linewidth,trim=0mm 0mm 17mm 8mm,clip]{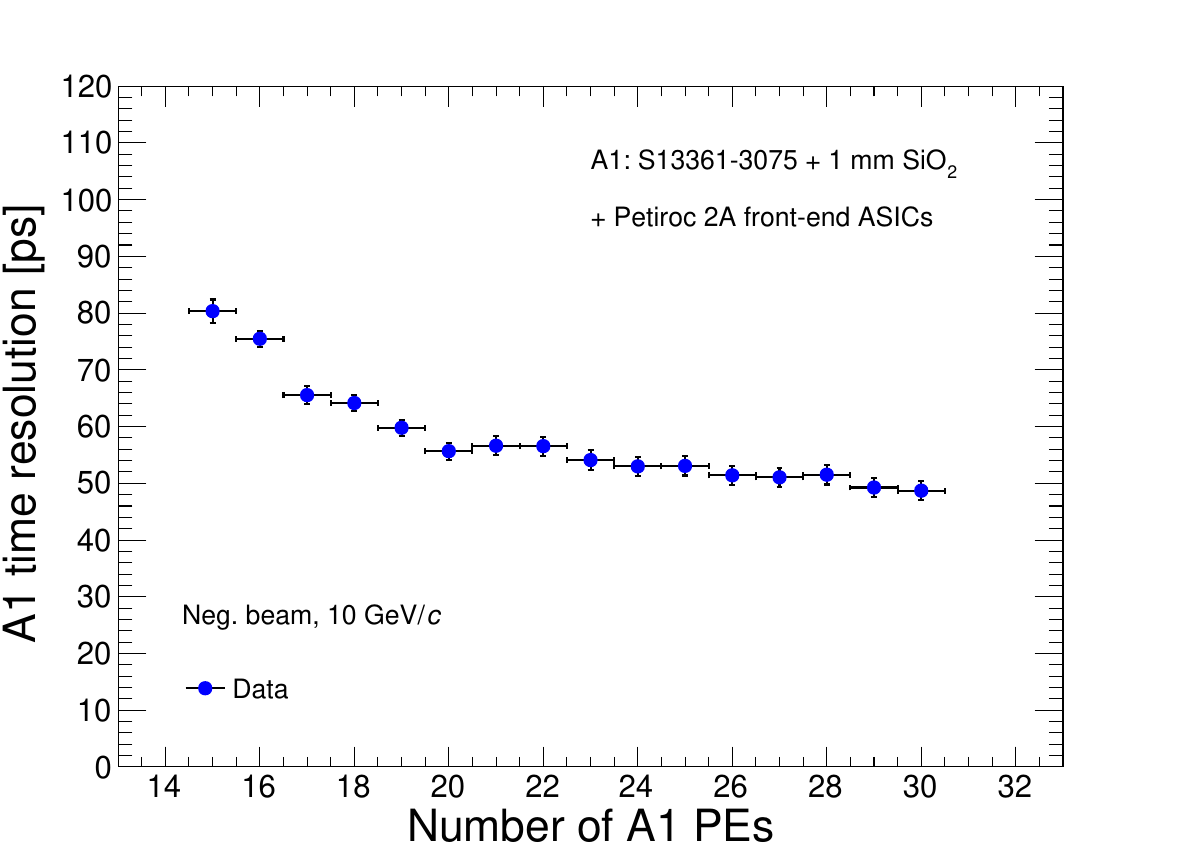}
    \caption{Left: Resolution of the time difference between the times of the SiPMs with maximum charge in A1 and A0 as a function of the minimum number of PEs required in both A0 and A1 for runs taken using Petiroc boards. Right: Extrapolation of the A1 time resolution as a function of the number of PEs in a single SiPM. The plots refer to the configuration with S13361-3075 arrays coupled to a 1~mm thick SiO$_2$ window (A1) and to a 1~mm thick MgF$_2$ (A0). } 
    \label{fig:timing_Petiroc_fine}
\end{figure}

From the left panel of Figure~\ref{fig:timing_Petiroc_fine}, which shows the measured resolution of the time difference as a function of the minimum number of photoelectrons required in both A0 and A1 for the same configuration as the left panel of Figure~\ref{fig:timing_Petiroc_all}, it can be observed that the contributions of the two S13361-3075 arrays with 1~mm SiO$_2$ and MgF$_2$ is almost identical. Since the measurements by the two arrays are independent, the observed 74.9~ps resolution on the time difference is, as a first approximation, the result of the sum in quadrature of two equal contributions. Therefore, this corresponds  to a  single-array time resolution $\sigma_{t,\text{maxq}}$ of approximately $\sigma_{t,\text{maxq}} = \sigma_{\Delta t,\text{maxq}}/\sqrt{2}=74.9/\sqrt{2}\approx 53.0$~ps overall for both the S13361-3075 arrays read out with the Petiroc boards.
Subtracting this contribution in quadrature from the $\sigma_{\Delta t,\text{maxq}}$ of the distribution in the right panel of Figure~\ref{fig:timing_Petiroc_all} it is possible to extrapolate a  $\sigma_{t,\text{maxq}}\approx 67.1$~ps for the S13361-1350 array with a 2~mm thick SiO$_2$ window.

The left panel of Figure~\ref{fig:timing_Petiroc_fine} also shows that, requiring more than 30 PEs in both the S13361-3075 arrays, a $\sigma_{\Delta t,\text{maxq}}\approx70$~ps is achieved, corresponding to a $\sigma_{\ t,\text{maxq}}(N_{\text{PE}}>30)\approx49.5$~ps for each of the S13361-3075 arrays. 
By scanning the $\sigma_{\Delta t,\text{maxq}}$ values obtained at fixed $N_{\text{PE}}$ in A0 (A1) while requiring $N_{\text{PE}}>30$ in A1 (A0), and then swapping the roles of A0 and A1, we found discrepancies in $\sigma_{\Delta t,\text{maxq}}$ below 2.0~ps. We therefore assigned a systematic uncertainty of $2.0/\sqrt{2}\simeq1.4$~ps to the extrapolated single array $\sigma_{t,\text{maxq}}(N_{\text{PE}}>30)$.

The right panel of Figure~\ref{fig:timing_Petiroc_fine} shows the extrapolated $\sigma_{t, \text{maxq}}$ as a function of the number of PEs for the S13361-3075 array coupled to a 1~mm thick SiO$_2$ window. We obtained the reported values
by first extracting the  $\sigma_{\Delta t, \text{maxq}}$ relative to the  S13361-3075 array coupled to a 1~mm thick window made of MgF$_2$ by scanning on the number of PEs in that array and requiring more than 30 PEs in the reference array. We finally subtracted in quadrature the reference  $\sigma_{t,\text{maxq}}(N_{\text{PE}}>30)$ from all the measured points to get the single-array resolution at a given number of PEs. 
We observed an overall improvement of $\sigma_{t,\text{maxq}}$ from 80~ps to 49~ps with the number of PEs increasing from 14 to 30.

\begin{figure}[!t]
    \centering   \includegraphics[height=0.42\linewidth,width=0.495\linewidth,trim=0mm 0mm 14mm 8mm,clip]{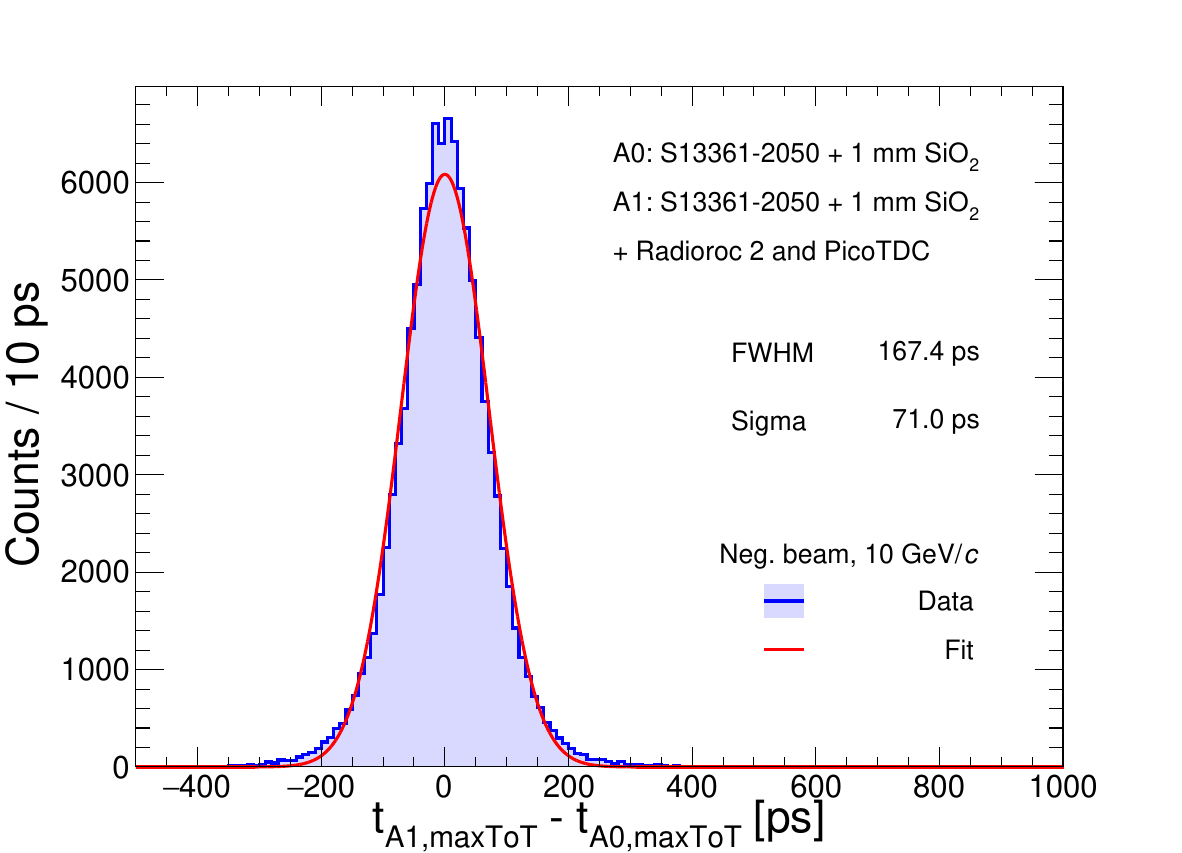}
    \includegraphics[height=0.42\linewidth,width=0.495\linewidth,trim=0mm 0mm 14mm 8mm,clip]{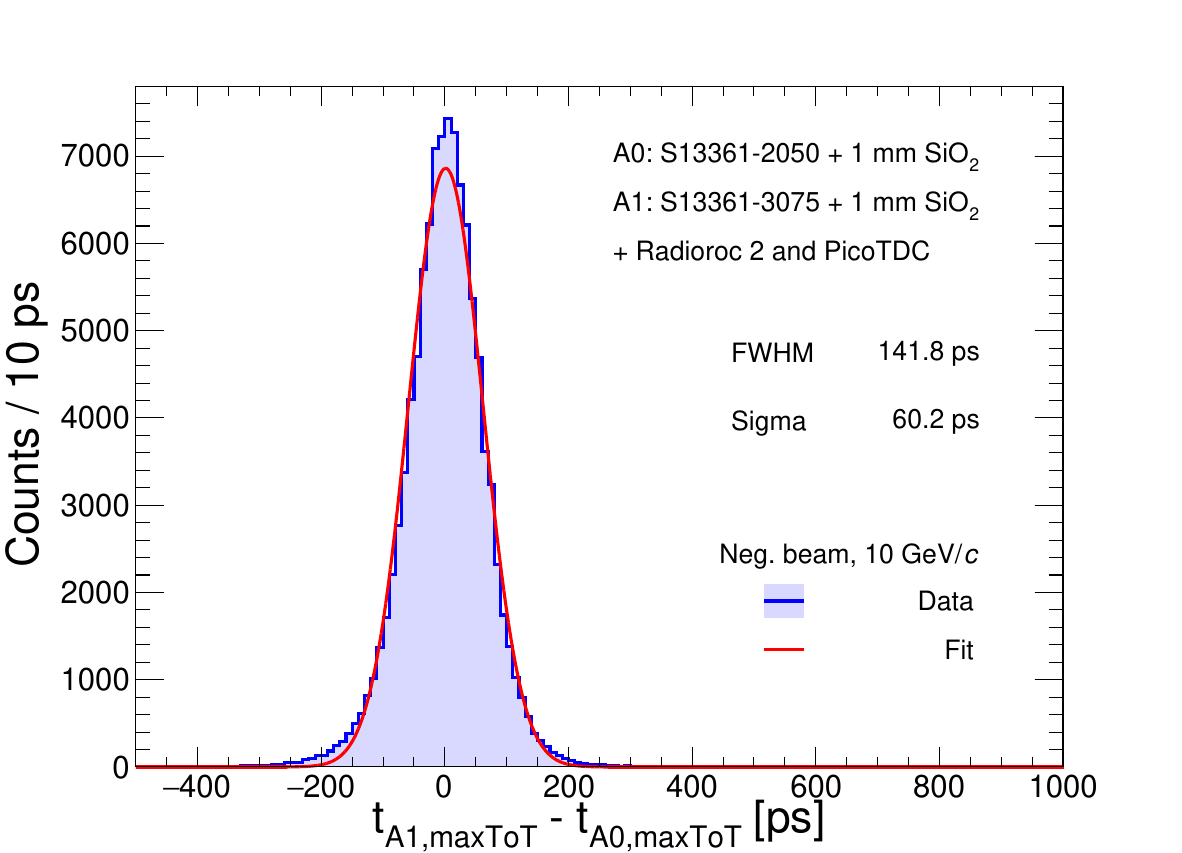}
    \caption{Measured distributions of the time difference between the SiPMs with maximum ToT in the clusters for runs taken using RadioPico boards. The left and right panels refer to the configurations with the S13361-2050 array coupled to a 1 mm thick SiO$_2$ window (A0), and either another S13361-2050 array coupled to a 1 mm thick SiO$_2$ or the S13361-3075 array coupled to a  1 mm thick  SiO$_2$ window (A1). The Gaussian fit to the distributions is also shown.
    }
    \label{fig:timing_picoTDC_2024}
\end{figure}

In order to highlight the impact of the front-end on the time resolution, Figure~\ref{fig:timing_picoTDC_2024} shows examples of the measured distributions on the time difference between the SiPMs with maximum ToT in the clusters for runs taken with the RadioPico boards.
The left panel refers to the configuration with
S13361-2050 arrays coupled to a 1 mm thick SiO$_2$ windows used as both A0 and A1.  The right panel of the same figure shows the corresponding distribution with the S13361-3075 array coupled to a 1 mm thick SiO$_2$ window used as A0.

In both cases, the expected resolution on the time difference would be intermediate relative to the configurations in Figure~\ref{fig:timing_Petiroc_all} if they were read out using the Petiroc boards. 
However, the Gaussian fit to the distributions yields a resolution on the time difference $\sigma_{\Delta t, \text{maxToT}}$ of about 71.0~ps and 60.2~ps, respectively.
Assuming equal contribution from the two S13361-2050 arrays, the measured 71.0~ps correspond to an overall single-array time resolution  $\sigma_{t, \text{maxToT}}\approx50.2$~ps. 
Subtracting this value in quadrature to the $\sigma_{\Delta t, \text{maxToT}}$ of the right panel of Figure~\ref{fig:timing_Petiroc_all}, we extrapolated a $\sigma_{t, \text{maxToT}}\approx33.2$~ps for the S13361-3075 array coupled to a 1~mm thick SiO$_2$ window.
Compared to the corresponding $\sigma_{t,\text{maxq}}\approx53.0$~ps measured with the Petiroc boards, this result highlights the significantly better timing performance of the RadioPico boards and represents the best resolution measured in this study.

\subsubsection{Analysis using the cluster mean time}

The presence of more fired channels in the same cluster improves the achievable time resolution by averaging the respective timestamps.
The effect is expected to be particularly relevant for windows thicker than the size of the considered SiPMs, as shown in Figure~\ref{fig:sigt_d}. 
In addition, the gain in time resolution also depends on the track impact region in the SiPMs of the array, as shown in Figure~\ref{figure_Petiroc_scanXY_topology}. 

\begin{figure}[!t]
\centering
\includegraphics[width=0.325\linewidth,trim=0mm 0mm 10mm 0mm,clip]{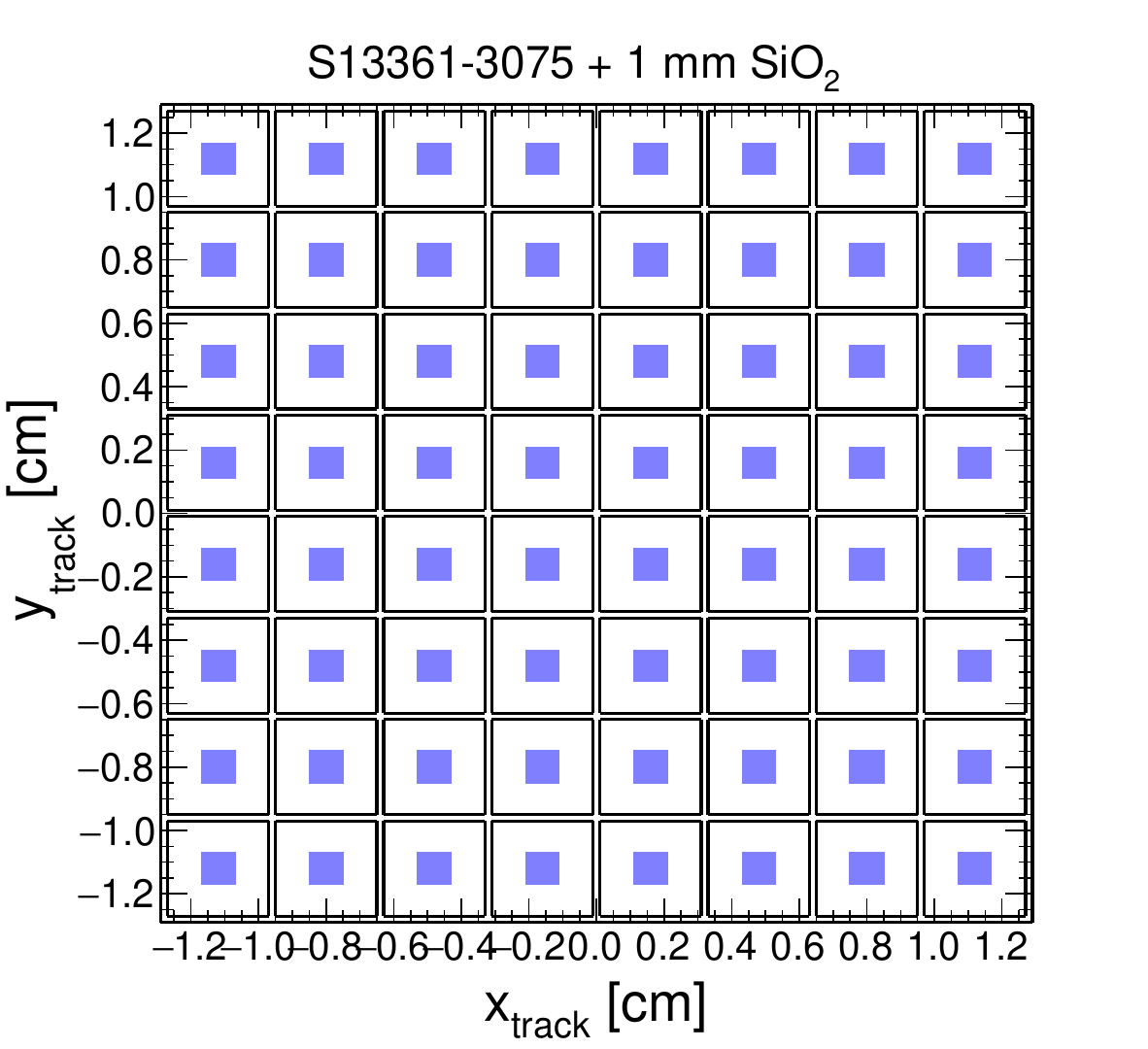}
\includegraphics[width=0.325\linewidth,trim=0mm 0mm 10mm 0mm,clip]{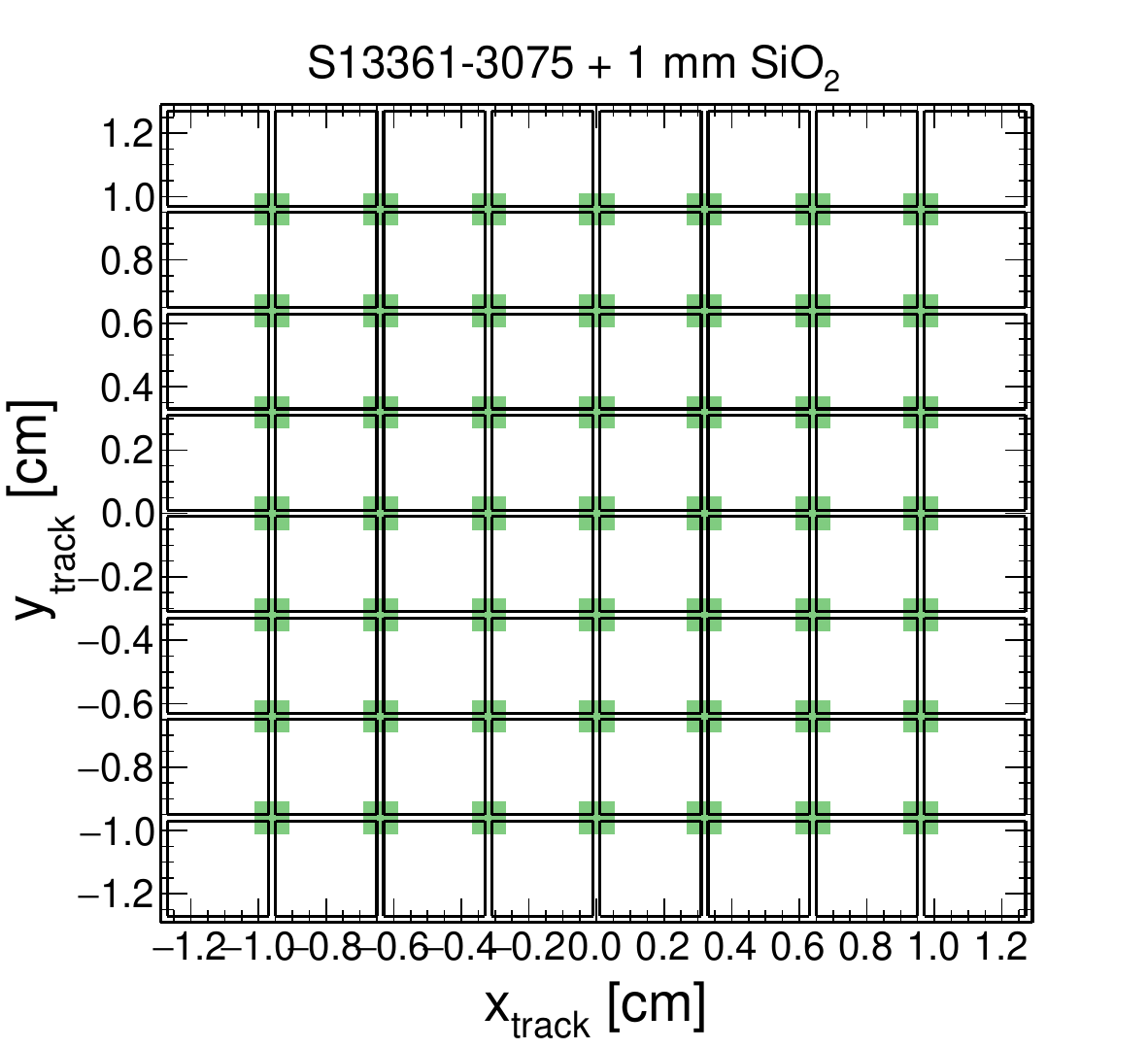}
\includegraphics[width=0.325\linewidth,trim=0mm 0mm 10mm 0mm,clip]{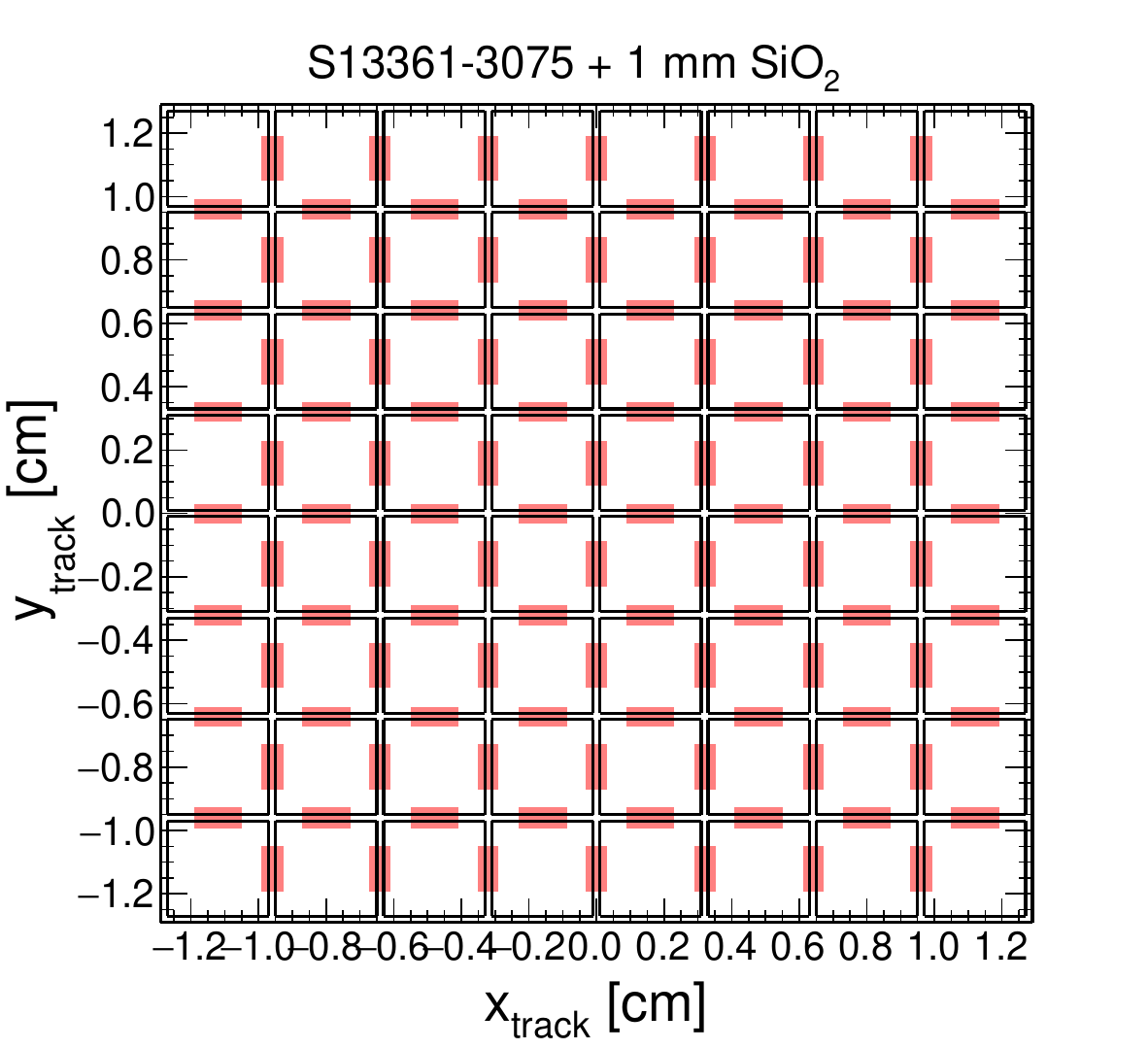}
\includegraphics[height=0.42\linewidth,width=0.49\linewidth,trim=0mm 0mm 15mm 7mm,clip]{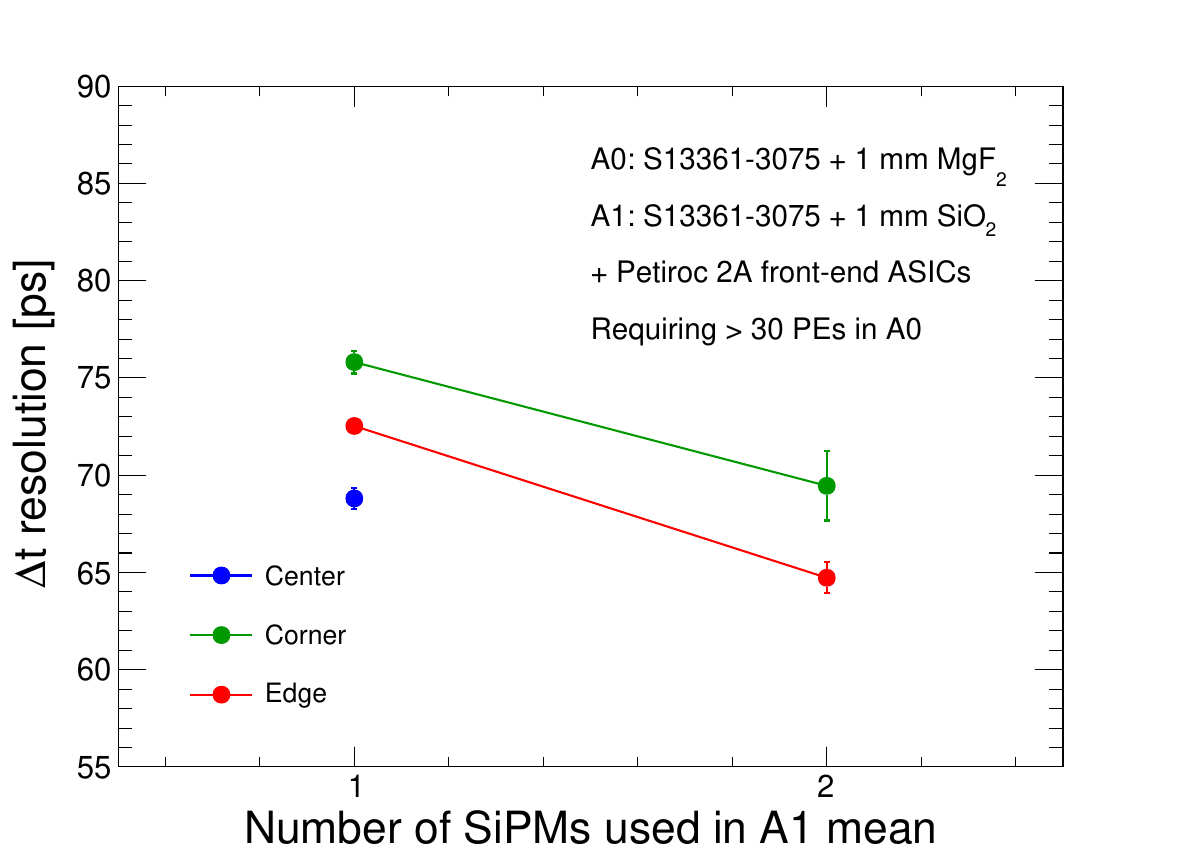}
\includegraphics[height=0.42\linewidth,width=0.49\linewidth,trim=0mm 0mm 15mm 7mm,clip]{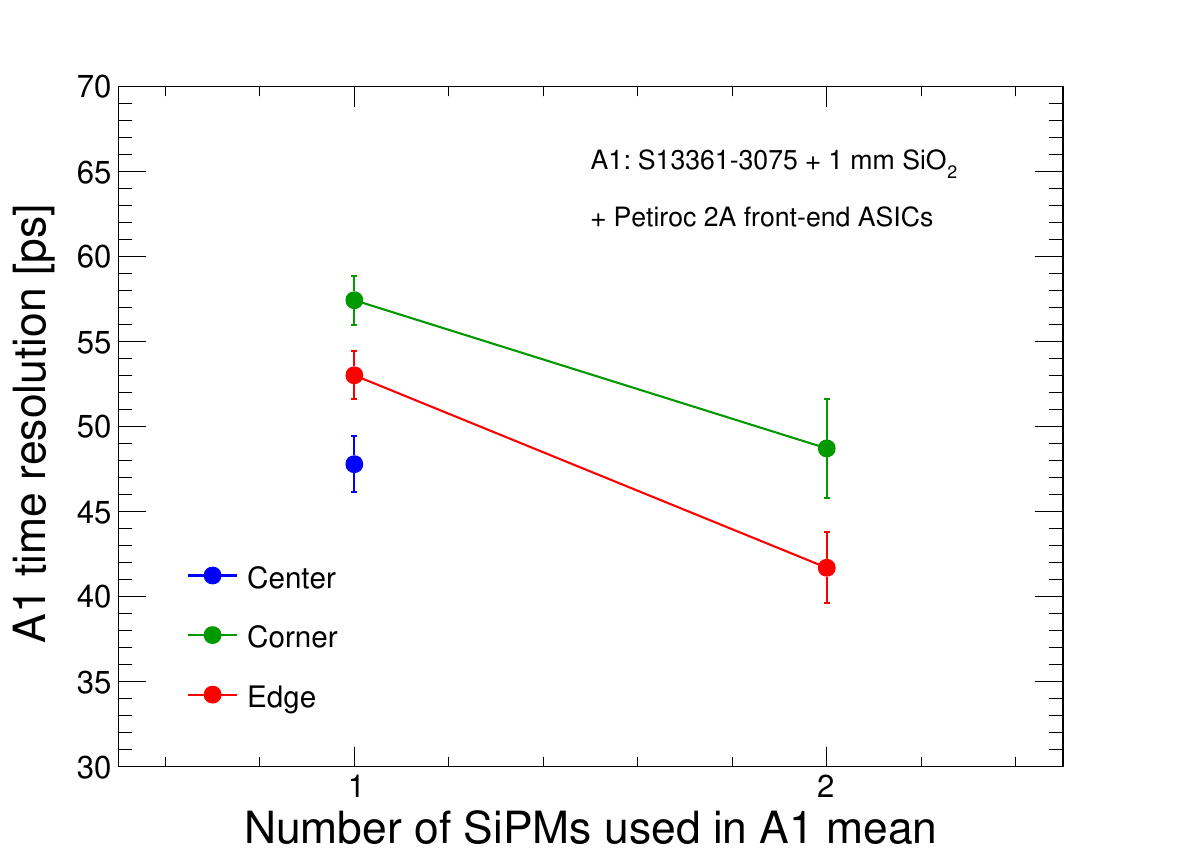}
\caption{Top: Illustration of the regions used for the time-resolution analysis as a function of the impact position in each SiPM: central region (left), corners shared by four adjacent SiPMs (middle), and edges shared by two adjacent SiPMs (right), for the S13361-3075 SiPM arrays.
Bottom: Resolution of the time difference between the A1 mean ToA and the A0 ToA of the maximum-charge SiPM as a function of the number $N$ of SiPMs used in the A1 mean. Events with a maximum charge of at least 30~PE in A0 and a minimum charge of 16~PEs in each SiPM entering the A1 mean are considered. The plots refer to runs taken using Petiroc boards for the configuration with S13361-3075 arrays coupled to a 1~mm thick SiO$_2$ window (A1) and to a 1~mm thick MgF$_2$ window (A0). The right panel reports the resulting A1 resolution after subtracting in quadrature the A0 contribution of $(49.5\pm1.4)$~ps.
}
\label{figure_Petiroc_scanXY_fiducial_regions_1}
\end{figure}

\begin{figure}[!t]
\centering
\includegraphics[width=0.32\linewidth,trim=0mm 0mm 10mm 0mm,clip]{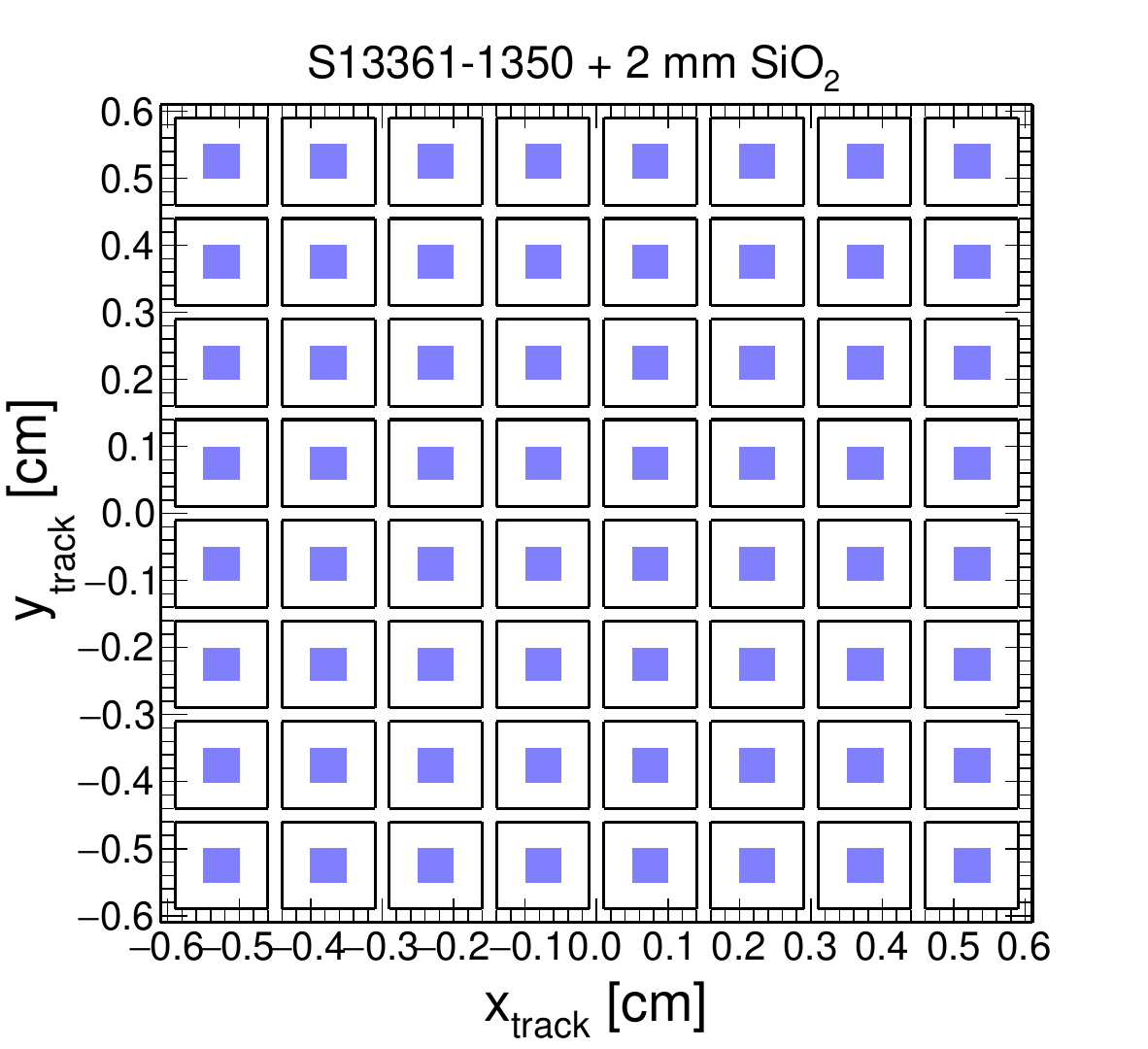}
\includegraphics[width=0.32\linewidth,trim=0mm 0mm 10mm 0mm,clip]{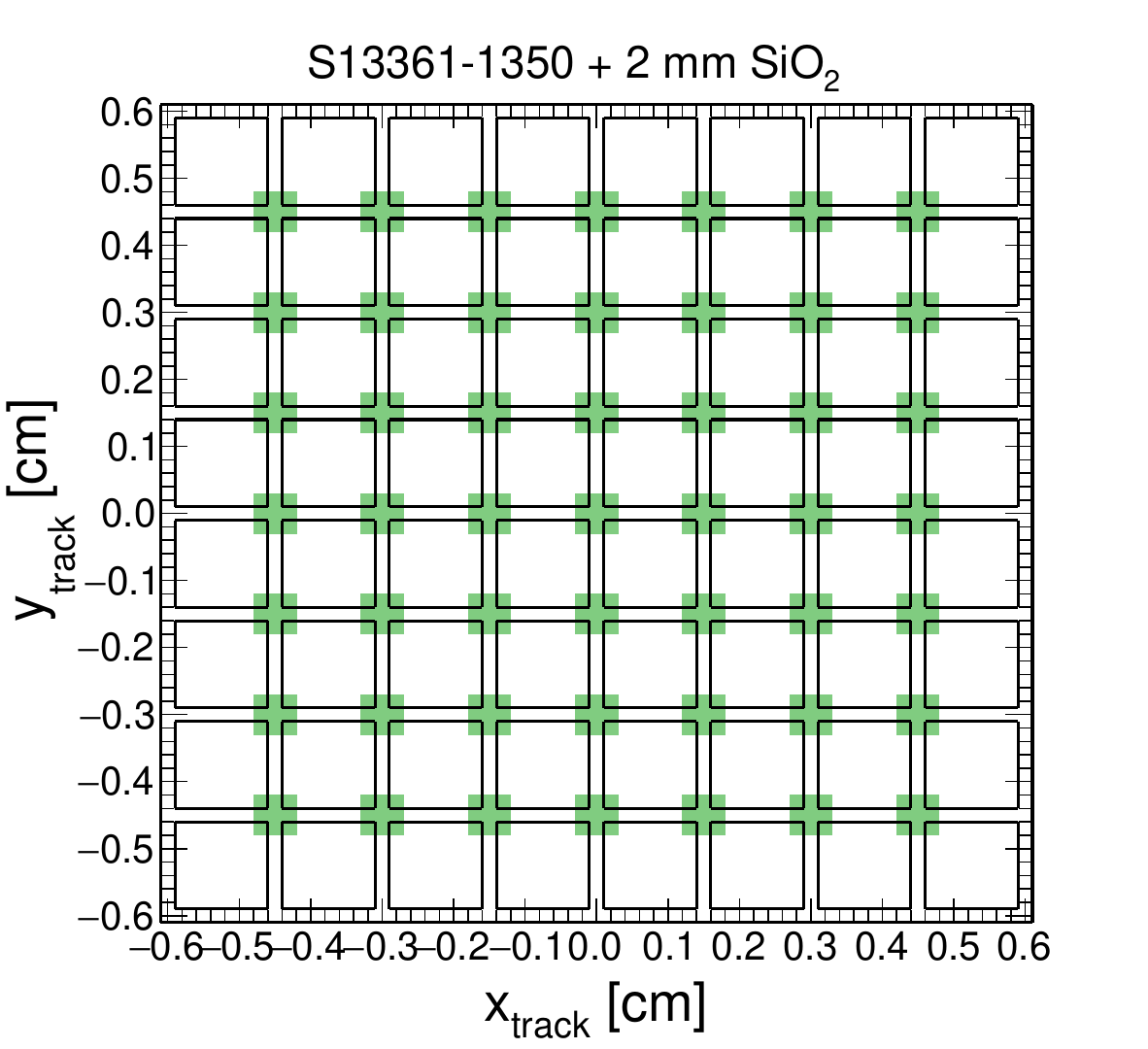}
\includegraphics[width=0.32\linewidth,trim=0mm 0mm 10mm 0mm,clip]{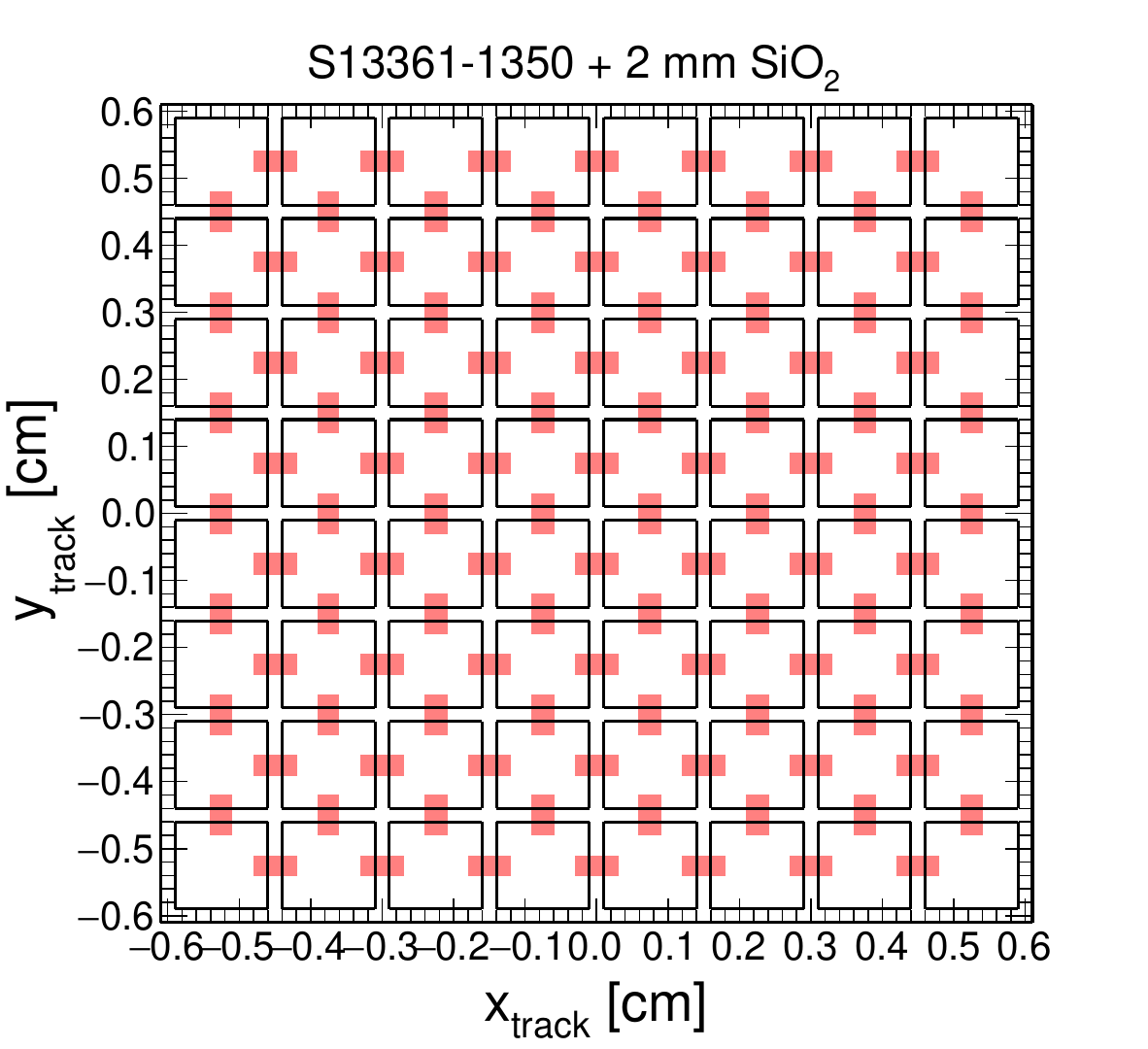}
\includegraphics[height=0.42\linewidth, width=0.49\linewidth,trim=0mm 0mm 15mm 7mm,clip]{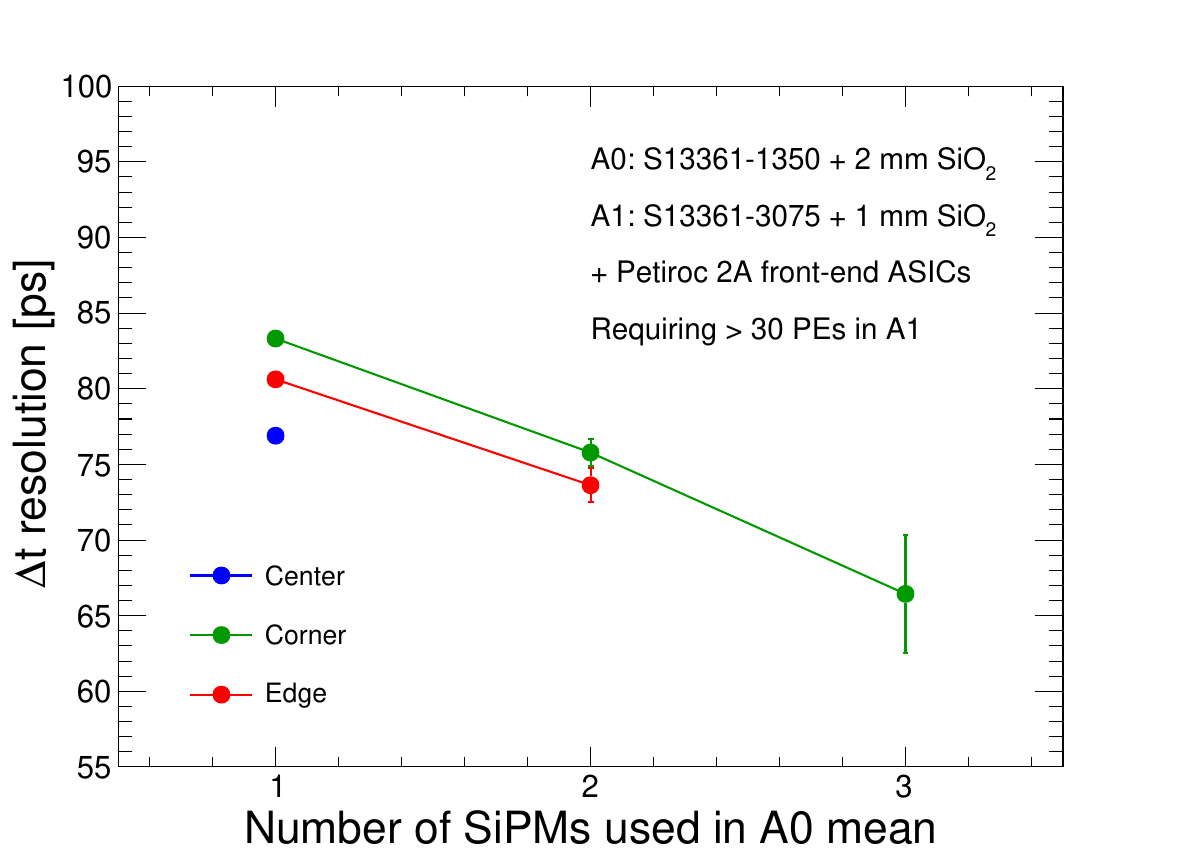}
\includegraphics[height=0.42\linewidth, width=0.49\linewidth,trim=0mm 0mm 15mm 7mm,clip]{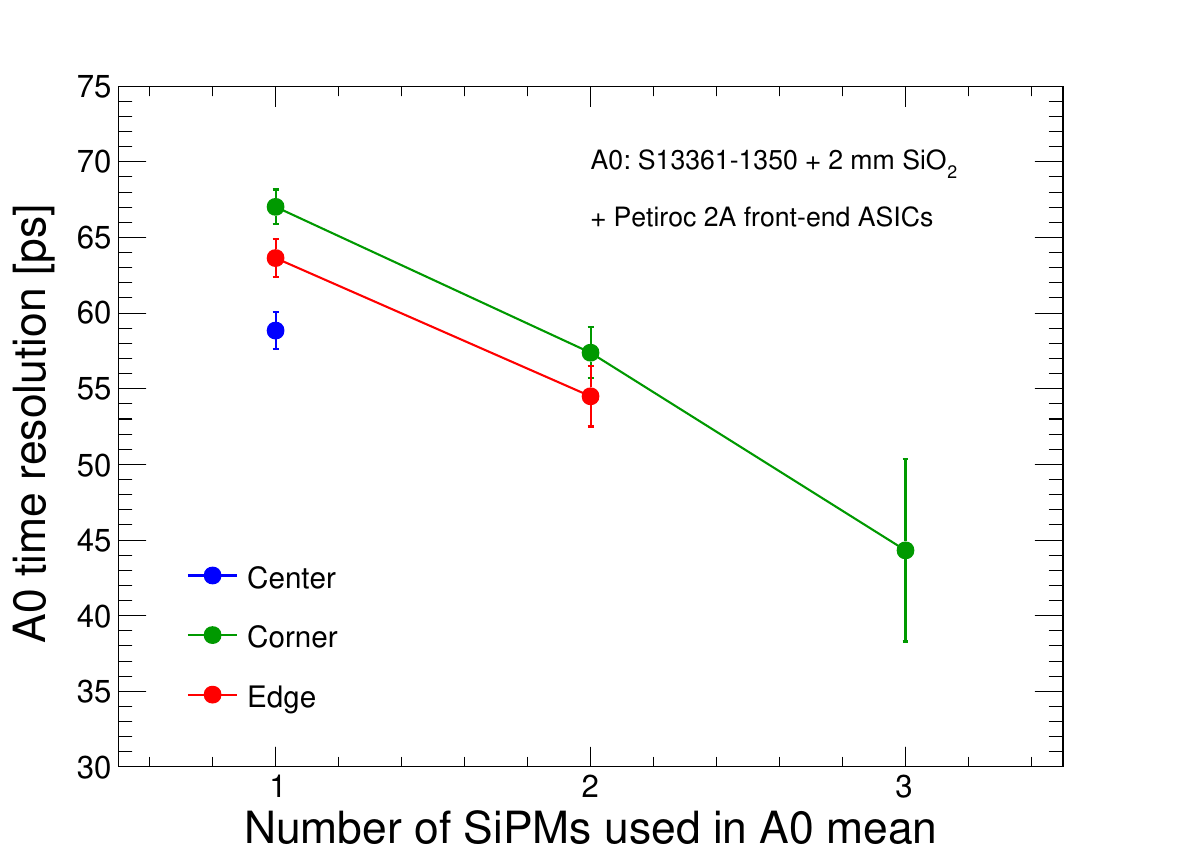}
\caption{Top: Illustration of the regions used for the time-resolution analysis as a function of the impact position in each SiPM: central region (left), corners shared by four adjacent SiPMs (middle), and edges shared by two adjacent SiPMs (right), for the S13361-1350 SiPM arrays. Bottom: Resolution of the time difference between the A0 mean ToA and the A1 ToA of the maximum-charge SiPM as a function of the number $N$ of SiPMs used in the A0 mean. Events with a maximum charge of at least 30~PEs in A1 and a minimum charge of 9~PEs in each SiPM entering the A0 mean are considered. The plots refer to runs taken using Petiroc boards for the configuration with the S13361-3075 array coupled to a 1~mm thick SiO$_2$ window (A1) and the S13361-1350 array coupled to a 2~mm thick SiO$_2$ window (A0).  The right panel reports the resulting A0 resolution after subtracting in quadrature the A1 contribution of $(49.5\pm1.4)$~ps.
} 
\label{figure_Petiroc_scanXY_fiducial_regions_2}
\end{figure}

We investigated the spatial dependence of the time resolution for clusters up to three fired channels.
In order to address the possibility of achieving an improvement of the resolution as $1/\sqrt{N_{\text{SiPM}}}$ for $N_{\text{SiPM}}$ fired channels with a sufficient number of PEs, we focused on runs taken with the Petiroc boards.
We studied the clusters in the S13361-3075 array coupled to a 1 mm thick SiO$_2$ window and the S13361-1350 array coupled to a 2 mm thick SiO$_2$ window  considering the three regions illustrated in the top panels of Figures~\ref{figure_Petiroc_scanXY_fiducial_regions_1} and~\ref{figure_Petiroc_scanXY_fiducial_regions_2}, respectively. They correspond to the central regions of the SiPMs, the corners shared by four adjacent SiPMs, and the edges shared by two adjacent SiPMs.

We considered runs taken with one of these arrays and the S13361-3075 array coupled to a 1 mm thick window made either of MgF$_2$ or of SiO$_2$ used as a reference.
We selected events requiring the charged track crossing the region of interest in the array under study and releasing a maximum number of PEs larger than 30 in the reference array.  
We subsequently analyzed the clusters and  calculated the mean time the $N$ highest-charge SiPMs up to $N=3$ requiring more than 16 PEs for the SiPMs of the S13361-3075 array and more than 9 PEs for the SiPMs of the S13361-1350 array to enter the calculation of the mean. 
Finally we evaluated the difference relative to the time of the SiPM with maximum charge in the reference array and we built the corresponding distributions. 

The bottom-left panels of Figures~\ref{figure_Petiroc_scanXY_fiducial_regions_1} and \ref{figure_Petiroc_scanXY_fiducial_regions_2} show the resulting obtained from the Gaussian fits to the corresponding  distributions as a function of the number of SiPMs used in the mean for the three considered regions. Only points with enough available statistics are shown. The bottom-right panels of the same figures show the corresponding resolutions at the single array level by subtracting in quadrature the contribution of the referene array, $(49.5\pm 1.4)$~ps, calculated as discussed above. 

As expected, when considering only the SiPM with maximum charge, the best resolution is achieved for tracks incident at the SiPM center, while it degrades for incidence at the edges and is worst at the corners.
Including more SiPMs in the mean, the resolution at the edges and at the corners improves. The improvement is particularly visible for the S13361-1350 array with 2~mm SiO$_2$ due to the more uniform charge sharing among the SiPMs.
In particular, the observed resolution scaling at the corners of this array approaches the ideal $1/\sqrt{N_{\text{SiPM}}}$ dependence, improving from about 67~ps to about 44~ps.
Similar considerations also apply to the edges. 
For the S13361-3075 array we observed a softer dependence, attributed to the unbalanced sharing of the charge in combination to the limited 1 mm thickness of the window. 

\subsection{Comparison with a bare SiPM array}

\begin{figure}[!t]
    \centering
    \includegraphics[width=1.0\linewidth,trim=0mm 0mm 0mm 0mm,clip]{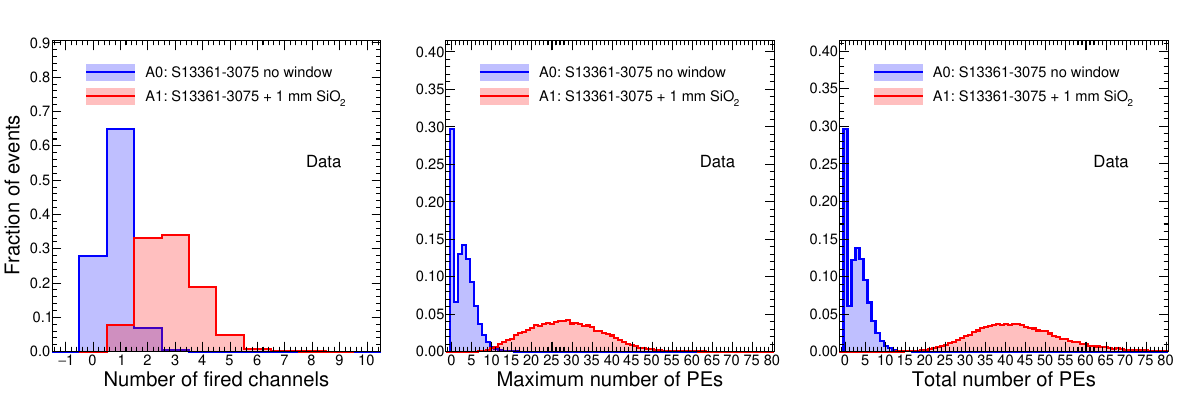}
    \includegraphics[height=0.42\linewidth,width=0.495\linewidth,trim=0mm 0mm 14mm 5mm,clip]{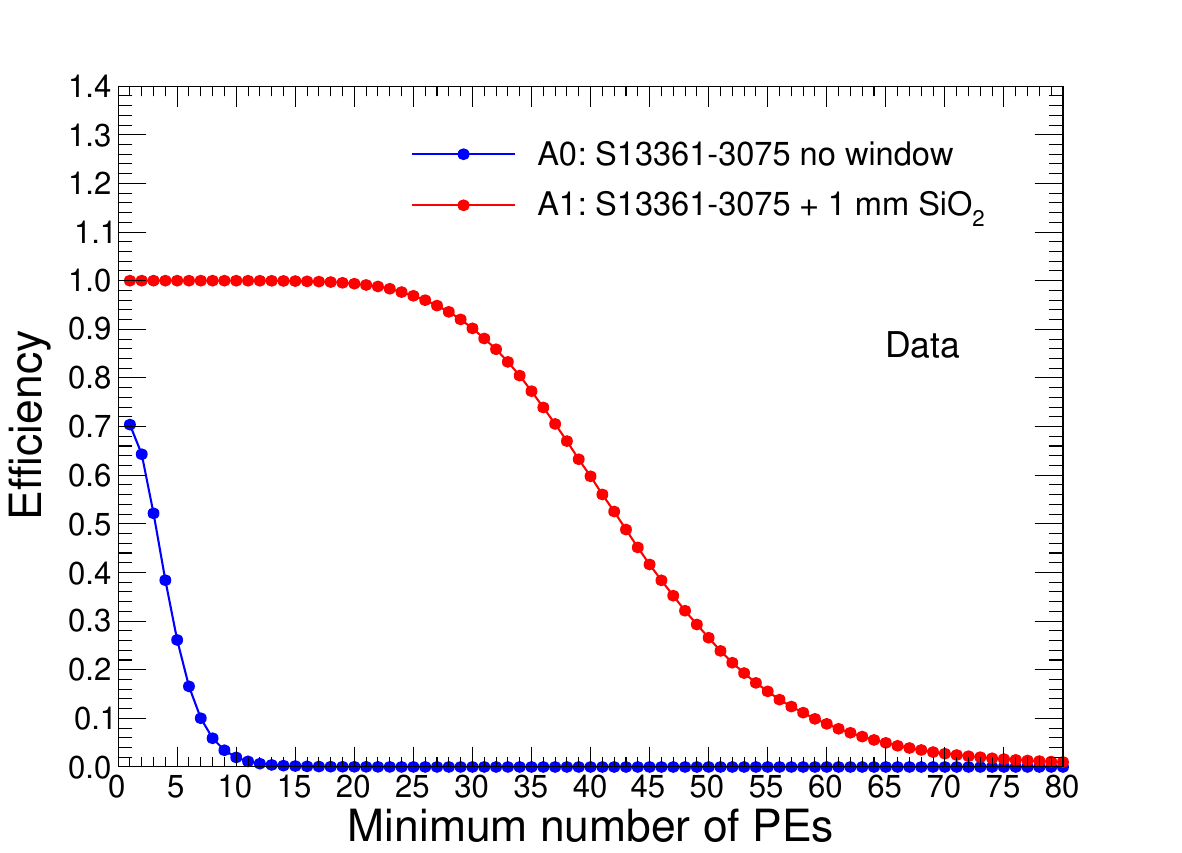}
    \includegraphics[height=0.42\linewidth,width=0.495\linewidth,trim=0mm 0mm 14mm 5mm,clip]{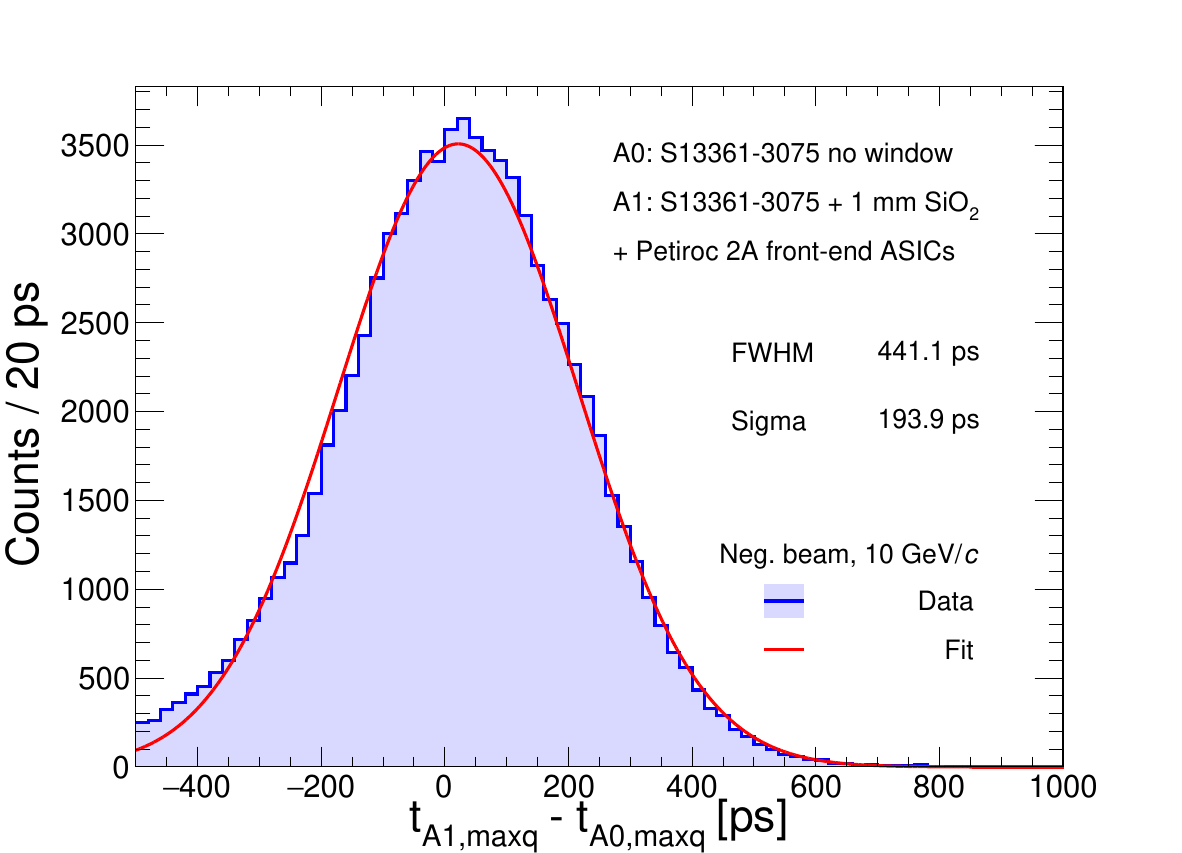}
    \caption{Measured charged-particle cluster topology (top), detection efficiency as a function of the minimum number of photoelectrons $(N_{\text{PE}} \geq1)$ required in the clusters (bottom left) and resolution of the time difference between the SiPMs with maximum charge (bottom right) for runs taken using Petiroc boards for the configuration with the S13361-3075 array coupled to a 1~mm thick SiO$_2$ window (A1) and the S13361-3075 array without any glued window (A0).
    }
    \label{fig:cell_and_charge_Petiroc_3x3vs3x3_nowindw}
\end{figure}

The impact of the Cherenkov signal from the tested windows, and the results discussed above, can be clearly shown by comparing them with the case of an array of bare SiPMs.
Figure~\ref{fig:cell_and_charge_Petiroc_3x3vs3x3_nowindw} shows the distributions of the number of fired channels and the charge in the clusters, the measured efficiency and the distribution of the time difference between the SiPMs with maximum charge for the configuration with S13361-3075 arrays coupled to a 1~mm thick SiO$_2$ window and no external window used as both A1 and A0, respectively. 
Despite the lower threshold set for the A0 array, as discussed in Section~\ref{sec:vov_and_th}, there are about 30\% of the events the array is not responding at all, and about 65\% where a single channel is fired, resulting in a limited charged particle detection efficiency.  
Nevertheless, a maximum charge corresponding to up to 10 PEs is still observed for the events where the array is responding. 
As confirmed by the comprehensive  Geant4 simulation of this configuration, these PEs result from both the Cherenkov emission in the 100~$\upmu$m thick built-in epoxy resin layer and the direct ionization loss in the silicon depleted region, as well as Cherenkov photons produced in the upstream Ar volume and the correlated cross-talk of the SiPMs.
The resolution of the time difference between the SiPMs with maximum charge in A0 and A1 is $\sigma_{\Delta t, \text{maxq}}\approx 193.9$~ps. By subtracting in quadrature the extrapolated contribution of A1, about 53.0~ps, we found a single-array time resolution $\sigma_{t, \text{maxq}}\approx186.5$~ps for the array without window. 
This result is significantly worse than the ones obtained for all configurations with a window, highlighting the substantial performance gains in terms of signal strength, efficiency and time resolution achievable with the considered timing approach when a radiator window is included.

\section{Discussion}

On the basis of the excellent performance in terms of efficiency and time resolution demonstrated in the previous section, the following considerations can be made.

The optimization of the sensor and radiator parameters (SiPM size and pitch, SPAD pitch, and window thickness) is crucial to achieve the best time resolution. 
Simulations and measurement results show that the optimal radiator configuration, both in spectral response and yield of detected photon yield, is a SiO$_2$ window coupled to SiPMs using silicone resin, which provides the best optical coupling and NUV transmittance.

Independently of the SiPM size, a window thickness of about 1 mm is sufficient to achieve full charged-particle detection efficiency operating with thresholds of up to tens of PEs, thereby suppressing background from the DCR even in high-radiation environments.

While smaller SiPMs are typically expected to provide better single-photon timing due to their lower terminal capacitance~\cite{Gundacker:2020cnv}, in the present multi-photon regime larger SiPMs  collect substantially more photoelectrons in the same pixel. 
Larger SiPMs yield the best time resolution when exploiting the channel with maximum charge in the clusters.

At the same time, smaller SiPMs result in a more uniform charge sharing across channels, which enables timing improvements by averaging the detection times of multiple SiPMs in the cluster. The observed time resolution exhibits the expected improvement with the number of fired SiPM channels, scaling approximately as $1/\sqrt{N_{\text{SiPM}}}$ within the experimental uncertainties. For larger SiPMs, instead, except for tracks incident near SiPM boundaries, neighbouring channels typically collect only a limited charge. Consequently, including their timing information could even worsen the time resolution.
A more uniform charge sharing is achieved by increasing the ratio between the window thickness and the SiPM pitch in the array.
The performance is also influenced by the larger SPAD pitch, which results in a larger fill factor and, consequently, a higher PDE. This effect can be clearly observed by comparing the results achieved with the S13361-3075 and S13361-1350 arrays.

The comparison between the results achieved with Petiroc and RadioPico boards shows that, to fully exploit the intrinsic timing potential of the proposed approach, the use of front-end electronics with very low jitter is mandatory.
Using the RadioPico boards, we achieved the best time resolution reported in this study, about 33.2~ps, for the S13361-3075 arrays coupled to a 1~mm thick SiO$_2$ window.
This result might be further improved with larger calibration statistics, which will enable a more precise offset equalization over the full array and the inclusion of the dependence on the track impact position in the considered lookup-table procedure.

The ultimate performance of the tested setup is affected by signal-integrity limitations of the test-beam cabling scheme, including the HLCD-20 cables and the feedthrough board, which degrade the analogue signal quality with respect to an ideal direct coupling, especially at low photoelectron yields.

Further improvements in both efficiency and timing are expected with dedicated developments, such as custom ARCs optimized for the wavelength-dependent Cherenkov spectrum, the polarization of Cherenkov photons, and the SiPM spectral response, designed to maximize photon collection over the relevant charged track and photon incidence-angle distributions by reducing reflections and increasing the effective SiPM PDE.

As a final consideration, it is worth highlighting that in order to preserve the overall performance while integrating TOF devices into a full HEP experiment requires the implementation of appropriate design optimizations.

\section{Conclusions}

We have shown that time resolutions of few tens of ps can be readily achieved with full charged particle detection efficiency by detecting the Cherenkov photons emitted by charged particles traversing a thin slab of fused silica viewed by a SiPM array.  Devices based on this approach represent a step forward in particle identification systems
for future HEP experiments planning to extend the identification of charged particles to higher momenta. Along the collider luminosity increase, the developed timing sensors will also be compulsory for pile-up suppression and 4-D tracking.



\vspace{6pt} 


\section*{Author Contributions}
Conceptualization, M.N.M., G.D.R., F.L., E.N. and N.N.; 
methodology, M.N.M., L.C., G.D.R., F.L., E.N., N.N. and R.P.;
software, L.C., G.D.R., M.G., A.L., L.L., N.N., G.P. and R.P.; 
validation, M.G., A.L., L.L., N.N., G.P. and R.P.;
formal analysis, M.G., A.L., L.L., N.N., G.P. and R.P.;
investigation, M.N.M., M.G., A.L., R.L., L.L., N.N., G.P. and R.P.; 
resources, A.D.M., L.C., M.N.M., G.D.R., M.G., F.L., A.L., L.L., N.N., G.P. and R.P.; 
data curation, M.N.M., M.G., A.L., L.L., N.N., G.P. and R.P.; 
writing---original draft preparation, M.N.M., E.N., N.N. and R.P.; 
writing---review and editing, M.N.M., E.N. and N.N.; 
visualization, M.N.M., N.N. and R.P.; 
supervision, M.N.M., R.P.; 
project administration, A.D.M., M.N.M., G.V.; 
funding acquisition, M.N.M.
All authors have read and agreed to the published version of the manuscript.

\section*{Funding}
This research received no external funding.

\section*{Data Availability Statement}
The datasets generated during and/or analyzed during the current study are available from the corresponding author on reasonable request.

\section*{Acknowledgments}
The authors would like to thank the INFN Bari staff for their contribution to the procurement and to the construction of the prototype. In particular, we thank D. Dell’Olio, M. Franco, N. Lacalamita, F. Maiorano, M. Mongelli, C. Pastore and R. Triggiani for their technical support. The authors would like to thank Weeroc for contributing to the development of the Radioroc~2/picoTDC board and providing support for the operation of Radioroc~2. They also thank the CERN beam team group for providing the facilities and logistical support for the test.
This work was carried out in the context of the DRD4 Collaboration based at CERN.

\section*{Conflicts of Interest}
The authors declare no conflicts of interest.

\section*{Abbreviations}
The following abbreviations are used in this manuscript:
\begin{center}
\begin{tabular}{@{}ll@{}}
ADC  & Analog to Digital Converter \\
DAQ  & Data Acquisition \\
FEB  & Front-End Board \\
FF   & Fill factor \\
IF   & Integration factor \\
LSB  & Least significant bit \\
LUT  & Lookup Table \\
MPPC & Multi-Pixel Photon Counter \\
NaF  & Sodium Fluoride \\
PE   & Photoelecton \\
PDE  & Photo Detection Efficiency \\
PID  & Particle Identification \\
RICH & Ring Imaging CHerenkov \\
SPAD & Single-Photon Avalanche Diode \\
SiPM & Silicon PhotoMultiplier \\
TOF  & Time of Flight \\
ToA  & Time of Arrival \\
ToT  & Time over Threshold \\
\end{tabular}
\end{center}

\appendix
\section{Modeling the SiPM reflectance}

\label{appending_model_reflections}

On the basis of hypotheses about the sources of specular and diffuse reflection in the tested arrays, we formulated a simple model for the specular and diffuse reflectance, $R_{\text{spec}}$ and $R_{\text{diff}}$.
As a first approximation, specular reflections are assumed within the SiPM active regions, while diffuse reflections are assumed in the dead regions of a SiPM or  between adjacent SiPMs in the array.
Let g denote the gas surrounding the array, w the window, and r the protective resin. Let $\mathrm{FF}$ be the fill factor, i.e. the fraction of the active area of a single SiPM, and $\mathrm{IF}$ the integration factor, i.e. the fraction of the array area covered by the SiPMs. Our model is based on the following assumptions:

\begin{itemize}
    \item Only specular reflection $R_{x\rightarrow y}$ with $\text{x},\text{y}=\text{g},\text{w},\text{r}$ occurs at the smooth interfaces between gas and resin, gas and window, and window and resin.
    \item Only specular reflection $R_{r\rightarrow \mathrm{ARC}\rightarrow \mathrm{Si}}$ occurs at the resin–ARC–passivation-silicon interfaces within the active area, corresponding to a fraction $\mathrm{IF}\cdot \mathrm{FF}$ of the array area.
    \item Only diffuse reflection $R_{\mathrm{diff,1}}$ occurs in the dead area within any SiPM, corresponding to a fraction $\mathrm{IF}\,(1-\mathrm{FF})$ of the array area.
    \item Only diffuse reflection $R_{\mathrm{diff,2}}$ occurs in the dead area between adjacent SiPMs, corresponding to a fraction $1-\mathrm{IF}$ of the array area.
\end{itemize}
The following expressions are derived for the reflectances $R_{\text{spec}}$ and  $R_{\text{diff}}$ for the wavelength region where absorption effects in the window or protective resin are negligible:
\begin{equation}
\begin{aligned}
R_{\text{spec}} ={}& R_{\text{g}\rightarrow\text{w}} + (1-R_{\text{g}\rightarrow\text{w}})R_{\text{w}\rightarrow\text{r}} + (1-R_{\text{g}\rightarrow\text{w}})(1-R_{\text{w}\rightarrow\text{r}}) \\ &  \cdot \text{IF}  \cdot \text{FF}\cdot R_{\text{r}\rightarrow\text{ARC}\rightarrow\text{Si}}\cdot (1-R_{\text{r}\rightarrow\text{w}})(1-R_{\text{w}\rightarrow\text{g}}),
\end{aligned}
\label{equation_specular_reflectance_model}
\end{equation}
\begin{equation}
\begin{aligned}
R_{\text{diff}} ={}& (1-R_{\text{g}\rightarrow\text{w}})(1-R_{\text{w}\rightarrow\text{r}}) \cdot [ (1-\text{IF})\cdot R_{\text{diff,2}} \\ & + \text{IF}\cdot(1-\text{FF})\cdot R_{\text{diff,1}}]\cdot(1- \langle R_{\text{diff,r}\rightarrow\text{w}\rightarrow{\text{g}}}\rangle  ).
\end{aligned}
\label{equation_diffuse_reflectance_model}
\end{equation}
The terms $R_{\text{x}\rightarrow\text{y}}$ are obtained from Fresnel equations~\cite{steck2006classical} using the known refractive indices of the media $\text{x}$ and $\text{y}$, the photon incidence angle, and the transmission angles in the considered media determined via Snell's law.
For arrays without resin, the corresponding equations are obtained by setting $\text{w}=\text{g}$ and $R_{\text{g}\rightarrow\text{g}}=0$.

The term $\langle R_{\text{diff,r}\rightarrow\text{w}\rightarrow\text{g}}\rangle$ represents the mean reflectance for photons that undergo a diffuse reflection in the dead area within a SiPM or between adjacent SiPMs, possibly followed by specular reflections at the resin, window, and gas interfaces. 
For the studies carried out using this model, we evaluated it with dedicated Monte Carlo simulations assuming, as a first approximation, an isotropic random reflection angle in the dead regions within the array.

The terms $R_{\text{r}\rightarrow\text{ARC}\rightarrow\text{Si}}$, $R_{\text{diff,1}}$, and $R_{\text{diff,2}}$ are not known a priori due to the limited information on the SiPM ARC and the microstructures in the dead regions. However, $R_{\text{r}\rightarrow\text{ARC}\rightarrow\text{Si}}$ can be easily inferred from the measured specular reflectance inverting Eq.~\ref{equation_specular_reflectance_model}:

\begin{equation}
 R_{\text{r}\rightarrow\text{ARC}\rightarrow\text{Si}} = \frac{R_{\text{spec}} - R_{\text{g}\rightarrow\text{w}} - (1-R_{\text{g}\rightarrow\text{w}})R_{\text{w}\rightarrow\text{r}} }{(1-R_{\text{g}\rightarrow\text{w}})(1-R_{\text{w}\rightarrow\text{r}})  \cdot \text{IF}  \cdot \text{FF} \cdot (1-R_{\text{r}\rightarrow\text{w}})(1-R_{\text{w}\rightarrow\text{g}})}
    \label{equation_ARC_extrapolation}
\end{equation}
The expression for arrays without resin is obtained by setting $\text{w}=\text{g}$, giving $R_{\text{g}\rightarrow\text{g}}=0$.

\bibliographystyle{unsrt} 
\bibliography{SiPM_ToF}

\end{document}